\pdfoutput=1

\documentclass[onluarkali,ingilizce,lisans,bez,elektrikelektronik]{thesis_itu}

\yazarbir{Mehmet Eymen}{ÜNAY}
\ogrencinobir{040190218}
\yazariki{Bora}{İNAN}
\ogrencinoiki{040190205}
\yazaruc{Emrecan}{YİĞİT}
\ogrencinouc{040190203}

\unvan{}

\anabilimdali{Elektronik ve Haberleşme Mühendisliği}{Electronics and Communication Engineering}
\programi{Elektrik Mühendisliği Programı}{Electrical Engineering Programme}

\tarih{Haziran 2023}{June 2023}
\tarihKucuk{Haziran 2023}{June 2023}

\tezyoneticisi{Dr. Tankut AKGÜL}{Istanbul Technical University}   

\baslik{LLVM DERLEYİCİSİYLE RISC-V İŞLEMCİ İÇİN}
{EK BUYRUKLARIN DESTEKLENMESİ}
{}

\title{SUPPORTING CUSTOM INSTRUCTIONS WITH THE LLVM COMPILER} %
{FOR RISC-V PROCESSOR}
{}

\tezvermetarih{Haziran 2023}{June 2023} 

\tezsavunmatarih{Haziran 2023}{June 2023}

\esdanismani{}{}   

\juriBir{Prof. Dr. Name SURNAME}{Middle East Technical University}  

\juriIki{Prof. Dr. Name SURNAME}{Boğaziçi University}  

\juriUc{Prof. Dr. Name SURNAME}{Bilkent University}  

\juriDort{Prof. Dr. Name SURNAME}{Sabancı University}  

\juriBes{Prof. Dr. Name SURNAME}{Koç University}

\usepackage[T1]{fontenc} 				%
\usepackage[utf8]{inputenc}  			%
\usepackage[open,openlevel=1]{bookmark} %
\usepackage{listings} %
\usepackage{xcolor}

\usepackage{color}
\usepackage{times}
\usepackage{amssymb,amsmath,mathptmx,amsbsy,bm}
\usepackage{caption}            
\usepackage{graphics}
\usepackage{wrapfig}
\usepackage{epsfig}
\usepackage{enumerate}
\usepackage{rotating}
\usepackage{multirow}					%
\usepackage{colortbl}
\usepackage{pstricks}
\usepackage{pst-plot}
\usepackage{cite}
\usepackage{latexsym}
\usepackage{hyperref}\hypersetup{hidelinks} %
\usepackage{fixltx2e} %
\usepackage{xpatch} 		%
\usepackage[normalem]{ulem} %
\usepackage[bottom,multiple]{footmisc} 	%
\usepackage{enumitem} 					%
\renewcommand\labelitemi{\normalsize$\bullet$} %
\usepackage{pdfpages} 					%
\usepackage{tikzpagenodes} 				%
\usepackage{etoolbox}					%

\usepackage{ragged2e}
\justifying
\usepackage{parskip}
\usepackage{multicol}
\usepackage{llvm-lang}  %
\usepackage{nasm-style} %
\usepackage{csquotes}

\definecolor{codegreen}{rgb}{0,0.6,0}
\definecolor{codegray}{rgb}{0.5,0.5,0.5}
\definecolor{codepurple}{rgb}{0.58,0,0.82}
\definecolor{backcolour}{rgb}{0.95,0.95,0.92}
\lstdefinestyle{mystyle}{
    backgroundcolor=\color{backcolour},
    commentstyle=\color{codegreen},
    keywordstyle=\color{magenta},
    numberstyle=\tiny\color{codegray},
    stringstyle=\color{codepurple},
    basicstyle=\ttfamily\footnotesize,
    breakatwhitespace=false,
    breaklines=true,
    captionpos=b,
    keepspaces=true,
    numbers=left,
    numbersep=5pt,
    showspaces=false,
    showstringspaces=false,
    showtabs=false,
    tabsize=2
}
\makeatletter 															%
\patchcmd{\@makechapterhead}{\vspace*{50\p@}}{}{}{}						%
\patchcmd{\@makeschapterhead}{\vspace*{50\p@}}{\vspace*{21.5mm}}{}{}	%
\makeatother
\usepackage{longtable} 	%
\usepackage{hhline} 	%
\usepackage{siunitx}
\usepackage[list=true,listformat=simple,position=below]{subcaption} %
\DeclareCaptionLabelFormat{subfig}{\normalsize\figurename #1~\arabic{chapter}.\arabic{chapter}\alph{subfigure} :}
\makeatletter
\def\tagform@#1{\maketag@@@{\ignorespaces#1\unskip\@@italiccorr}} %
\renewcommand{\eqref}[1]{\textup{\bf(\ref{#1})}} %
\makeatother

\def\be{\begin{equation}} %
\def\ee{\end{equation}}%
\def\beq{\begin{eqnarray}}%
\def\eeq{\end{eqnarray}}%
\def\bse{\begin{subequations}}%
\def\ese{\end{subequations}}%
\def\[{\left[}
\def\]{\right]}
\def\({\left(}
\def\){\right)}

\ithaf{}

\kisaltmalistesi{\begin{tabular}{@{}p{2cm}l}
{\bf{ABI}} & {\bf:} Application Binary Interface\\
{\bf ALU} & {\bf:} Arithmetic Logic Unit\\
{\bf ASIP} & {\bf:} Application Specific Integrated Processor\\
{\bf AST} & {\bf:} Abstract Syntax Tree\\
{\bf CPU} & {\bf:} Central Processing Unit\\
{\bf DAG} & {\bf:} Directed Acyclic Graph\\
{\bf DCE} & {\bf:} Dead Code Elimination\\
{\bf DSE} & {\bf:} Dead Store Elimination\\
{\bf FPGA} & {\bf:} Field Programmable Gate Arrays\\
{\bf GCC} & {\bf:} GNU Compiler Collection\\
{\bf IC} & {\bf:} Integrated Circuit\\
{\bf IR} & {\bf:} Intermediate Representation\\
{\bf ISA} & {\bf:} Instruction Set Architecture\\
{\bf IoT} & {\bf:} Internet of Things\\
{\bf LLVM} & {\bf:} Low Level Virtual Machine\\
{\bf LSB} & {\bf:} Least Significant Bit\\
{\bf MC} & {\bf:} Machine Code\\
{\bf MSB} & {\bf:} Most Significant Bit\\
{\bf RISC} & {\bf:} Reduced Instruction Set Computer\\
{\bf SDnode} & {\bf:} SelectionDAG node\\
{\bf SDvalue} & {\bf:} SelectionDAG value\\
{\bf SSA} & {\bf:} Static Single Assignment\\
\end{tabular}

}
\sembollistesi{\begin{tabular}{@{}p{2cm}l}
{$\boldsymbol\oplus$} & {\bf:} XOR\\
{$\boldsymbol\land$} & {\bf:} AND\\
{$\boldsymbol\lor$} & {\bf:} OR\\
{$\boldsymbol\lnot$} & {\bf:} NOT\\
\end{tabular}

}

\onsoz{\vspace*{-6pt}
At the outset we would like to express our sincere thanks and gratitude to our project advisor Dr. Tankut Akgül who kindly provided us  academic support and guidance throughout our efforts. Without his support and encouragement we would not be able to overcome the difficulties encountered over the course of our studies.  We would like to express our gratitude to him for helping us make progress with the project. We would also like to take this opportunity to express our sincere gratitude and appreciation to Prof. Sıddıka Berna Örs Yalçın for shaping our targets and supporting us. LLVM community was greatly helpful and thus, we are thankful to them as well. Finally, we would like to thank our families and friends who did not spare their moral support during our university education.

}             
\ozet{Özel buyruklara sahip donanım hızlandırıcılarının yükselişi, bu hızlandırıcıları destekleyen özel derleyici arka uçlarını gerektirmektedir. Bu çalışma, LLVM ve LLVM RISC-V arka ucunun ayrıntılı analizlerini sunmakta ve söz konusu dönüşümlere uçtan uca genel bakış sağlayan vaka çalışmalarıyla desteklenmektedir. 

Buyruk tasarımının hem donanım hem de yazılım tasarım alanında dikkate alınması gerektiğini düşünüyoruz. Gerekli derleyici değişiklikleri, buyruğun iyi tasarlanmadığı ve yeniden gözden geçirilmesi gerektiği anlamına gelebilir. RISC-V standart uzantılarının buyruk tasarımcılarına rehberlik edebilecek örnek buyruklar sağladığını tartışıyoruz.

Bu çalışmada, derleyiciye özel bir buyruk ekleme süreci çevirici desteği ve örüntü eşleştirme desteği olarak iki kısma ayrılmıştır. Örüntü eşleştirme desteği olmadan, geleneksel yazılımlar hızlandırıcı için elle satır arası çevirme dili girişleri gerektirir ve bu da ölçeklenebilir değildir. Buyruk semantiğinden bağımsız olarak çevirici desteği eklemek basit olsa da, örüntü eşleştirme desteğinde durum tam tersidir. Örüntü eşleştirme desteği ve değişiklik için doğru derleyici aşamasını seçmek, derleyicideki iç dönüşümlerin bilinmesini gerektirir. Bu çalışma, örüntü eşleştirme konusunu derinlemesine incelemekte ve örüntü eşleştirme desteği sorununa çeşitli yaklaşımlar sunmaktadır. Örüntünün karmaşıklığına bağlı olarak, daha yüksek seviyeli dönüşümlerin, örneğin Ara Form seviyesinin, Buyruk Seçimi aşamasına kıyasla daha sürdürülebilir olabileceği tartışılmaktadır.

}               
\summary{
The rise of hardware accelerators with custom instructions necessitates custom compiler backends supporting these accelerators. This study provides detailed analyses of LLVM and its RISC-V backend, supplemented with case studies providing end-to-end overview of the mentioned transformations. 

We discuss that instruction design should consider both hardware and software design space. The necessary compiler modifications may mean that the instruction is not well designed and need to be reconsidered. We discuss that RISC-V standard extensions provide exemplary instructions that can guide instruction designers.

In this study, the process of adding a custom instruction to compiler is split into two parts as Assembler support and pattern matching support. Without pattern matching support, conventional software requires manual entries of inline Assembly for the accelerator which is not scalable. While it is trivial to add Assembler support regardless of the instruction semantics, pattern matching support is on the contrary. Pattern matching support and choosing the right stage for the modification, requires the knowledge of the internal transformations in the compiler. This study delves deep into pattern matching and presents multiple ways to approach the problem of pattern matching support. It is discussed that depending on the pattern's complexity, higher level transformations, e.g. IR level, can be more maintainable compared to Instruction Selection phase.

}             

\begin{document}

\chapter{INTRODUCTION}\label{Ch1}

Recent advances and studies on Integrated Circuits (IC) caused technology to produce application-specific circuits for various areas of usage. Extensions for the open source processor architectures became a part of the industrial development. More custom accelerators are developed, especially with RISC-V open and modular instruction set architecture (ISA). Hardware accelerators have the promise of being fast and efficient.

However, loading new abilities to an extended processor comes with a problem. Programming languages and their compilers are developed for common architectures. A compiler targeting standard ISA will not produce the custom instructions for the accelerator. A compiler modification is needed to be able to introduce the accelerator to the high-level languages. In this thesis, we show various ways to approach the problem and present the best practices for it.

For the research, several accelerators with specific custom instructions are targeted \cite{Sairoglu, eryilmaz}. Instructions which are targeted to hardware are SHLXOR, RORI and S-box. The encodings and instruction operations were mostly designed by the hardware developers. The process of required compiler modifications for SHLXOR and RORI are demonstrated in Sections \ref{sec:shlxor} and \ref{sec:rori}. S-box instruction, due to its non-linearity, was a complicated instruction to characterize. It is a good example that not every instruction can be added in a similar process and instruction-specific design can be required. 
Also similar to the design of ISAs, instructions should be designed by considering both hardware and software. 

S-box instruction is analyzed from several aspects. Firstly, the Intermediate Representation (IR) optimizations it gets through are demonstrated in depth in Section \ref{sbox-case}. Secondly, the limitations of TableGen which was a sufficient system for the previous instructions, are discussed and C++ pattern matching is explained in Section \ref{sec:cpp}. 
Thirdly, pattern matching in IR and MCInst level are discussed in Section \ref{sec:patmatchdisc}. Finally, we proposed two new instructions that can be implemented in hardware that can accelerate S-box operation as LXR and NAXOR. A simplified version of LXR which has independent Load addresses is demonstrated in Section \ref{sec:lxr}. The S-box case where the load addresses are dependent is presented in C++ pattern matching in Section \ref{sec:cpp}. The second proposed instruction, NAXOR, is shown in Section \ref{sec:naxor}.

In conjunction with LXR and NAXOR which do not have target hardware, MLA instruction is also presented without target hardware. MLA is discussed in detail in Chapter \ref{Ch4} where it is traced from the C code to Assembly in steps of compilation and Section \ref{sec:MLA_add_section} where its support was added with TableGen.

\section{Purpose of Project}
Application-Specific Instruction Set Processors (ASIP) are becoming more popular with the development of embedded systems. The specialization of the core causes a tradeoff between flexibility and performance. For special purposes, using ASIPs increases efficiency, however, we can program a custom ASIP only by using assembly instructions that we defined. Programming custom processors with assembly language is not a preferred way of coding. We are also not able to use high-level languages because compiling tools are designed for common architectures with certain instructions. The ability to add custom instructions to compilers will enable us to make more use of custom hardware designs.

ASIPs are feasible for all application-specific embedded systems like consumer, industrial, automotive, home appliances, cryptology, medical, telecommunication, commercial, aerospace, and military applications. The custom back-end that we will design under the supervision of Dr. Tankut Akgül, is going to serve the processor designed by Prof. Dr. Sıddıka Berna Örs Yalçın’s research team. When the project is completed, Prof. Yalçın is going to be able to produce the assembly codes that are compatible with the processor’s extended instruction set in addition to RISC-V.

Prof. Yalçın and her team are designing application-specific instruction set processors. The purpose of this project is to create a compiler back-end for a processor that supports custom instructions on top of RISC-V instructions. This compiler is going to help to program the custom processor by using high-level languages. Existing RISC-V compilers are not able to produce efficient assembly codes for ASIPs. Therefore a need arose for a compiler back-end.
The main reason for choosing this project is that we wanted to meet an actual need for a critical existing problem. The project has the potential to be the bridge between hardware and software of custom hardware projects in research, enabling them to be candidates for production use cases. 

\cleardoublepage
\clearpage
\chapter{BASICS OF A COMPILER}\label{Ch2}
\begin{figure}
    \centering
    \includegraphics[scale=0.25]{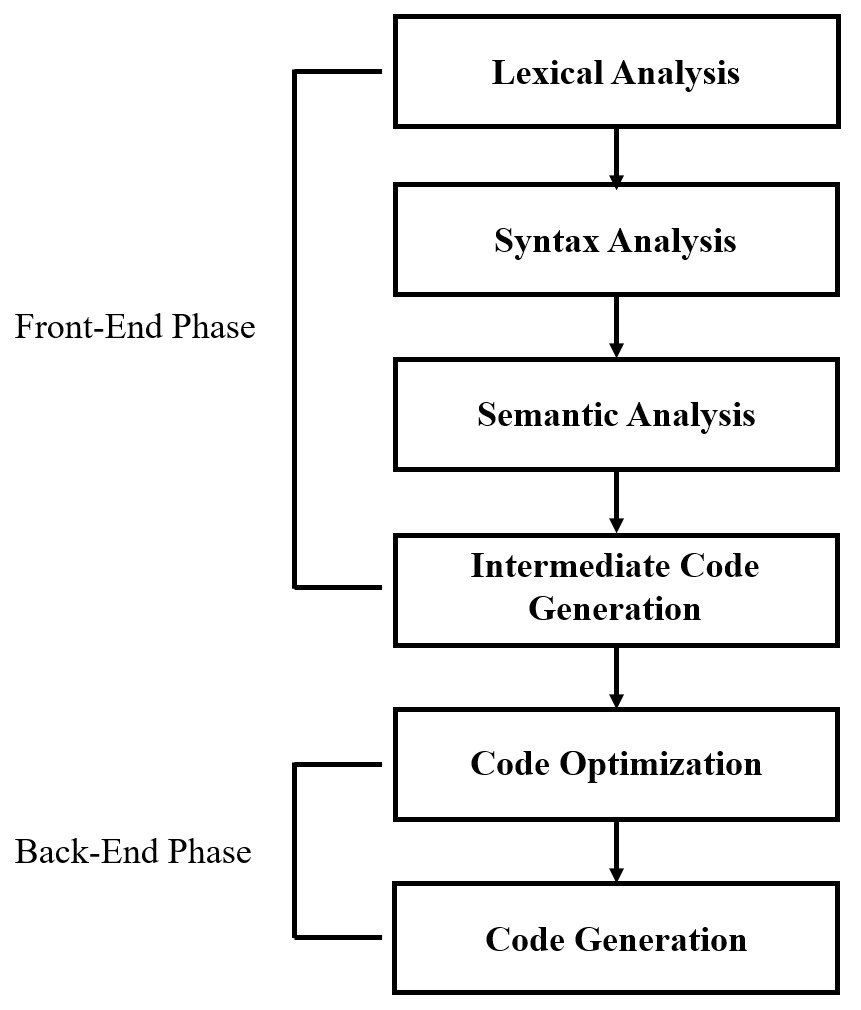}
    \caption{Compiler Stages}
    \label{fig:comp_stages}
\end{figure}

A compiler is a software that converts source code written in a high-level programming language into machine code (MC) appropriate for a particular computer architecture.
There are different stages of a compiler but they can be grouped into two main parts such as “Front-End” and “Back-End”. %
These parts of the compiler are also called the analysis and synthesis parts of the compiler. The analysis stage separates the source program into its individual components and applies a grammatical structure to them. The source code is then represented in an intermediate stage using this structure. The synthesis phase creates the final target program by using the IR. We can think of the compilation process as a series of phases, each of which takes the source program and transforms it into another representation \cite{compileralfredaho}. These phases can be seen in Figure \ref{fig:comp_stages}.

\section{Front-End}

\subsection{Lexical Analysis}
The compiler breaks down the source code into smaller units called lexemes, which are pieces of code that correspond to specific patterns in the code. These lexemes are then converted into tokens that can be used for syntax and semantic analyses.

\subsection{Syntax Analysis}
The compiler checks that the code follows the proper syntax for the programming language it is written in. This process is also called parsing. As part of this step, the compiler often creates abstract syntax trees (AST) in order to represent the logical structure of different parts of the code.

\subsection{Semantic Analysis}
The compiler checks that the code makes logical sense, going beyond syntax analysis by ensuring that the code is correct. For example, the compiler might check that variables have been declared correctly and given the appropriate data types. This process is known as semantic analysis.

\subsection{IR Code Generation}
After the source code has been analyzed for lexemes, syntax, and semantics, the compiler creates an IR of the code.
This intermediate code is going to be converted to machine code in the last two phases. These two phases are platform-dependent, meaning they are specific to a particular hardware architecture but the previous phases were not. Therefore, to create a new compiler, it is not necessary to start from scratch. Instead, it is possible to use the intermediate code from an existing compiler and build the final stages of the process for a specific platform. Because of that, we are interested in the back-end part for our project.

\section{Middle-End}

\subsection{Optimization}
The intermediate code is prepared for the final code generation step. This process does not change the meaning or functionality of the code, but it can make the program run faster and more efficiently.
A Directed Acyclic Graph (DAG) used in the compiler design process might represent the dependencies between different instructions in the IR code, such as the order in which those instructions need to be executed or the data dependencies between them. The DAG can be used by the compiler to identify opportunities for optimization, such as removing unnecessary instructions, combining some of them, or rearranging the order of execution to reduce the number of resources required by the code.
Static single assignment (SSA) is also an important part of optimization. It is a technique that is used to organize the IR in a way that ensures each variable is assigned a value only once and that each variable is defined before it is used.

\section{Back-End}

\subsection{Target Code Generation}
The target code generator is the final stage of the compilation process, and its main function is to convert the optimized code into a form that the machine can understand. The optimized code is turned into a relocatable machine code. The relocatable machine code is the input to the linker and loader, which are responsible for combining the code with other necessary resources and preparing it for execution %
Target code generation can be divided into different parts:

\subsection{Instruction Selection} 
IR is the input of the code generation step, and it maps the IR into the target machine’s instruction set. There may be multiple ways for converting one representation, so the code generator tries to select the most suitable instructions.

\subsection{Register Allocation}
There may be many different variables/values in a program. The code generator decides which registers to use to keep these values.

\subsection{Instruction Scheduling}
The code generator determines the sequence in which instructions will be executed and creates schedules for the execution of those instructions.

\cleardoublepage
\clearpage
\chapter{THE LLVM COMPILER}\label{ch:Ch3}

LLVM is a collection of modular and flexible libraries and a toolchain software that can be used to build a wide variety of compilers and other tools. LLVM compilers consist of a set of libraries that implement the parts of a compiler. There are different front-end libraries a for every language and different back-end libraries for every architecture. There is only one common IR optimizer that connects specific front end and back end.

\begin{figure}
    \centering
    \includegraphics[scale = 0.8]{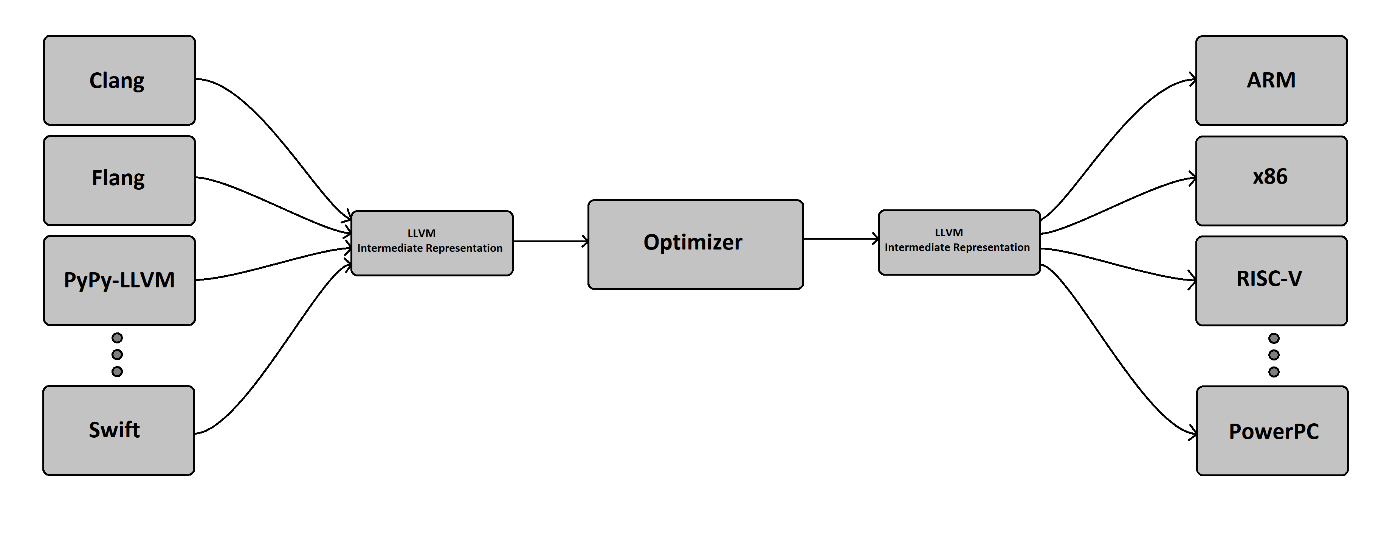}
    \caption{Front-end and Back-end libraries connected by LLVM}
    \label{fig:llvm_diagram}
\end{figure}

LLVM IR is the common target of programming languages and is the source for every target hardware. Various LLVM front-ends translate related languages into IR. Related back-end compiles IR into assembly according to the target hardware. This structure helps to increase flexibility between front-ends and back-ends. With this structure, we are able to have compilers for every combination of $M$ source codes and $N$ targets with $M$ front-end and $N$ back-end instead of $MxN$ compilers. In our case, we do not have to deal with the front end as our customised back end will be lowering any input programming language thanks to LLVM IR. The front-end we use in the development process will be Clang which is the LLVM C/C++ front-end \cite{clang}.

In LLVM, intrinsic functions are internal functions and they have their semantics directly defined by LLVM itself. LLVM provides both target-independent and target-specific intrinsics \cite{llvmdevmeeting}. These intrinsics have well-known semantics and names and they must adhere to certain restrictions. In general, these intrinsics serve as an expansion mechanism for the LLVM language that does not necessitate modifying all of the transformations in LLVM when introducing changes to the language. Intrinsic function names start with “llvm.” \cite{googlesiteintrinsic}.

Clang supports the notion of builtin functions used in GNU Compiler Collection (GCC). Some of these have the same syntax as in GCC to not disrupt portability. In addition to these, Clang supports other builtin functions that GCC doesn’t. Some of these are \_\_builtin\_shufflevector, \_\_builtin\_unreachable etc. As we can see, these builtin functions start with double underscores. 

Intrinsic functions and builtin functions are two separate things and should not be confused. Builtins are at C level (source code) while intrinsics are LLVM IR level and are not exposed to the user. A builtin function may or may not be expanded into intrinsic calls. It is possible to connect a new builtin function in the front-end to an intrinsic in the middle-end.

\section{Parts of the Clang Front-end}

\subsection{Clang Lex Library}
Clang Lex Library is a typical lexer implemented as finite state machines that read source code one character at a time and transition between different states based on the characters read. The Clang lexer, which is a front-end compiler for the C, C++, and Objective-C programming languages, uses this approach to filter out comments and white space, recognize and tokenize language elements such as keywords, identifiers, and operators, and handle escape sequences and string literals. The implementation files of the Clang lexer can be found in the llvm-project/clang/lib/lex directory within the LLVM infrastructure.

\subsection{Clang Parse Library}

Clang Parse Library is the parser that takes the tokens produced by the lexer and constructs an AST to represent the structure and meaning of the source code. The Clang parser checks the source code for proper syntax and resolves symbols and identifiers. It also performs type-checking to ensure the source code follows the rules of the programming language. It creates the AST, a tree-like structure, that represents the source code in a way that is easily processed by the compiler. The implementation files of the Clang lexer can be found in the llvm-project/clang/lib/parse directory within the LLVM infrastructure.

\subsection{Clang Sema Library}

Clang Sema Library is a semantic analyzer that involves examining the meaning and context of the source code in a program. In Clang, semantic analysis is a phase in the compilation process that analyzes the AST generated by the parser to verify that the source code conforms to the rules of the programming language and is properly constructed. Semantic analysis performs various checks and transformations on the AST to ensure the source code is correct. The implementation files of the Clang semantic analyzer can be found in the llvm-project/clang/lib/sema directory within the LLVM infrastructure.

\subsection{Clang CodeGen Library}

Clang CodeGen is the code generation library that takes the AST as input which is generated by the parser and corrected by the semantic analyzer. It generates the IR code and produces a .ll file which will be used in the back end. The implementation files of the Clang lexer can be found in the llvm-project/clang/lib/CodeGen directory within the LLVM infrastructure. 

\section{LLVM IR Optimizer}\label{sec:opt}
LLVM IR is a representation which serves as a common ground for front-ends and back-ends. LLVM IR is not as high level as programming languages but it provides more information than assembly by having types or more expressive functions. LLVM IR instructions are stored in Basic Block structures which contain sequential IR instructions with an entry and exit. 

LLVM Optimizer is a common optimization medium used for every possible source-target combination of a compiler. It takes the output file of CodeGen as input and runs three types of passes:

\begin{enumerate}
    \item Analysis passes: These passes analyze the IR and collect information about the IR without modifying the IR. 
    \item Transformation passes: These passes modify the IR by using the information gathered from Analysis passes. The optimizations are the product of these transformations. 
    \item Utility passes: These passes are used to perform tasks such as printing the IR or verifying the IR. 
\end{enumerate}

The output of the optimizer becomes the input for the target back-end which lowers the LLVM IR to the target Assembly. As the generated LLVM IR at the end of the optimizations is the object of pattern matching and assembly support for any custom instruction, it is a critical part of the design process.

\subsection{Analysis Passes}
There are almost 40 analysis passes. The significant documented analysis passes are listed below:

\begin{multicols}{3}
\begin{itemize}
\scriptsize
     \item Exhaustive Alias Analysis Precision Evaluator
     \item Basic Alias Analysis (stateless AA impl)
     \item Basic CallGraph Construction
     \item Count Alias Analysis Query Responses
     \item Dependence Analysis
     \item AA use debugger
     \item Dominance Frontier Construction
     \item Dominator Tree Construction
     \item Simple mod/ref analysis for globals
     \item Counts the various types of Instructions
     \item Interval Partition Construction
     \item Induction Variable Users
     \item Lazy Value Information Analysis
     \item LibCall Alias Analysis
     \item Statically lint-checks LLVM IR
     \item Natural Loop Information
     \item Memory Dependence Analysis
     \item Decodes module-level debug info
     \item Post-Dominance Frontier Construction
     \item Post-Dominator Tree Construction
     \item Alias Set Printer
     \item Find Used Types
     \item Detect single entry single exit regions
     \item Scalar Evolution Analysis
     \item ScalarEvolution-based Alias Analysis
     \item Stack Safety Analysis
     \item Target Data Layout

\end{itemize}
\end{multicols}

\subsection{Transformation Passes}
There are almost 60 transformation passes. The documented transformation passes are listed below:

\begin{multicols}{3}
\begin{itemize}
\scriptsize
     \item Aggressive Dead Code Elimination
     \item Inliner for always\_inline functions
     \item Promote ‘by reference’ arguments to scalars
     \item Basic-Block Vectorization
     \item Profile Guided Basic Block Placement
     \item Break critical edges in CFG
     \item Optimize for code generation
     \item Merge Duplicate Global Constants
     \item Dead Code Elimination
     \item Dead Argument Elimination
     \item Dead Type Elimination
     \item Dead Instruction Elimination
     \item Dead Store Elimination
     \item Deduce function attributes
     \item Dead Global Elimination
     \item Global Variable Optimizer
     \item Global Value Numbering
     \item Canonicalize Induction Variables
     \item Function Integration/Inlining
     \item Combine redundant instructions
     \item Combine expression patterns
     \item Internalize Global Symbols
     \item Interprocedural Sparse Conditional Constant Propagation
     \item Jump Threading
     \item Loop-Closed SSA Form Pass
     \item Loop Invariant Code Motion
     \item Delete dead loops
     \item Extract loops into new functions
     \item Extract at most one loop into a new function
     \item Loop Strength Reduction
     \item Rotate Loops
     \item Canonicalize natural loops
     \item Unroll loops
     \item Unroll and Jam loops
     \item Unswitch loops
     \item Lower global destructors
     \item Lower atomic intrinsics to non-atomic form
     \item Lower invokes to calls, for unwindless code generators
     \item Lower SwitchInsts to branches
     \item Promote Memory to Register
     \item MemCpy Optimization
     \item Merge Functions
     \item Unify function exit nodes
     \item Partial Inliner
     \item Remove unused exception handling info
     \item Reassociate expressions
     \item Relative lookup table converter
     \item Demote all values to stack slots
     \item Scalar Replacement of Aggregates
     \item Sparse Conditional Constant Propagation
     \item Simplify the CFG
     \item Code sinking
     \item Strip all symbols from a module
     \item Strip debug info for unused symbols
     \item Strip Unused Function Prototypes
     \item Strip all llvm.dbg.declare intrinsics
     \item Strip all symbols, except dbg symbols, from a module
     \item Tail Call Elimination

\end{itemize}
\end{multicols}

\cite{passes}

\subsection{Case Study: Optimizations on S-box }\label{sbox-case}
One of the research topics of this study was to observe the changes to a function shown in Code \ref{lst:sbox-c} performing S-box with bitwise operations. As the pattern is large hundreds of lines of LLVM IR and RISC-V Assembly are generated without enabling optimizations. The unoptimized LLVM IR is given in Appendices in Code shown in Code \ref{lst:unopt-sbox}. However, when the optimizations are enabled the final LLVM IR file shown in Code \ref{lst:sbox-ir} and the RISC-V Assembly is significantly smaller. 

\begin{minipage}{\linewidth}
\begin{lstlisting}[caption={S-box C code},language=C, label={lst:sbox-c}]
typedef struct {
  int x[5];
} ascon_state_t;

void sbox(ascon_state_t state) {
     int t0, t1, t2, t3, t4;
     state.x[0] ^= state.x[4];    
     state.x[4] ^= state.x[3];    
     state.x[2] ^= state.x[1];
     t0  = state.x[0];    
     t1  = state.x[1];    
     t2  = state.x[2];    
     t3  = state.x[3];    
     t4  = state.x[4];
     t0 =~ t0;    
     t1 =~ t1;    
     t2 =~ t2;    
     t3 =~ t3;    
     t4 =~ t4;
     t0 &= state.x[1];   
     t1 &= state.x[2];    
     t2 &= state.x[3];    
     t3 &= state.x[4];    
     t4 &= state.x[0];
     state.x[0] ^= t1;   
     state.x[1] ^= t2;    
     state.x[2] ^= t3;    
     state.x[3] ^= t4;    
     state.x[4] ^= t0;
     state.x[1] ^= state.x[0];
     state.x[0] ^= state.x[4]; 
     state.x[3] ^= state.x[2]; 
     state.x[2] =~ state.x[2];

     return;
}

\end{lstlisting}
\end{minipage}

At the end of optimization passes the following IR will be generated:
\begin{minipage}{\linewidth}
\begin{lstlisting}[caption={Optimized S-box LLVM IR}, label={lst:sbox-ir},language=llvm,style=nasm]
; Function Attrs: mustprogress nofree norecurse nosync nounwind willreturn memory(argmem: readwrite) uwtable
define dso_local void @sbox(ptr nocapture noundef %state) local_unnamed_addr #0 {
entry:
  %arrayidx = getelementptr inbounds [5 x i32], ptr %state, i32 0, i32 4
  %0 = load i32, ptr %arrayidx, align 4, !tbaa !7
  %1 = load i32, ptr %state, align 4, !tbaa !7
  %xor = xor i32 %1, %0
  %arrayidx4 = getelementptr inbounds [5 x i32], ptr %state, i32 0, i32 3
  %2 = load i32, ptr %arrayidx4, align 4, !tbaa !7
  %xor7 = xor i32 %2, %0
  %arrayidx9 = getelementptr inbounds [5 x i32], ptr %state, i32 0, i32 1
  %3 = load i32, ptr %arrayidx9, align 4, !tbaa !7
  %arrayidx11 = getelementptr inbounds [5 x i32], ptr %state, i32 0, i32 2
  %4 = load i32, ptr %arrayidx11, align 4, !tbaa !7
  %xor12 = xor i32 %4, %3
  %not = xor i32 %xor, -1
  %not23 = xor i32 %3, -1
  %not24 = xor i32 %xor12, -1
  %not25 = xor i32 %2, -1
  %not26 = xor i32 %xor7, -1
  %and = and i32 %3, %not
  %and31 = and i32 %4, %not23
  %and34 = and i32 %2, %not24
  %and37 = and i32 %0, %not25
  %and40 = and i32 %xor, %not26
  %xor43 = xor i32 %and31, %xor
  %xor46 = xor i32 %and34, %3
  %xor49 = xor i32 %xor12, %and37
  %xor52 = xor i32 %and40, %2
  %xor55 = xor i32 %and, %xor7
  store i32 %xor55, ptr %arrayidx, align 4, !tbaa !7
  %xor60 = xor i32 %xor46, %xor43
  store i32 %xor60, ptr %arrayidx9, align 4, !tbaa !7
  %xor65 = xor i32 %xor43, %xor55
  store i32 %xor65, ptr %state, align 4, !tbaa !7
  %xor70 = xor i32 %xor49, %xor52
  store i32 %xor70, ptr %arrayidx4, align 4, !tbaa !7
  %not73 = xor i32 %xor49, -1
  store i32 %not73, ptr %arrayidx11, align 4, !tbaa !7
  ret void
}

\end{lstlisting}
\end{minipage}
LLVM optimization passes are responsible for the simplification of IR. In this case, the following passes were the passes changing the IR and were run sequentially.
\begin{enumerate}
    \item InferFunctionAttrsPass 
    \item SROAPass 
    \item EarlyCSEPass 
    \item GlobalOptPass 
    \item InstCombinePass 
    \item EarlyCSEPass 
    \item InstCombinePass 
    \item ReassociatePass 
    \item InstCombinePass 
    \item DSEPass 
    \item PostOrderFunctionAttrsPass 
\end{enumerate}
 As it can be observed some passes can run several times. For example, InstCombinePass runs to canonicalize and prepare the expressions for the following pass which is the reason why it is running prior to three distinct passes.

 \subsubsection{Infer Function Attributes - InferFunctionAttrsPass}\label{sec:funcAttrs}

 This pass adds metadata to LLVM IR, by analyzing it. Function attributes are used to pass information about functions between LLVM passes.

\begin{lstlisting}[caption={LLVM IR Before InferFunctionAttrsPass},linerange={181-185} , label={lst:sbox-preinfFuncAttr},language=llvm, style=nasm]
; Function Attrs: nocallback nofree nosync nounwind willreturn memory(argmem: readwrite)
declare void @llvm.lifetime.start.p0(i64 immarg, ptr nocapture) #1

; Function Attrs: nocallback nofree nosync nounwind willreturn memory(argmem: readwrite)
declare void @llvm.lifetime.end.p0(i64 immarg, ptr nocapture) #1
\end{lstlisting}
\begin{lstlisting}[caption={LLVM IR After InferFunctionAttrsPass},linerange={181-185} , label={lst:sbox-infFuncAttr},language=llvm, style=nasm]
; Function Attrs: mustprogress nocallback nofree nosync nounwind willreturn memory(argmem: readwrite)
declare void @llvm.lifetime.start.p0(i64 immarg, ptr nocapture) #1

; Function Attrs: mustprogress nocallback nofree nosync nounwind willreturn memory(argmem: readwrite)
declare void @llvm.lifetime.end.p0(i64 immarg, ptr nocapture) #1
\end{lstlisting}

 Function attribute is inferred as "mustprogress" as the lifetime starting function is interacting with its environment in an observable way making memory access \cite{llvmref-funcAttrs}. 
 \par
The lifetime function decides the accessibility of the pointer to the memory. When memory is allocated the lifetime of the pointer to the memory starts and ends when deallocated \cite{llvmref-objectLifetime}.

\subsubsection{Scalar Replacement of Aggregates - SROAPass}

Aggregate IR instructions such as "alloca" are promoted to registers. The promotion to registers also means the lifetime is under control and the explicit lifetime intrinsic calls can be removed.

\begin{lstlisting}[caption={Alloca and Lifetime Start Lines Removed From LLVM IR Before SROAPass},linerange={11-22} ,language=llvm, style=nasm]
  %state.indirect_addr = alloca ptr, align 4
  %t0 = alloca i32, align 4
  %t1 = alloca i32, align 4
  %t2 = alloca i32, align 4
  %t3 = alloca i32, align 4
  %t4 = alloca i32, align 4
  store ptr %state, ptr %state.indirect_addr, align 4, !tbaa !7
  call void @llvm.lifetime.start.p0(i64 4, ptr %t0) #2
  call void @llvm.lifetime.start.p0(i64 4, ptr %t1) #2
  call void @llvm.lifetime.start.p0(i64 4, ptr %t2) #2
  call void @llvm.lifetime.start.p0(i64 4, ptr %t3) #2
  call void @llvm.lifetime.start.p0(i64 4, ptr %t4) #2
\end{lstlisting}
\begin{lstlisting}[caption={Lifetime End Lines Removed From LLVM IR Before SROAPass},linerange={172-177} ,language=llvm, style=nasm]
  store i32 %not73, ptr %arrayidx75, align 4, !tbaa !11
  call void @llvm.lifetime.end.p0(i64 4, ptr %t4) #2
  call void @llvm.lifetime.end.p0(i64 4, ptr %t3) #2
  call void @llvm.lifetime.end.p0(i64 4, ptr %t2) #2
  call void @llvm.lifetime.end.p0(i64 4, ptr %t1) #2
  call void @llvm.lifetime.end.p0(i64 4, ptr %t0) #2
\end{lstlisting}

An important transformation SROA does is promoting the use of registers instead of using the stack for local variables and using "Load/Store" operations to use them in the unoptimized IR \cite{llvmcode-sroa}. "Store/Load" operations are reduced significantly in this stage, especially for intermediate variables where the C code is not referring to the array directly. 

SROA pass relies on the analysis passes of Alias Analysis through the collection of analysis passes for Loads.

\begin{lstlisting}[linerange={15-19}, caption={NOT operations between Intermediate Variables} ,language=C, label={lst:sbox-c-nots}]
     t0 =~ t0;    
     t1 =~ t1;    
     t2 =~ t2;    
     t3 =~ t3;    
     t4 =~ t4;
\end{lstlisting}

\begin{lstlisting}[caption={Intermediate Load and Stores in LLVM IR Before SROAPass},linerange={66-81} , language=llvm, style=nasm]
  store i32 %10, ptr %t4, align 4, !tbaa !11
  %11 = load i32, ptr %t0, align 4, !tbaa !11
  %not = xor i32 %11, -1
  store i32 %not, ptr %t0, align 4, !tbaa !11
  %12 = load i32, ptr %t1, align 4, !tbaa !11
  %not23 = xor i32 %12, -1
  store i32 %not23, ptr %t1, align 4, !tbaa !11
  %13 = load i32, ptr %t2, align 4, !tbaa !11
  %not24 = xor i32 %13, -1
  store i32 %not24, ptr %t2, align 4, !tbaa !11
  %14 = load i32, ptr %t3, align 4, !tbaa !11
  %not25 = xor i32 %14, -1
  store i32 %not25, ptr %t3, align 4, !tbaa !11
  %15 = load i32, ptr %t4, align 4, !tbaa !11
  %not26 = xor i32 %15, -1
  store i32 %not26, ptr %t4, align 4, !tbaa !11
\end{lstlisting}

\begin{lstlisting}[caption={Load and Stores Promoted to Registers in LLVM IR After SROAPass},linerange={45-49} ,label={sbox-sroa} , language=llvm, style=nasm]
  %not = xor i32 %6, -1
  %not23 = xor i32 %7, -1
  %not24 = xor i32 %8, -1
  %not25 = xor i32 %9, -1
  %not26 = xor i32 %10, -1
\end{lstlisting}

Similar to the dramatic change in the previous example, operations between the array elements and the intermediate variables are optimized so that the registers are used instead of the stack.

According to the statistics obtained from the "opt" tool of LLVM:

\begin{displayquote}
 5 mem2reg - Number of alloca's promoted within one block \\
 1 mem2reg - Number of alloca's promoted with a single store \\
 1 sroa    - Maximum number of partitions per alloca \\
 8 sroa    - Maximum number of uses of a partition \\
41 sroa    - Number of alloca partition uses rewritten \\
 6 sroa    - Number of alloca partitions formed \\
 6 sroa    - Number of allocas analyzed for replacement \\
41 sroa    - Number of instructions deleted \\
 6 sroa    - Number of allocas promoted to SSA values \\
\end{displayquote}

\subsubsection{Early Common Subexpression Elimination - EarlyCSEPass}\label{sec:earlyCSE1}
Performs a simple dominator tree walk, eliminating trivially redundant instructions. A dominator tree is a type of tree where every parent node dominates the child node. The definition of dominance from graph theory is that every path to the dominated node passes through the dominator node \cite{prosserDom}.

Early CSE pass relies on MemorySSA analysis which analyses by representing memory operations in SSA form \cite{llvmdoc-memoryssa, novillo2007memory}.

\begin{lstlisting}[caption={Redundant Load Instructions in LLVM IR Before EarlyCSEPass},linerange={4-11},label={lst:sbox-cse1}, language=llvm, style=nasm]
define dso_local void @sbox(ptr noundef %state) #0 {
entry:
  %x = getelementptr inbounds %struct.ascon_state_t, ptr %state, i32 0, i32 0
  %arrayidx = getelementptr inbounds [5 x i32], ptr %x, i32 0, i32 4
  %0 = load i32, ptr %arrayidx, align 4, !tbaa !7
  %x1 = getelementptr inbounds %struct.ascon_state_t, ptr %state, i32 0, i32 0
  %arrayidx2 = getelementptr inbounds [5 x i32], ptr %x1, i32 0, i32 0
  %1 = load i32, ptr %arrayidx2, align 4, !tbaa !7
\end{lstlisting}
In Code \ref{lst:sbox-cse1} you can see that "\%x1" and "\%arrayidx2" are equal to the function argument "\%state". EarlyCSE pass detects this redundant condition and uses the already present "\%state" pointer in the output.

\begin{lstlisting}[caption={Optimized Load Instructions in LLVM IR After EarlyCSEPass},linerange={2-6} , label={lst:sbox-cseLoad}, language=llvm, style=nasm]
define dso_local void @sbox(ptr noundef %state) #0 {
entry:
  %arrayidx = getelementptr inbounds [5 x i32], ptr %state, i32 0, i32 4
  %0 = load i32, ptr %arrayidx, align 4, !tbaa !7
  %1 = load i32, ptr %state, align 4, !tbaa !7
\end{lstlisting}

Another remark from this example is that the pointer calculation is done in two instructions by "getelementptr" LLVM instruction which accesses the struct's address and then the element's address in it. EarlyCSE combines these two instructions outputting the offset calculated pointers.

A natural result of these simple optimizations is that the section which makes use of registers has increased. It can be seen in Code \ref{lst:sbox-cse2} that the recalculation of pointers by getelementptr is removed as they are used at the beginning of the function, as shown partly in Code \ref{lst:sbox-cseLoad}.

\begin{minipage}{\linewidth}
\begin{lstlisting}[caption={Redundant Instructions in LLVM IR Before EarlyCSEPass},linerange={30-69},label={lst:sbox-cse2}, language=llvm, style=nasm-small]
  %x13 = getelementptr inbounds %struct.ascon_state_t, ptr %state, i32 0, i32 0
  %arrayidx14 = getelementptr inbounds [5 x i32], ptr %x13, i32 0, i32 0
  %6 = load i32, ptr %arrayidx14, align 4, !tbaa !7
  %x15 = getelementptr inbounds %struct.ascon_state_t, ptr %state, i32 0, i32 0
  %arrayidx16 = getelementptr inbounds [5 x i32], ptr %x15, i32 0, i32 1
  %7 = load i32, ptr %arrayidx16, align 4, !tbaa !7
  %x17 = getelementptr inbounds %struct.ascon_state_t, ptr %state, i32 0, i32 0
  %arrayidx18 = getelementptr inbounds [5 x i32], ptr %x17, i32 0, i32 2
  %8 = load i32, ptr %arrayidx18, align 4, !tbaa !7
  %x19 = getelementptr inbounds %struct.ascon_state_t, ptr %state, i32 0, i32 0
  %arrayidx20 = getelementptr inbounds [5 x i32], ptr %x19, i32 0, i32 3
  %9 = load i32, ptr %arrayidx20, align 4, !tbaa !7
  %x21 = getelementptr inbounds %struct.ascon_state_t, ptr %state, i32 0, i32 0
  %arrayidx22 = getelementptr inbounds [5 x i32], ptr %x21, i32 0, i32 4
  %10 = load i32, ptr %arrayidx22, align 4, !tbaa !7
  %not = xor i32 %6, -1
  %not23 = xor i32 %7, -1
  %not24 = xor i32 %8, -1
  %not25 = xor i32 %9, -1
  %not26 = xor i32 %10, -1
  %x27 = getelementptr inbounds %struct.ascon_state_t, ptr %state, i32 0, i32 0
  %arrayidx28 = getelementptr inbounds [5 x i32], ptr %x27, i32 0, i32 1
  %11 = load i32, ptr %arrayidx28, align 4, !tbaa !7
  %and = and i32 %not, %11
  %x29 = getelementptr inbounds %struct.ascon_state_t, ptr %state, i32 0, i32 0
  %arrayidx30 = getelementptr inbounds [5 x i32], ptr %x29, i32 0, i32 2
  %12 = load i32, ptr %arrayidx30, align 4, !tbaa !7
  %and31 = and i32 %not23, %12
  %x32 = getelementptr inbounds %struct.ascon_state_t, ptr %state, i32 0, i32 0
  %arrayidx33 = getelementptr inbounds [5 x i32], ptr %x32, i32 0, i32 3
  %13 = load i32, ptr %arrayidx33, align 4, !tbaa !7
  %and34 = and i32 %not24, %13
  %x35 = getelementptr inbounds %struct.ascon_state_t, ptr %state, i32 0, i32 0
  %arrayidx36 = getelementptr inbounds [5 x i32], ptr %x35, i32 0, i32 4
  %14 = load i32, ptr %arrayidx36, align 4, !tbaa !7
  %and37 = and i32 %not25, %14
  %x38 = getelementptr inbounds %struct.ascon_state_t, ptr %state, i32 0, i32 0
  %arrayidx39 = getelementptr inbounds [5 x i32], ptr %x38, i32 0, i32 0
  %15 = load i32, ptr %arrayidx39, align 4, !tbaa !7
  %and40 = and i32 %not26, %15
\end{lstlisting}
\end{minipage}

\begin{lstlisting}[caption={Optimized LLVM IR After EarlyCSEPass},linerange={20-33} ,label={lst:sbox-cse3}, language=llvm, style=nasm]
  %6 = load i32, ptr %state, align 4, !tbaa !7
  %7 = load i32, ptr %arrayidx9, align 4, !tbaa !7
  %8 = load i32, ptr %arrayidx4, align 4, !tbaa !7
  %9 = load i32, ptr %arrayidx, align 4, !tbaa !7
  %not = xor i32 %6, -1
  %not23 = xor i32 %7, -1
  %not24 = xor i32 %xor12, -1
  %not25 = xor i32 %8, -1
  %not26 = xor i32 %9, -1
  %and = and i32 %not, %7
  %and31 = and i32 %not23, %xor12
  %and34 = and i32 %not24, %8
  %and37 = and i32 %not25, %9
  %and40 = and i32 %not26, %6
\end{lstlisting}

Register-based operations increased because redundant load operations from the same pointers are removed. For example, in Code \ref{lst:sbox-cse2} to obtain "\%and", register operation result "\%not" and loaded value "\%11" are used. In the output of EarlyCSE, Code \ref{lst:sbox-cse3}, we can see that instead of reloading to register the loaded register is used, "\%7" in this case.

According to the statistics obtained from the "opt" tool of LLVM:
\begin{displayquote}
19 early-cse - Number of instructions Common Subexpression Eliminated \\
 7 early-cse - Number of load instructions Common Subexpression Eliminated \\
35 early-cse - Number of instructions simplified or Dead Code Eliminated \\
\end{displayquote}

\subsubsection{Optimize Global Variables - GlobalOpt}
This pass aims to optimize global variables and transforms them into constants if necessary. This pass did not significantly change the IR. It only added an attribute to the function, "local\_unnamed\_addr" meaning that the address of the function is not significant in the module.

\begin{lstlisting}[caption={"local\_unnamed\_addr" Attribute Added to LLVM IR After GlobalOpt},linerange={7-7} ,label={lst:sbox-globvar}, language=llvm, style=nasm]
define dso_local void @sbox(ptr noundef %state) local_unnamed_addr #0 {
\end{lstlisting}

\subsubsection{Combine Redundant Instructions - InstCombinePass}
Combines redundant instructions and canonicalizes them. Canonicalization is the form in which a single way of commutability is preferred. For example, if a binary operator has a constant operand it is moved to the right. Canonic instructions can then be used by other passes which can assume the instructions to be in the canonic form \cite{llvmpass-instcombine}.
\par 

\begin{lstlisting}[caption={Redundant Load Instruction in LLVM IR Before InstCombine},linerange={10-17} ,label={lst:sbox-instc1_1}, language=llvm, style=nasm]
  %0 = load i32, ptr %arrayidx, align 4, !tbaa !7
  %1 = load i32, ptr %state, align 4, !tbaa !7
  %xor = xor i32 %1, %0
  store i32 %xor, ptr %state, align 4, !tbaa !7
  %arrayidx4 = getelementptr inbounds [5 x i32], ptr %state, i32 0, i32 3
  %2 = load i32, ptr %arrayidx4, align 4, !tbaa !7
  %3 = load i32, ptr %arrayidx, align 4, !tbaa !7
  %xor7 = xor i32 %3, %2
\end{lstlisting}

\begin{minipage}{\linewidth}
\begin{lstlisting}[caption={Removed Load Instruction in LLVM IR After InstCombine},linerange={4-11} , language=llvm, style=nasm]
  %arrayidx = getelementptr inbounds [5 x i32], ptr %state, i32 0, i32 4
  %0 = load i32, ptr %arrayidx, align 4, !tbaa !7
  %1 = load i32, ptr %state, align 4, !tbaa !7
  %xor = xor i32 %1, %0
  store i32 %xor, ptr %state, align 4, !tbaa !7
  %arrayidx4 = getelementptr inbounds [5 x i32], ptr %state, i32 0, i32 3
  %2 = load i32, ptr %arrayidx4, align 4, !tbaa !7
  %xor7 = xor i32 %0, %2
\end{lstlisting}
\end{minipage}

According to the statistics obtained from the "opt" tool of LLVM:
\begin{displayquote}
  5 aa             - Number of NoAlias results \\
109 assume-queries - Number of Queries into an assume assume bundles \\
 10 basicaa        - Number of times a GEP is decomposed \\
 11 instcombine    - Number of insts combined \\
  1 instcombine    - Number of expansions \\
  2 instcombine    - Number of instruction combining iterations performed \\
\end{displayquote}

\subsubsection{Early Common Subexpression Elimination - 2nd Run of EarlyCSEPass}
Similar to the previous EarlyCSE run in Section \ref{sec:earlyCSE1}, load instructions to registers are reused in the subsequent instructions.
\begin{lstlisting}[caption={Redundant Load Instructions in LLVM IR Before 2nd EarlyCSEPass},linerange={34-48} ,label={lst:sbox-cse2_1}, language=llvm, style=nasm]
  %8 = load i32, ptr %arrayidx9, align 4, !tbaa !7
  %xor46 = xor i32 %8, %and34
  store i32 %xor46, ptr %arrayidx9, align 4, !tbaa !7
  %9 = load i32, ptr %arrayidx11, align 4, !tbaa !7
  %xor49 = xor i32 %9, %and37
  store i32 %xor49, ptr %arrayidx11, align 4, !tbaa !7
  %10 = load i32, ptr %arrayidx4, align 4, !tbaa !7
  %xor52 = xor i32 %10, %and40
  store i32 %xor52, ptr %arrayidx4, align 4, !tbaa !7
  %11 = load i32, ptr %arrayidx, align 4, !tbaa !7
  %xor55 = xor i32 %11, %and
  store i32 %xor55, ptr %arrayidx, align 4, !tbaa !7
  %12 = load i32, ptr %state, align 4, !tbaa !7
  %13 = load i32, ptr %arrayidx9, align 4, !tbaa !7
  %xor60 = xor i32 %13, %12
\end{lstlisting}

\begin{minipage}{\linewidth}
\begin{lstlisting}[caption={Removed Load Instructions in LLVM IR After 2nd EarlyCSEPass},linerange={31-39} ,label={lst:sbox-cse2_2}, language=llvm, style=nasm]
  %xor46 = xor i32 %3, %and34
  store i32 %xor46, ptr %arrayidx9, align 4, !tbaa !7
  %xor49 = xor i32 %xor12, %and37
  store i32 %xor49, ptr %arrayidx11, align 4, !tbaa !7
  %xor52 = xor i32 %2, %and40
  store i32 %xor52, ptr %arrayidx4, align 4, !tbaa !7
  %xor55 = xor i32 %xor7, %and
  store i32 %xor55, ptr %arrayidx, align 4, !tbaa !7
  %xor60 = xor i32 %xor46, %xor43
\end{lstlisting}
\end{minipage}

\subsubsection{Combine Redundant Instructions - 2nd Run of InstCombinePass}

\begin{lstlisting}[caption={XOR Instruction in LLVM IR Before InstCombine},linerange={4-5,9-11,22-22,27-27} ,label={lst:sbox-instc2-1}, language=llvm, style=nasm]
  %arrayidx = getelementptr inbounds [5 x i32], ptr %state, i32 0, i32 4
  %0 = load i32, ptr %arrayidx, align 4, !tbaa !7
  %arrayidx4 = getelementptr inbounds [5 x i32], ptr %state, i32 0, i32 3
  %2 = load i32, ptr %arrayidx4, align 4, !tbaa !7
  %xor7 = xor i32 %0, %2
  %not25 = xor i32 %2, -1
  %and37 = and i32 %xor7, %not25
\end{lstlisting}

\begin{lstlisting}[caption={XOR Instruction in LLVM IR After InstCombine},linerange={4-5,9-10,22-22,27-27} ,label={lst:sbox-instc2_2}, language=llvm, style=nasm]
  %arrayidx = getelementptr inbounds [5 x i32], ptr %state, i32 0, i32 4
  %0 = load i32, ptr %arrayidx, align 4, !tbaa !7
  %arrayidx4 = getelementptr inbounds [5 x i32], ptr %state, i32 0, i32 3
  %2 = load i32, ptr %arrayidx4, align 4, !tbaa !7
  %not25 = xor i32 %2, -1
  %and37 = and i32 %0, %not25
\end{lstlisting}

The first algebraic optimization in the process can be observed in this example. In Code \ref{lst:sbox-instc2-1}, to obtain "\%and37" the boolean operations of the following must be performed:

$$ (\%0 \oplus \%2)  \land ( \%2 \oplus -1)  $$
It can be shown that a simpler boolean form can be obtained by transitioning equivalent boolean equations:
$$ (\%0 \oplus \%2)  \land  \lnot \%2   $$
$$ ((\%0 \land \lnot \%2) \lor (\lnot \%0 \land \%2))  \land  \lnot \%2   $$
$$ (\%0 \land \lnot \%2\land  \lnot \%2) \lor (\lnot \%0 \land \%2\land  \lnot \%2)     $$
$$ (\%0 \land \lnot \%2) \lor (0)     $$
$$ \%0 \land (\%2 \oplus -1) $$

In Code \ref{lst:sbox-instc2_2}, the redundant operation $\%0 \oplus \%2$ is removed and the result is: 

$$ \%0  \land ( \%2 \oplus -1)  $$

According to the statistics obtained from the "opt" tool of LLVM:
\begin{displayquote}
60 assume-queries - Number of Queries into an assume assume bundles \\
 5 instcombine    - Number of insts combined \\
 1 instcombine    - Number of expansions \\
 2 instcombine    - Number of instruction combining iterations performed \\ 
12 instsimplify   - Number of reassociations \\
\end{displayquote}

\subsubsection{Reassociate Expressions - ReassociatePass}
Reassociates associative expressions, to promote better constant propagation and simplify expression graph to reduce instruction count. It implements an algorithm where the constants have the least rank and the rank increases with the expression reverse post-order traversal \cite{llvmcode-reassoc}. 

\begin{lstlisting}[caption={Instructions in LLVM IR Before ReassociatePass},linerange={26-29} ,label={lst:sbox-reass}, language=llvm, style=nasm]
  %and34 = and i32 %2, %not24
  %and37 = and i32 %0, %not25
  %and40 = and i32 %xor, %not26
  %xor43 = xor i32 %xor, %and31
\end{lstlisting}

\begin{lstlisting}[caption={Instructions with Reassociated Arguments in LLVM IR After ReassociatePass},linerange={26-29} ,label={lst:sbox-reass2}, language=llvm, style=nasm]
  %and34 = and i32 %not24, %2
  %and37 = and i32 %not25, %0
  %and40 = and i32 %not26, %xor
  %xor43 = xor i32 %and31, %xor
\end{lstlisting}

A basic glance at the debug output of the pass gives more idea about how the reassociation works.
\clearpage
\begin{lstlisting}[language=llvm, style=nasm, caption={Debug Output of Reassociate Pass from LLVM opt tool}]
Calculated Rank[state] = 3
Combine negations for:   %
LINEARIZE:   %
OPERAND:   %
ADD USES LEAF:   %
OPERAND:   %
ADD LEAF:   %
RAIn:	xor i32	[ %
RAOut:	xor i32	[ %
RA:   %
TO:   %
Combine negations for:   %
LINEARIZE:   %
OPERAND:   %
ADD USES LEAF:   %
OPERAND:   %
ADD USES LEAF:   %
RAIn:	xor i32	[ %
RAOut:	xor i32	[ %
RA:   %
TO:   %
\end{lstlisting}

The debug output deals with the beginning of the function which is given below.

\begin{lstlisting}[caption={Instructions in LLVM IR Before ReassociatePass},linerange={5-11} ,label={lst:sbox-reass}, language=llvm, style=nasm]
  %0 = load i32, ptr %arrayidx, align 4, !tbaa !7
  %1 = load i32, ptr %state, align 4, !tbaa !7
  %xor = xor i32 %1, %0
  store i32 %xor, ptr %state, align 4, !tbaa !7
  %arrayidx4 = getelementptr inbounds [5 x i32], ptr %state, i32 0, i32 3
  %2 = load i32, ptr %arrayidx4, align 4, !tbaa !7
  %xor7 = xor i32 %0, %2
\end{lstlisting}
The pass computed that the reassociation would result in the instruction "\%xor = xor i32 \%1, \%0" which is already how the IR is so it is not changed. However, as it can be seen in Code \ref{lst:sbox-reass2}, "\%xor7 = xor i32 \%2, \%0" replaced its alternative representation as the rank of "\%0" is less than "\%2".

According to the statistics obtained from the "opt" tool of LLVM:
\begin{displayquote}
    16 reassociate - Number of insts reassociated
\end{displayquote}

\subsubsection{Combine Redundant Instructions - 2nd Run of InstCombinePass}
In this last run of InstCombinePass instruction count is not changed. Some reassociations by the previous pass are reversed.

\begin{lstlisting}[caption={Instructions in LLVM IR Before the 3rd InstCombinePass},linerange={26-29} ,label={lst:sbox-instc3_1}, language=llvm, style=nasm]
  %and34 = and i32 %not24, %2
  %and37 = and i32 %not25, %0
  %and40 = and i32 %not26, %xor
  %xor43 = xor i32 %and31, %xor
\end{lstlisting}

\begin{lstlisting}[caption={Instructions with Reassociated Arguments in LLVM IR After InstCombinePass},linerange={26-29} ,label={lst:sbox-instc3_2}, language=llvm, style=nasm]
  %and34 = and i32 %2, %not24
  %and37 = and i32 %0, %not25
  %and40 = and i32 %xor, %not26
  %xor43 = xor i32 %and31, %xor
\end{lstlisting}

Though it may seem wasteful, it is a common theme in LLVM that some transformations may be done and be completely reversed by another pass.

According to the statistics obtained from the "opt" tool of LLVM:
\begin{displayquote}
60 assume-queries - Number of Queries into an assume assume bundles \\
 7 instcombine    - Number of insts combined \\
 2 instcombine    - Number of instruction combining iterations performed \\
12 instsimplify   - Number of reassociations \\
\end{displayquote}

\subsubsection{Dead Store Elimination - DSEPass}
Dead code in the Dead Code Elimination (DCE) pass refers to the variables in any point of the program which are not used in the future. DCE does not eliminate control flow and store instructions, for this reason, Dead Store Elimination (DSE) pass is used to simplify the store instructions of the given program.

Trivial dead stores are eliminated. As the load operations are optimized, most of the store instructions become dead meaning that they do not affect the flow in any way.

Similar to Early CSE pass in Section \ref{sec:earlyCSE1}, DSE pass relies on Memory SSA analysis.

\begin{lstlisting}[caption={Redundant Store Instructions in LLVM IR Before DSEPass},linerange={30-36} ,label={lst:sbox-dse1}, language=llvm, style=nasm]
  store i32 %xor43, ptr %state, align 4, !tbaa !7
  %xor46 = xor i32 %and34, %3
  store i32 %xor46, ptr %arrayidx9, align 4, !tbaa !7
  %xor49 = xor i32 %xor12, %and37
  store i32 %xor49, ptr %arrayidx11, align 4, !tbaa !7
  %xor52 = xor i32 %and40, %2
  store i32 %xor52, ptr %arrayidx4, align 4, !tbaa !7
\end{lstlisting}

\begin{lstlisting}[caption={Removed Store Instructions in LLVM IR After DSEPass},linerange={27-29} ,label={lst:sbox-dse2}, language=llvm, style=nasm]
  %xor46 = xor i32 %and34, %3
  %xor49 = xor i32 %xor12, %and37
  %xor52 = xor i32 %and40, %2
\end{lstlisting}

To see how the DSE works we can observe two dead Store's and a killer Store, unnecessary code is stripped away.

\begin{lstlisting}[caption={Store Instructions to the Same Address in LLVM IR Before DSEPass},linerange={2-2, 8-8, 30-30, 42-42} ,label={lst:sbox-dse3}, language=llvm, style=nasm]
define dso_local void @sbox(ptr noundef %state) local_unnamed_addr #0 {
  store i32 %xor, ptr %state, align 4, !tbaa !7
  store i32 %xor43, ptr %state, align 4, !tbaa !7
  store i32 %xor65, ptr %state, align 4, !tbaa !7
\end{lstlisting}

Here is the debug output of the DSE pass.

\begin{lstlisting}[language=llvm, style=nasm, caption={Debug Output of DSE Pass from LLVM opt tool}]
Trying to eliminate MemoryDefs killed by 4 = MemoryDef(3) (
store i32 %
  trying to get dominating access
   visiting 3 = MemoryDef(2)->liveOnEntry (store i32 %
   visiting 2 = MemoryDef(1)->liveOnEntry (store i32 %
   visiting 1 = MemoryDef(liveOnEntry) (store i32 %
  Checking for reads of 1 = MemoryDef(liveOnEntry) (store i32 %
   4 = MemoryDef(3)->1 (  store i32 %
    ... skipping killing def/dom access
   2 = MemoryDef(1)->liveOnEntry (store i32 %
   3 = MemoryDef(2)->liveOnEntry (store i32 %
 Checking if we can kill 1 = MemoryDef(liveOnEntry) (store i32 %
DSE: Remove Dead Store:
  DEAD:   store i32 %
  KILLER:   store i32 %
  trying to get dominating access
   visiting 0 = MemoryDef(liveOnEntry)
   ...  found LiveOnEntryDef
  finished walk
  .
  .
  .
  Trying to eliminate MemoryDefs killed by 10 = MemoryDef(9) (
  store i32 %
  trying to get dominating access
   visiting 9 = MemoryDef(8) (store i32 %
   visiting 8 = MemoryDef(7) (store i32 %
   visiting 7 = MemoryDef(6)->liveOnEntry (store i32 %
   visiting 6 = MemoryDef(4) (store i32 %
   visiting 4 = MemoryDef(liveOnEntry) (store i32 %
  Checking for reads of 4 = MemoryDef(liveOnEntry) (store i32 %
   10 = MemoryDef(9)->4 (store i32 %
    ... skipping killing def/dom access
   6 = MemoryDef(4) (store i32 %
   9 = MemoryDef(8) (store i32 %
   7 = MemoryDef(6)->liveOnEntry (store i32 %
   8 = MemoryDef(7) (store i32 %
 Checking if we can kill 4 = MemoryDef(liveOnEntry) (
 store i32 %
DSE: Remove Dead Store:
  DEAD:   store i32 %
  KILLER:   store i32 %
  trying to get dominating access
   visiting 0 = MemoryDef(liveOnEntry)
   ...  found LiveOnEntryDef
  finished walk
\end{lstlisting}

It can be observed that whenever the DSE encounters a Store instruction with the same address as a previous Store instruction, kills the previous instruction. The last Store instruction survives DSE.

According to the statistics obtained from the "opt" tool of LLVM:
\begin{displayquote}
14 aa              - Number of MustAlias results \\
62 aa              - Number of NoAlias results \\
30 basicaa         - Number of times a GEP is decomposed \\
29 dse             - Number iterations check for reads in getDomMemoryDef \\
 0 dse             - Number of other instrs removed \\
 7 dse             - Number of stores deleted \\
 7 dse             - Number of times a valid candidate is returned from getDomMemoryDef \\
 5 dse             - Number of stores remaining after DSE \\
 1 ir              - Number of renumberings across all blocks \\
71 memory-builtins - Number of arguments with unsolved size and offset \\
\end{displayquote}

\subsubsection{Post-Order Function Attributes Pass - PostOrderFunctionAttrsPass}
This pass is similar to the InferFunctionAttrsPass in Section \ref{sec:funcAttrs}. It does not change the IR, adds metadata to it for the other passes.

\begin{minipage}{\linewidth}
\begin{lstlisting}[caption={Function Attributes After PostOrderFunctionAttrsPass},linerange={1-2} ,label={lst:sbox-post2}, language=llvm, style=nasm]
; Function Attrs: nounwind uwtable
define dso_local void @sbox(ptr noundef %state) local_unnamed_addr #0 {
\end{lstlisting}
\end{minipage}

\begin{lstlisting}[caption={Function Attributes Before PostOrderFunctionAttrsPass},linerange={1-2} ,label={lst:sbox-post1}, language=llvm, style=nasm]
; Function Attrs: mustprogress nofree norecurse nosync nounwind willreturn memory(argmem: readwrite) uwtable
define dso_local void @sbox(ptr nocapture noundef %state) local_unnamed_addr #0 {
\end{lstlisting}

The added attributes signal that the function does not deallocate memory, does not recurse by calling itself, never raises an exception, will continue execution at the end according to the call stack, and may read or write any memory. 

In the end of optimization passes, the IR at Code \ref{lst:sbox-ir} will be generated.

\subsection{Clang Optimization Levels}
It should be noted that in order to observe an optimised Assembly, the LLVM IR should be generated by enabling optimizations.
Clang can be invoked with optimization levels deciding which optimization passes are going to be run. The optimizations can target speed or code size. Speed optimizing options range from "-O1" to "-O3". "-O2" enables most of the optimizations. "-O3" enables optimizations that can increase the compile time and generate larger code. The main code optimizing options are "-Os" and "-Oz". "-Os" is similar to "-O2" but runs extra optimizations to reduce code size. "-Oz" runs more code-reducing optimizations compared to "-Os" and is similar to "-O2" again \cite{clangCommands}.
Caution must be taken as when no arguments are given to Clang, at the time of writing, Clang uses the "-O0" optimization level. Implementing pattern matching on unoptimized LLVM IR is not feasible for several reasons. Firstly, the IR is more sensitive to changes in the front end. Changing the code style in the front end can cause CodeGen to produce a slightly different IR which makes it less predictable. Secondly, the code size can be too large with redundant code which makes pattern matching large instructions cumbersome. We recommend using "-O2" or "-Os" optimization levels while developing instruction selection patterns. 

LLVM optimizations can be performed with an LLVM IR input by using the "opt" tool \cite{optimizer}. It is possible to experiment with different optimizations and observe their results on the output.

\section{Stages of the LLVM RISC-V Back-end}
LLVM RISC-V back-end is responsible for compiling optimized IR down to RISC-V assembly or object code. LLVM back-end consists of libraries for the code generation steps\cite{llvmbackend}.

\subsection{Instruction Selection}
SelectionDAG is the default instruction selector of LLVM RISC-V back-end which is responsible for selecting the appropriate RISC-V instructions for a given IR instruction. It takes the target-independent LLVM code as input and generates the target-dependent DAG of instructions. SelectionDAG is at the core of this study since we will be dealing with adding new instructions to the RISC-V back end.

\subsubsection{SelectionDAG construction}
After IR generation is done, SelectionDAG gets the optimized IR and converts it into a target-independent SelectionDAG representation. SelectionDAG consists of SelectionDAG nodes (SDnode) which are created by SelectionDAGBuilder class. SelectionDAGIsel visits all the IR instructions and uses the SelectionDAGBuilder class. The relevant instruction method requests an SDNode to the DAG and assigns its opcode. Every SDNode has an opcode for the operation it represents. SDNodes have multiple values to return as the result. SDValues (SelectionDAG value) hold the information to determine which number to return. SelectionDAGBuilder class reshapes the linear IR input to a SelectionDAG tree form. At the end of the construction, SelectionDAG is a target-independent and illegal DAG.

\subsubsection{SelectionDAG legalization}
\begin{figure}
  \centering
  \includegraphics[scale=0.5]{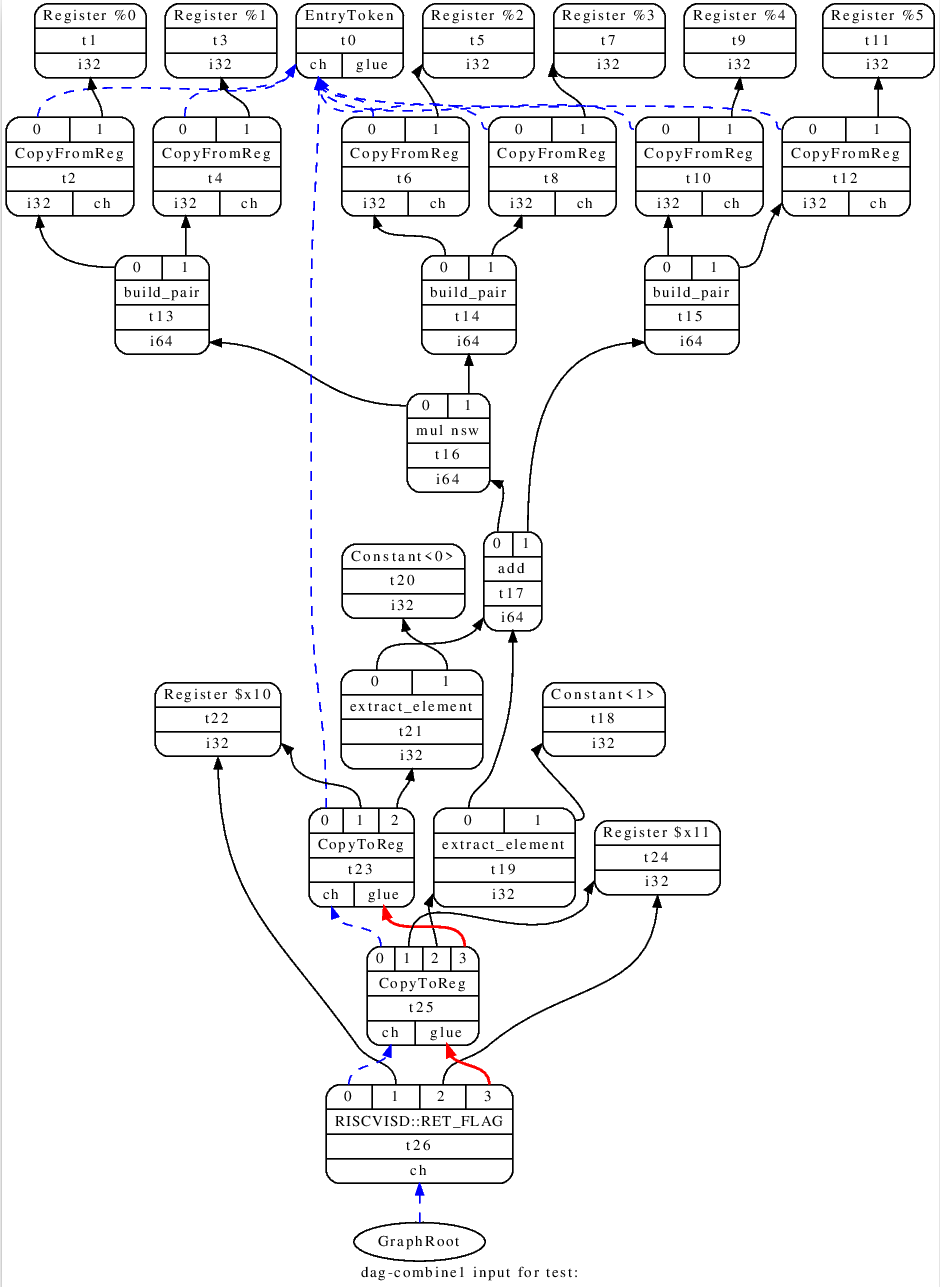}
  \caption{DAG diagram before the legalization stage}
  \label{fig:dag_before_legalization}
\end{figure}

\begin{figure}
  \centering
  \includegraphics[scale=0.35]{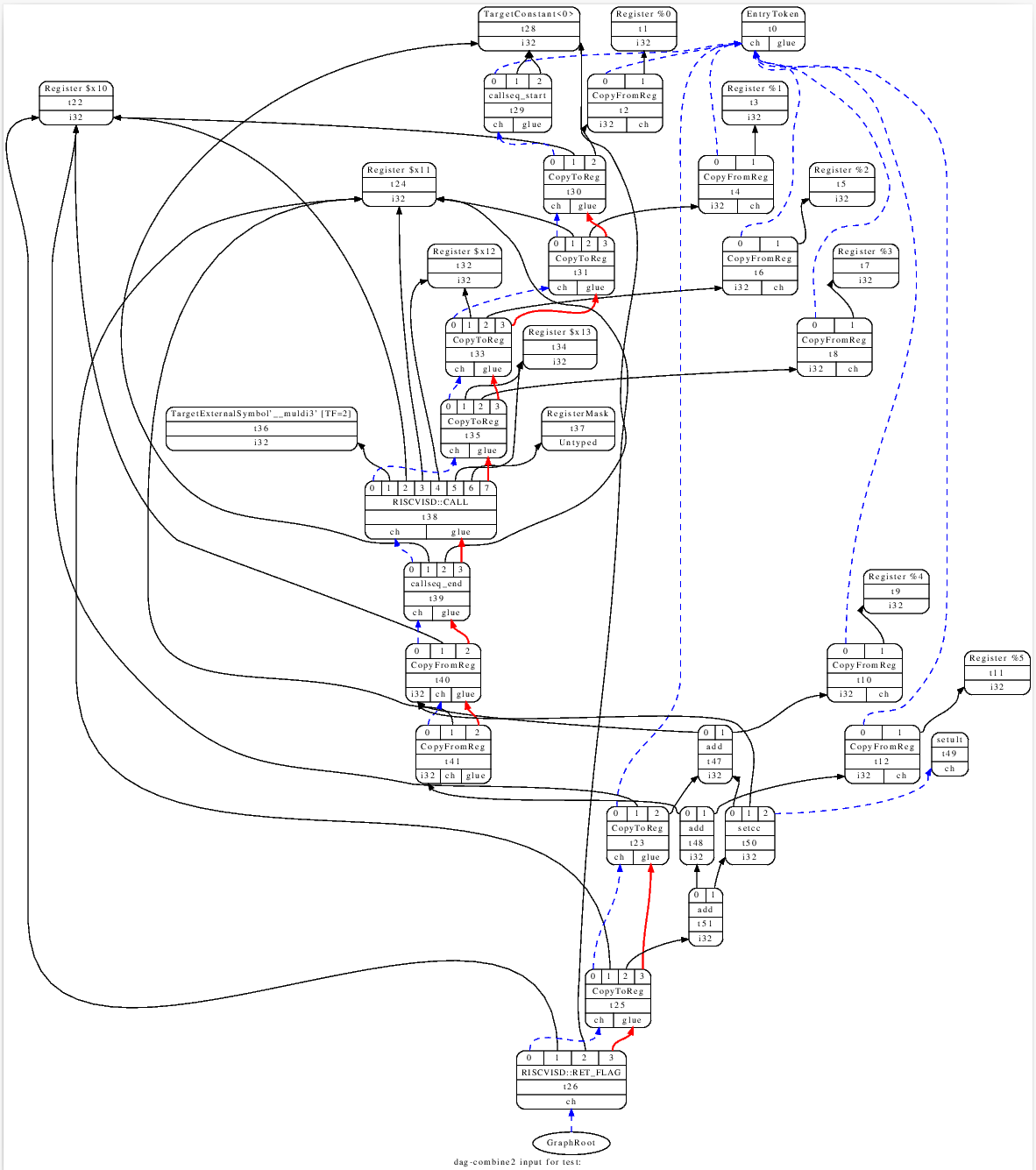}
  \caption{DAG diagram after the legalization stage}
  \label{fig:dag_after_legalization}
\end{figure}

SelectionDAG is a target-dependent representation after the construction stage of the instruction selection. Before creating a target-specific code, SelectionDAG checks if the DAG is legal because the constructed DAG may include incompatible instructions and data types to the target architecture \cite{legalizer}. SelectionDAG legalization refers to the process of transforming the SelectionDAG according to the constraints and requirements of the target architecture. Legalization may involve adding, removing, splitting or merging the nodes, targeting to match the register file and instruction set of the target architecture \cite{llvmcookbook}. SelectionDAG legalization also ensures that the data type of the target architecture is compatible with the target architecture by truncating or promoting the data types. For example, if SelectionDAG includes 32-bit integer (i32) data type nodes targeted to an i64 architecture, SDlegalizer promotes the i32 nodes to 64-bit integer (i64) data type. For every target architecture type, IselLowering.cpp files are responsible for legalizing the SelectionDAG. SDlegalizer legalizes the illegal DAG into a supported form and ensures that the generated code is efficient and compatible with the target architecture.\\
An example of legalizing the SelectionDAG by truncating the i64 data type DAG into i32 data type target architecture is shown in Code \ref{lst:ir_legalization}. IR code includes i64 data type variables however, target architecture supports only i32 data type. Before the legalization (Figure \ref{fig:dag_before_legalization}) DAG is not converted to the target data type yet and it needed to become i64 compatible. It can be seen that after legalization (Figure \ref{fig:dag_after_legalization}), i32 nodes are truncated and there are no i64 nodes in the DAG.

\begin{lstlisting}[caption={IR code input for legalization example}, label={lst:ir_legalization}, language=llvm, style=nasm]
  define i64 @test(i64 %
    ret i64 %
    }
  \end{lstlisting}

\clearpage
\subsubsection{SelectionDAG optimization}
The DAG should be optimized after legalization because the legalization phase may create unnecessary DAG nodes and the reducible nodes are not combined yet. SelectionDAG optimizer minimizes the DAG nodes before creating the target-specific instructions.

\subsubsection{SelectionDAG target-dependent instruction selection}
At the last phase of the instruction selection, SelectionDAG selects the suitable instructions for the target architecture. SelectionDAG uses the relevant TableGen target description (.td) files or C++ logic to match the patterns and replaces the patterns with the target-specific instructions.

\subsection{Scheduling and Formation}
Scheduling is the phase of assigning an order to the DAG form of RISC-V instructions. The formation phase is responsible for converting the DAG into a list of machine instructions.  

\subsection{SSA-based Machine Code Optimizations}
LLVM uses SSA-based optimizations before register allocation. SSA optimizations ensure that each variable is assigned and defined only once before it is used.

\subsection{Register Allocation}
The register allocation is responsible for assigning physical registers to virtual registers in the IR. Each target has a specific register count and order. The register allocator maps the registers by taking the RISC-V architecture registers into account. It uses the relevant TargetRegisterInfo, and MachineOperand classes. 

\subsection{Prologue/Epilogue Code Insertion}
Prologue and epilogue code insertion is another optimization phase that is responsible for frame-pointer elimination and stack packing.

\pagebreak
\subsection{Code Emission}
The code emission stage is responsible for lowering the code generator abstractions down to the Machine Code layer abstractions. It takes the assembly as input and creates the final RISC-V machine codes. 

\subsection{Linking}
LLD is the LLVM linker library that is responsible for combining multiple object files into a single executable file. LLD is invoked after the code emission and generates a file by resolving symbol references, adjusting addresses, and performing other tasks as necessary.

\cleardoublepage
\clearpage
\chapter{RISC-V}\label{ch:riscv}
In this project, our target is a 32-bit RISC-V core. RISC stands for reduced instruction set computer and RISC-V is an open standard ISA. \cite{riscvorgabout} It is structured as a small base ISA and it has different additional extensions. The base ISA is straightforward, rendering RISC-V appropriate for academic and learning purposes, yet extensive enough to function as a cost-effective and energy-efficient ISA for embedded systems \cite{watermanriscv}. Being open-source and royalty-free is another significant advantage and is an important reason why RISC-V is being commonly used. 
RISC-V was developed by Prof. Krste Asanović and his students Andrew Waterman and Yunsup Lee. They started working on this project in 2010 as a part of as part of the Parallel Computing Laboratory which was in UC Berkeley. Par Lab was sponsored by several companies and worked on advancing parallel computing.

\section{RISC-V ISA}
\begin{figure}[h!]
    \centering
    \includegraphics[scale=0.25]{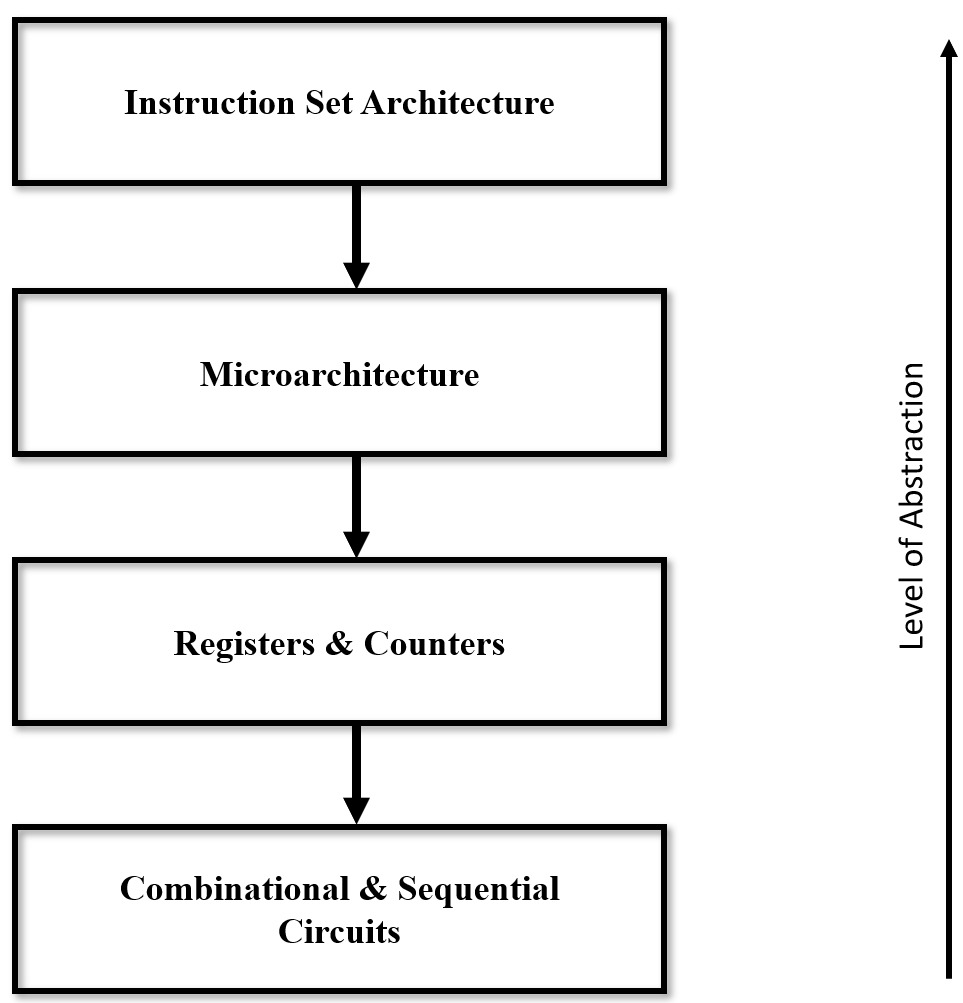}
    \caption{Level of abstraction diagram \cite{levelofabstrac}}
    \label{fig:level_of_abstraction_diagram}
\end{figure}

The ISA constitutes a part of a computer’s abstract design that defines how the Central Processing Unit (CPU) is managed by the software. It serves as a bridge between the software and hardware, defining the processor’s abilities and the methods by which it performs tasks. Its level in the system can be seen in Figure \ref{fig:level_of_abstraction_diagram}.

\begin{figure}[h!]
    \centering
    \includegraphics{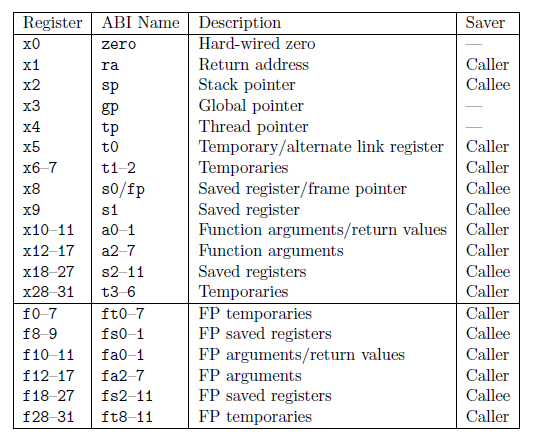}
    \caption{RISC-V registers \cite{rvregisters}}
    \label{fig:riscv_registers}
\end{figure}

There are different base integer variants of RISC-V such as RV32I, RV64I, and RV128I. These have address spaces of 32, 64, and 128 bits respectively \cite{Altinayozlem}. In our project, we are interested in 32 bits. RISC-V has 32 general-purpose registers. Their Application Binary Interface (ABI) names and purposes can be seen in Figure \ref{fig:riscv_registers}. Also in the Figure, we can see a different set of registers. These registers are used for floating point operations. Their ABI and purposes are also given.

\pagebreak
\section{RISC-V Base Instructions}
\begin{figure}
    \centering
    \includegraphics{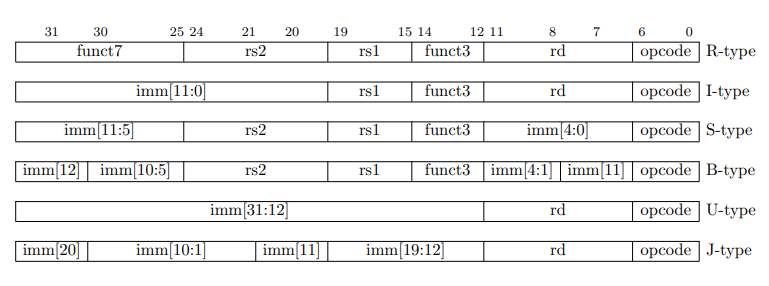}
    \caption{RISC-V base instruction formats \cite{rvmanual}}
    \label{fig:risc-v_base_instruction_formats}
\end{figure}

There are four basic instruction formats in the base RV32I ISA. These are named R, I, S, U and all of these are 32-bits in length. There are two more additional variants named B and J as well \cite{rvmanual}. These formats are given in Figure \ref{fig:risc-v_base_instruction_formats}.

\begin{figure}[h!]
    \centering
    \includegraphics{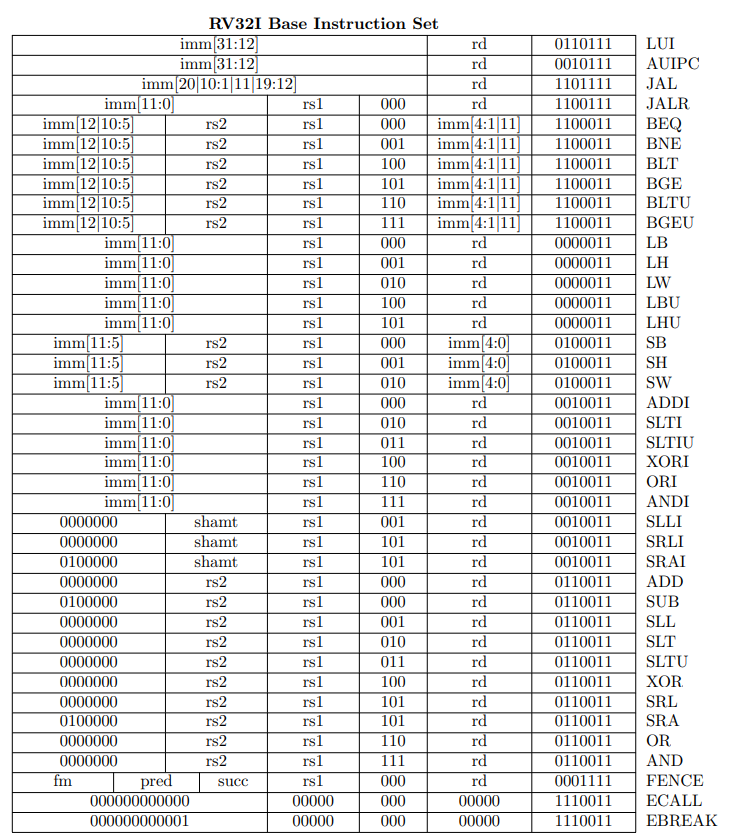}
    \caption{RV32I base instruction set \cite{rvmanual}}
    \label{fig:rv32i_base_instruction_set}
\end{figure}

$R_{S1}$ and $R_{S2}$ are the source registers and $R_d$ is the destination register. An immediate value can also be used in some of the formats. 
The base instructions of the RV32I are given in Figure \ref{fig:rv32i_base_instruction_set}. By inspecting their formats, we can see which type the instructions belong to. For example, the ADDI instruction is an I-type instruction and XOR is an R-type instruction.

\section{RISC-V Extensions}
\begin{figure}[h!]
    \centering
    \includegraphics{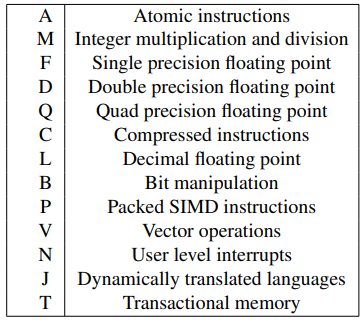}
    \caption{List of standard extension sets \cite{erfan}}
    \label{fig:list_of_standard_extension_sets}
\end{figure}

We had mentioned the extensions previously. Abbreviations for these extensions and what they are for are given in Figure \ref{fig:list_of_standard_extension_sets}.

\begin{figure}[h!]
    \centering
    \includegraphics{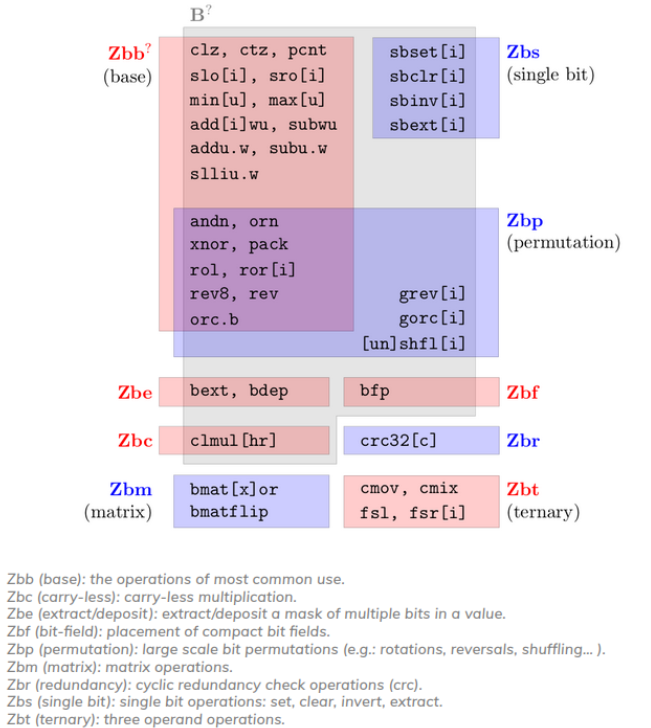}
    \caption{Bit manipulation extension groupings \cite{bitmanipgroups}}
    \label{fig:bit_manipulation_extension_groupings}
\end{figure}

Thanks to these instruction extensions, more specific tasks can be implemented since we are not limited by the base instructions.
Among these, the bit manipulation (B) standard extension contains numerous instructions that can be useful in a wide range of applications. This extension's instructions mainly operate on bits. These extensions are also divided into several groups according to common properties. These subgroups and their purposes can be seen in Figure \ref{fig:bit_manipulation_extension_groupings}.

\begin{figure}[h!]
    \centering
    \includegraphics{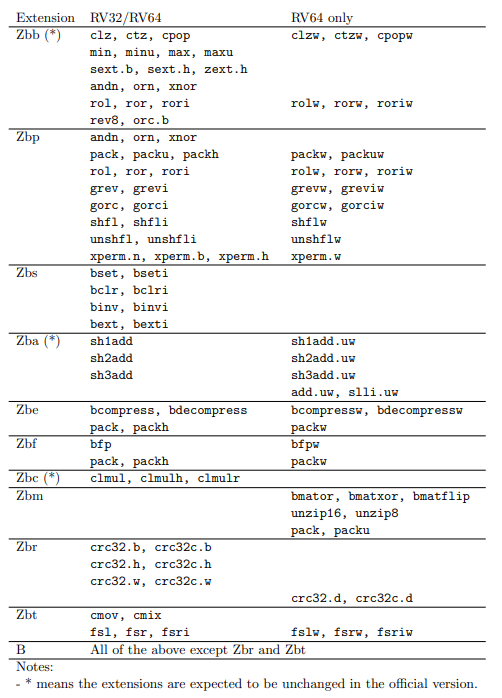}
    \caption{Bit RV32/RV64 compatibilities and groups \cite{bitmanipdraft}}
    \label{fig:rv32_rv64_compatibilities_and_groups}
\end{figure}

Grouping these instructions according to how commonly they are used and the similarity of the operations that they perform makes them more organized and easier to work on with hardware and software. 
Some of these extensions are compatible with RV64 only. The compatibilities and the groups the instructions belong to are given in Figure \ref{fig:rv32_rv64_compatibilities_and_groups}.

\clearpage

To give a clearer image of what bit manipulation (B) instructions do, a few of them will be explained. For example, “CLZ” is an instruction for counting the leading zeros. Its purpose is to find out how many zeros are there before encountering a 1, starting from the most significant bit. Another example is “ORN” instruction. It negates the second operand and performs bitwise or with the first one. 
\par
Zba is also a subgroup of the bit manipulation extensions. Shift and add instructions are included in this group and they perform a left shift by 1, 2, or 3 bits since they are commonly used in codes and also because they require only a minimal amount of extra hardware beyond that of a basic adder. This way, lengthening the critical path in implementations can be avoided. For example, SH1ADD is a part of this group and it shifts the operand by 1 and adds 1 \cite{bitmanipulationisaextensions}.
\par
\begin{figure}
    \centering
    \includegraphics{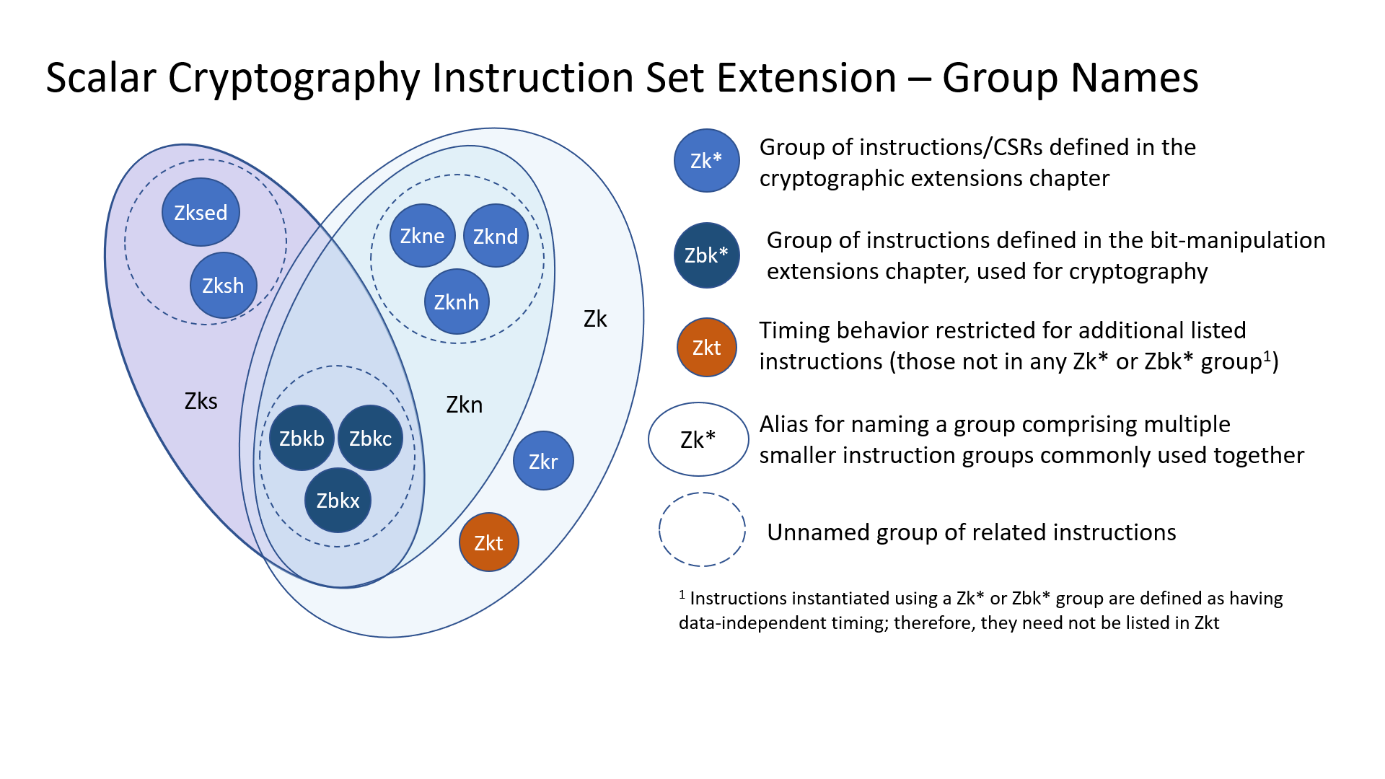}
    \caption{Cryptography extension subgroups \cite{cryptogroupsdiag}}
    \label{fig:cryptography_extension_subgroups}
\end{figure}

There is also a scalar cryptography instruction set extension for RISC-V. The RISC-V Scalar Cryptography extensions allow cryptographic tasks to be completed more quickly. Furthermore, these extensions significantly reduce the difficulty of implementing fast and secure cryptography in embedded devices and IoT \cite{cryptogroups}. This instruction set extension is also divided into subgroups according to the purpose and similarity of the instructions. The groups are given in Figure \ref{fig:cryptography_extension_subgroups}. These groups and their purposes can be explained briefly.

\begin{itemize}
    \item Zbkb contains bit manipulation instructions for cryptography. These are a selection of the bit manipulation extension Zbb that have specific applications in cryptography. 
    \item Zbkc contains carry-less multiply instructions.
    \item Zbkx instructions can be useful for implementing s-boxes in constant time.
    \item Zknd contains instructions that help speed up the decryption and key schedule functions of the AES block cipher and Zkne does the same for encryption.
    \item Zknh has some instructions that can help speed up the SHA2 family of cryptographic hash functions.
    \item Zksed contains instructions that speed up the SM4 block cipher.
    \item Zksh instructions help accelerate the SM3 hash function.
    \item Zkr can be useful to seed cryptographic random bit generators \cite{cryptoextensiondoc}.
\end{itemize}

These extensions are supported by the compiler but pattern matching support for Zbkb and Zbkx is incomplete in the LLVM RISC-V backend. Also for Zknd, Zkne, Zknh, Zksed and Zksh, no pattern matching exists. Therefore, these instructions can be only used via builtin functions or from the assembler \cite{llvmextensionspage}.

The modular structure of these extensions is useful for hardware and software developers. For example, the B extension is built in Clang so only an extra argument will provide the necessary instructions from the input extension.

It is important for hardware developers to consider that developing accelerators targeting instructions in standard extensions will reduce the software workload significantly. The reason for this is LLVM supports RISC-V standard extensions and follows updates closely. Corner cases are thought out and optimization opportunities are utilized. RISC-V standard extensions are comprehensive and may already contain the extensions that we want to implement. After making sure that the extension we want is not present, we may try to implement non-standard extensions.

\cleardoublepage
\clearpage
\chapter{ASCON CRYPTOGRAPHIC ALGORITHM}\label{Chascon}
ASCON (Authenticated Encryption with Associated Data) is a lightweight encryption algorithm that is a family of lightweight authenticated ciphers. ASCON is designed to have both authenticity and confidentiality for transmitted data and it is efficient in terms of both speed and code size. It has a clean and simple design, making it suitable for resource-constrained environments. 
ASCON was designed by Christoph Dobraunig, Maria Eichlseder, Florian Mendel, and Martin Schläffer in 2014. It was proposed as a candidate for the lightweight authenticated encryption competition (CAESAR) in 2014 and was selected as one of the finalists.

\section{ASCON Structure}
\begin{figure}
    \centering
    \includegraphics[scale = 0.4]{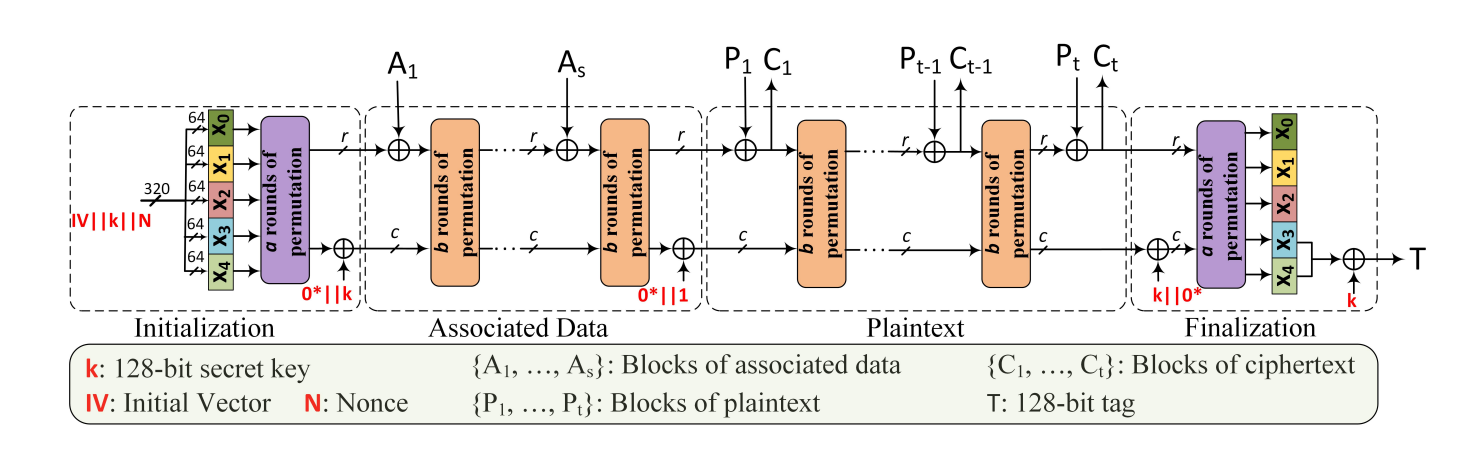}
    \caption{Associated data and plaintext are absorbed into the sponge-based structure}
    \label{fig:sbox_structure}
\end{figure}

Ascon is based on the Sponge structure that is shown in Figure \ref{fig:sbox_structure}. ASCON gets an initial input to start and encrypt the algorithm. The length of the initial input is 320 bits which consists of five 64-bit words. Initial input includes a secret key, initial vector and nonce. A secret key is used to encrypt and decrypt the transmitted information. Information can be read if the key is known thus it must be kept secret. The initial vector is a random value to start an iterated process. Nonce increases the protection of the cipher against cryptanalysis techniques. 

Concatenated input consists of two parts. The first r bits of the input are the rate bits. The last c = 320 – r bits of the input are called capacity bits. In the initialization stage, “a” rounds of permutation functions are implemented to concatenated input. After permutation, the last 128 bits of the capacity bits are XORed with the 128-bit secret key. 

At the beginning of the associated data stage, rate bits are XORed with the first block of the associated data then “b” rounds of permutation are implemented to the output. This step is repeated with the previous output and the next block of the associated data until all the blocks are covered. Associated data is absorbed into the sponge structure. At the end of the associated stage, capacity bits are XORed with 1’s.

In the Plaintext stage, the plaintext blocks are absorbed into the sponge-like the associated stage and the ciphertext blocks are obtained. At the beginning of the finalization stage, the secret key is XOR'ed with the first 128 bits of the capacity bits. “a” rounds of permutations are implemented and the last 128 bits of the capacity bits are XORed with the secret key. The output of the finalization stage is called the 128-bit tag. 

\section{Permutation Function of the ASCON Algorithm}

\begin{figure}
    \centering
    \includegraphics[scale = 0.6]{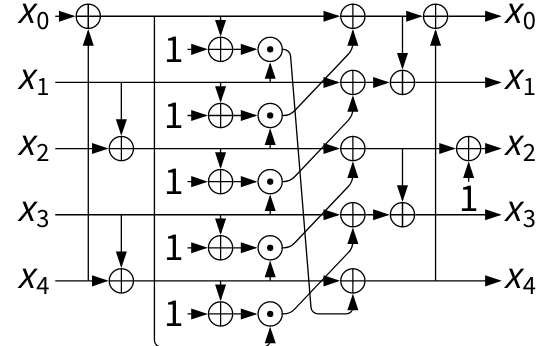}
    \caption{S-box operations}
    \label{fig:sbox_operations}
\end{figure}

ASCON’s permutation function consists of a nonlinear substitution layer and a linear diffusion layer. The substitution layer performs a 5-bit S-box. S-box takes five 64-bit concatenated words as input. 64 S-boxes are performed for every bit of the words in a single permutation function. S-box operations are shown in Figure \ref{fig:sbox_operations}. The linear diffusion layer performs the rotations and XORs shown in Equation \ref{eq:ascon_linear}.

\begin{equation}
\label{eq:ascon_linear}
\begin{aligned}
& x_0 \leftarrow \Sigma_0\left(x_0\right)=x_0 \oplus\left(x_0 \ggg 19\right) \oplus\left(x_0 \ggg 28\right) \\
& x_1 \leftarrow \Sigma_1\left(x_1\right)=x_1 \oplus\left(x_1 \ggg 61\right) \oplus\left(x_1 \ggg 39\right) \\
& x_2 \leftarrow \Sigma_2\left(x_2\right)=x_2 \oplus\left(x_2 \ggg 1\right) \oplus\left(x_2 \ggg 6\right) \\
& x_3 \leftarrow \Sigma_3\left(x_3\right)=x_3 \oplus\left(x_3 \ggg 10\right) \oplus\left(x_3 \ggg 17\right) \\
& x_4 \leftarrow \Sigma_4\left(x_4\right)=x_4 \oplus\left(x_4 \ggg 7\right) \oplus\left(x_4 \ggg 41\right)
\end{aligned}
\end{equation}

\cleardoublepage

\clearpage
\chapter{PATH OF AN INSTRUCTION}\label{Ch4}
In this chapter, the path of an instruction will be demonstrated and the corresponding DAG input of the most critical phases of SelectionDAG will be shown. We selected the input program as a function that performs multiplication and addition. This was our litmus test code used while adding MLA (Multiply and Add) instruction to the LLVM back-end with TableGen. We explained how to modify the compiler so that it recognises MLA instruction thoroughly in Section \ref{sec:MLA_add_section}. 

\begin{lstlisting}[language=C, caption=madd.c program, label=maddc]
int a,b,c;
void maddFunc() {
	a = 3;
	b = 103;
	
	c = 127;
	a = a * b + c;
}
\end{lstlisting}

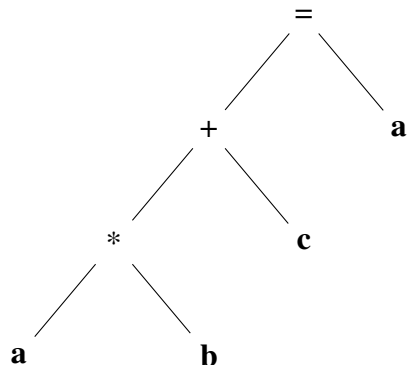
\begin{figure}
    \centering
\begin{tikzpicture}
    \node {=} [sibling distance = 2.5cm]
    child {node {+}
    child {node {*}
    child {node {\bf a}}
    child {node {\bf b}}}
    child {node {\bf c}}}
    child {node {\bf a}};
\end{tikzpicture}
    \caption{AST of MLA operation}
    \label{fig:ast_mla}
\end{figure}

\section{Clang AST}
\begin{figure}
    \centering
    \includegraphics[width=\textwidth]{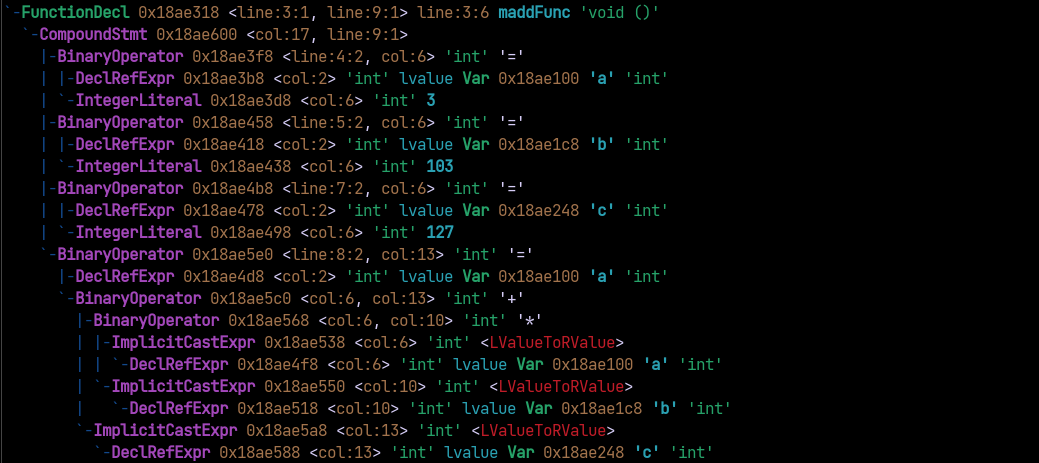}
    \caption{AST generated by Clang}
    \label{fig:clang_ast}
\end{figure}

The simplified AST of the expression is shown in Figure \ref{fig:ast_mla}. The AST consists of an expression tree with three levels. At the highest level, there is an expression tree of multiplication between variables 'a' and 'b'. This expression tree's result becomes an argument for another expression tree with the addition operator. The second argument at this addition subtree is the variable 'c'. The expression tree at the root has assignment as an operator. The first argument to this tree is 'a' and the second argument is the result of multiplication and addition. Figure \ref{fig:clang_ast} shows the AST output of Clang for Code \ref{maddc}. 

\section{LLVM IR}
Clang CodeGen produces LLVM IR with the AST as the input. Figure \ref{fig:llvm_ir} shows the produced LLVM IR. The optimized LLVM IR is the input to SelectionDAG to generate target-specific instructions.  

\begin{lstlisting}[caption={LLVM IR file generated at the output of Clang},label={fig:llvm_ir}, language=llvm, style=nasm]
define void @maddFunc() {
  store i32 3, i32* @a
  store i32 103, i32* @b
  store i32 127, i32* @c
  %1 = load i32, i32* @a
  %2 = load i32, i32* @b
  %5 = mul nsw i32 %1, %2
  %4 = load i32, i32* @c
  %5 = add nsw i32 %3, %4
  store i32 %5, i32* @a
  ret void
}
\end{lstlisting}

\section{SelectionDAG}
 Input DAGs to SelectionDAG's passes will be demonstrated so on. The following phases will be demonstrated:
\begin{enumerate}
    \item First Optimization
    \item Legalization
    \item Second Optimization
    \item Instruction Selection
    \item Instruction Scheduling 
    \item Register Allocation
\end{enumerate}

\subsection{First Optimization Pass}
\begin{figure}
    \centering
    \includegraphics[width=0.9\textwidth]{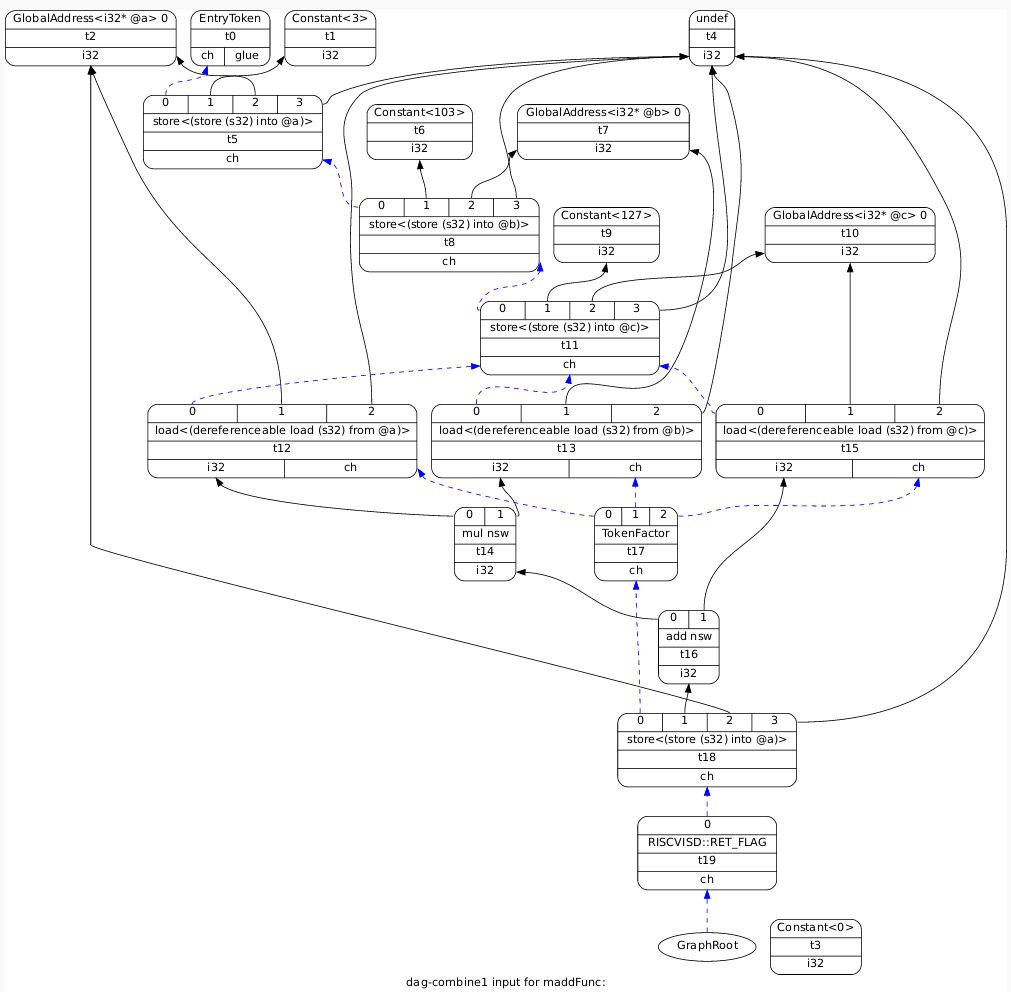}
    \caption{DAG before first optimization pass}
    \label{fig:combine1}
\end{figure}

Figure \ref{fig:combine1} shows the DAG before the first optimization pass. It is the direct translation of LLVM IR to DAG form. After optimization, redundant nodes will be removed such as "Constant<0>" node.

\begin{figure}
    \centering
    \includegraphics[width=0.9\textwidth]{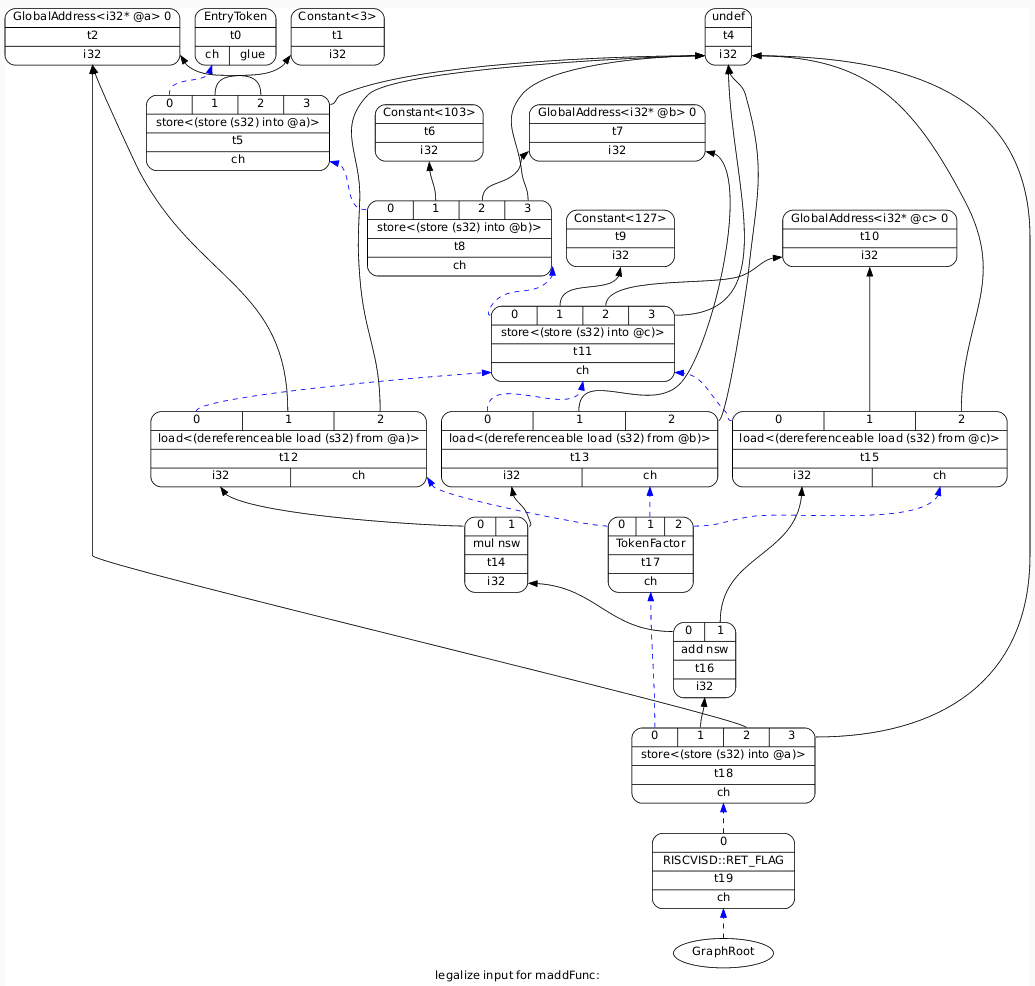}
    \caption{DAG before Legalization}
    \label{fig:legalize}
\end{figure}

Figure \ref{fig:legalize} shows the DAG before legalization. The first optimization took place by removing nodes that do not contribute to the DAG. However, the instructions are not, in LLVM terms, "legal" as these general SDNodes do not map directly to every target's instructions.  

\subsection{Instruction Legalization}

Figure \ref{fig:combine2} shows the DAG before the second optimization pass. The DAG is legalized by introducing RISCVISD::ADD\_LO and RISCVISD::HI nodes. These SDnodes act as flags to give target-specific information to target-independent algorithms. These definitions are introduced at lib/Target/RISCV/RISCVISelLowering.h file \cite{riscvIselh}. It is the RISCV DAG lowering interface. 
\par
According to the interface file, RISCVISD::ADD\_LO is meant to add Lo 12 bits from an address and to be replaced by ADDI (Add Immediate) at Instruction Selection. Similarly, RISCVISD::HI is meant to get Hi 20 bits from an address and to be replaced by LUI (Load Upper Immediate). With a legalized DAG the second optimization pass begins.
\subsection{Second Optimization Pass}
\begin{figure}
    \centering
    \includegraphics[width=0.9\textwidth]{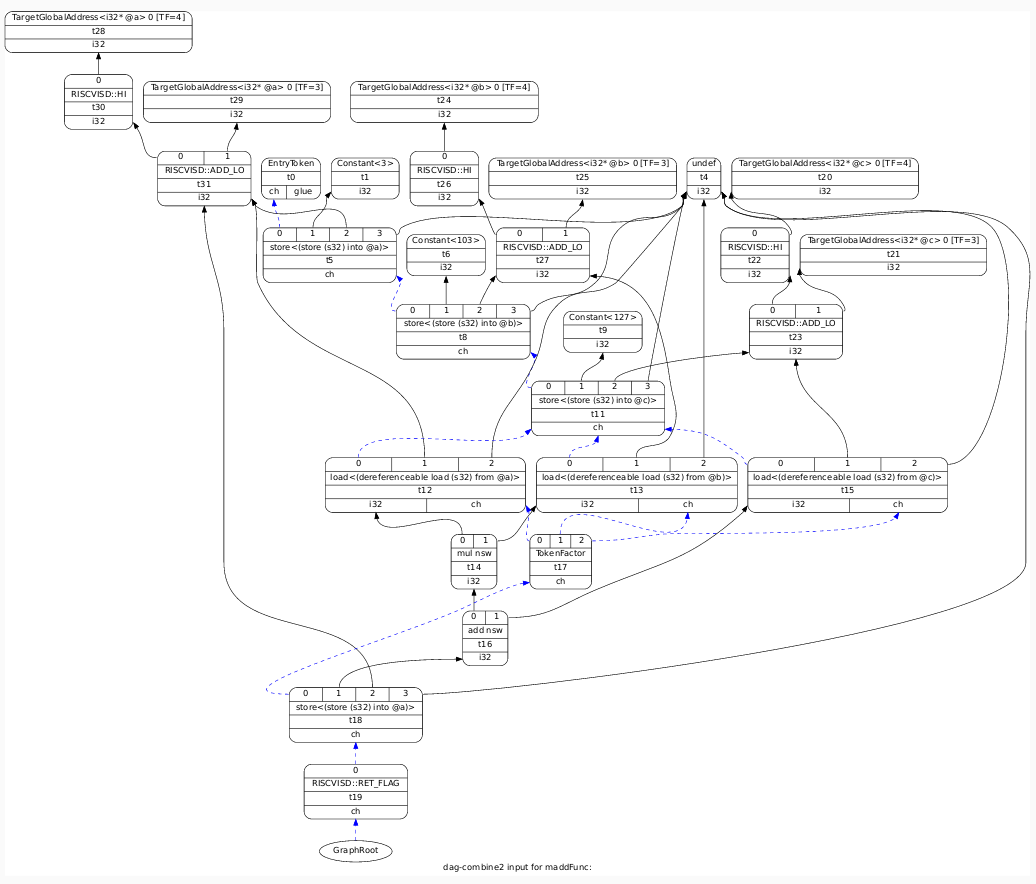}
    \caption{DAG before the second optimization}
    \label{fig:combine2}
\end{figure}

\begin{figure}
    \centering
    \includegraphics[width=0.9\textwidth]{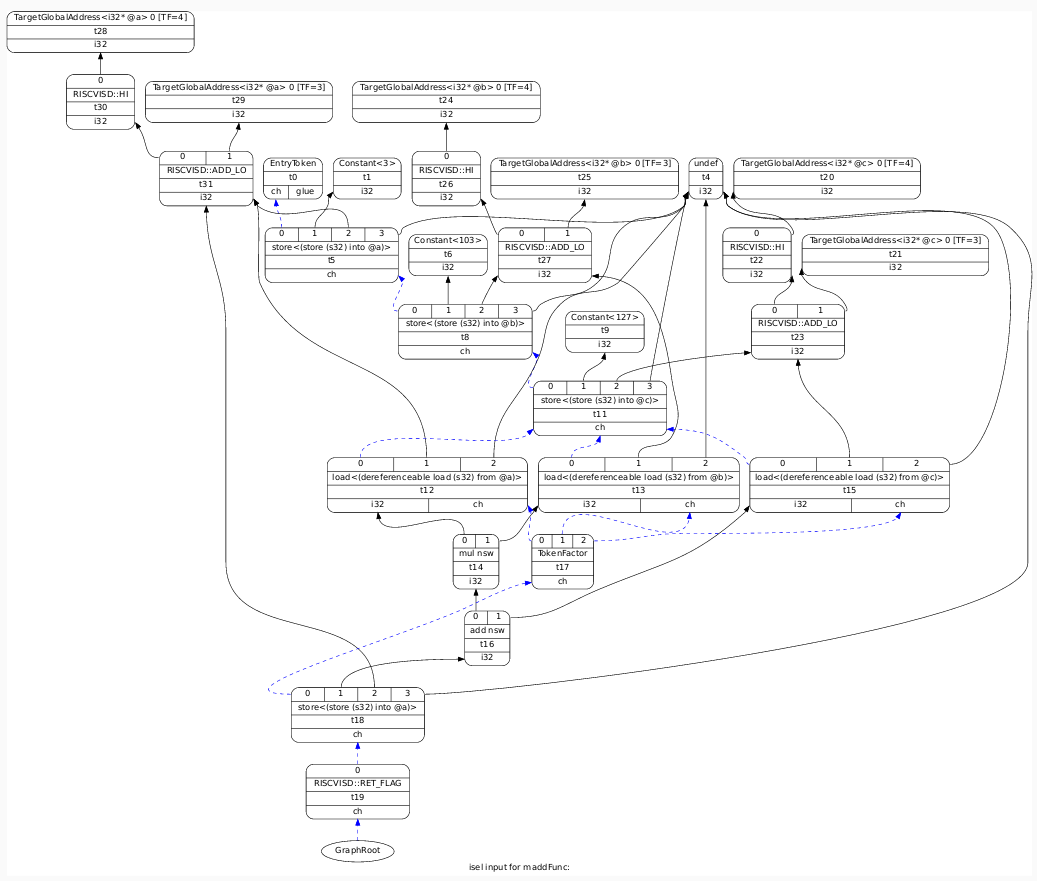}
    \caption{DAG before Instruction Selection}
    \label{fig:isel}
\end{figure}

Figure \ref{fig:isel} shows the DAG before the Instruction Selection phase. A comparison of Figure \ref{fig:combine2} and \ref{fig:isel} indicates that the second optimization did not change the DAG. This may be due to the reason that the subgraphs including the legalized nodes are not complex enough as the input C code is minimal.
\par
The DAG nodes up until Instruction Selection are instances of SDNode class which are target-independent nodes.

\subsection{Instruction Selection}
\begin{figure}
    \centering
    \includegraphics[width=0.9\textwidth]{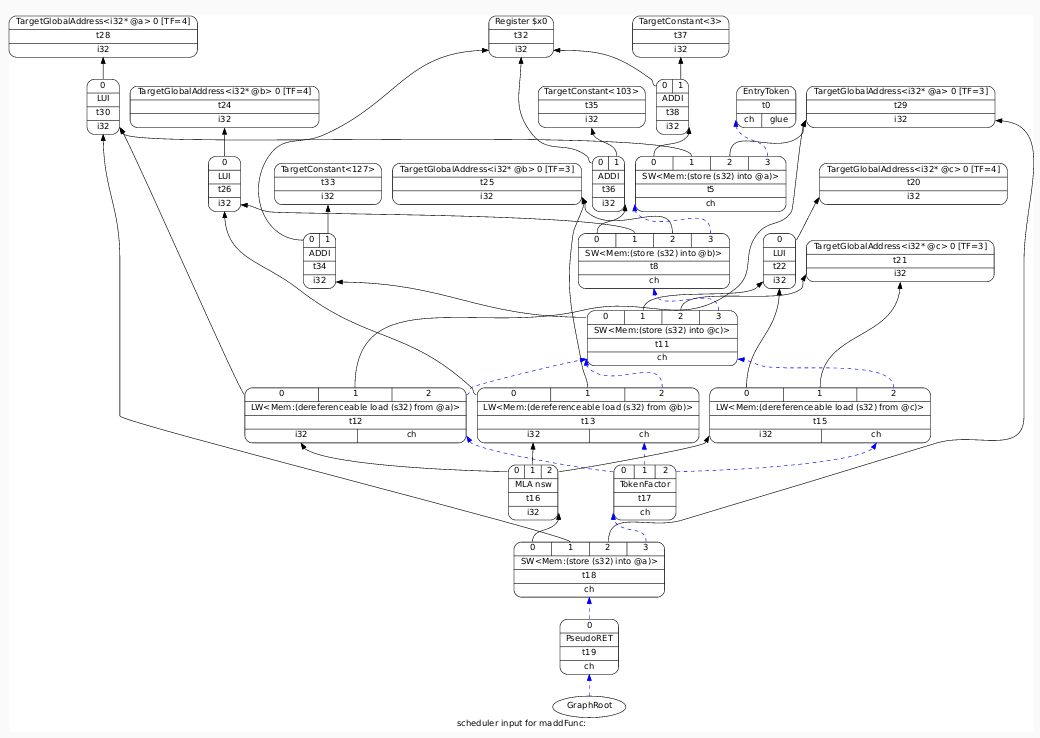}
    \caption{DAG before Instruction Scheduling}
    \label{fig:dag_sched}
\end{figure}

Figure \ref{fig:dag_sched} shows the DAG before the Instruction Scheduling phase. You can see that the instructions are selected according to the RISC-V target. SDNode class nodes are replaced by MachineSDNode class nodes which are target-specific.
\par
RISCVISD nodes are replaced by their counterparts. The general Load and Store instructions are replaced by their type-aware corresponding LW (Load Word) and SW (Store Word) instructions. Most importantly the MLA instruction is selected replacing the subgraph of 'mul' and 'add' LLVM instructions. 
\par
Our pattern definition of MLA instruction declares operand relations as in the subgraph. The instruction selection phase took it as a reference, detected the pattern inside the global DAG, and used it to place the MLA node. The pattern definition process is explained thoroughly in Section \ref{sec:MLA_add_section}.

\subsection{Instruction Scheduling}
\begin{figure}
    \centering
    \includegraphics[width=0.9\textwidth]{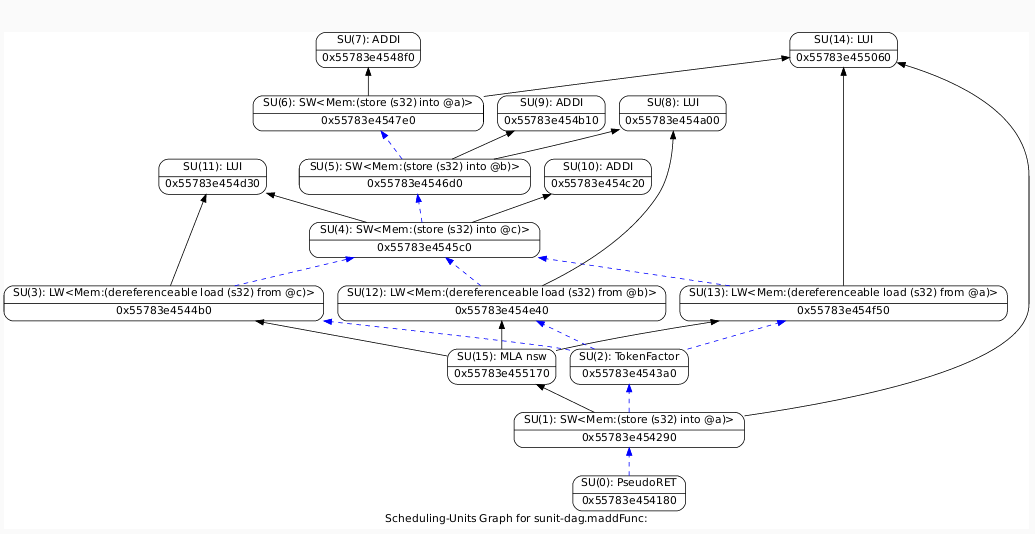}
    \caption{Scheduling Dependency Graph}
    \label{fig:sunit}
\end{figure}

The DAG is transformed into a target-specific DAG with the result of legalization and selection phases. However, to generate a linear byte sequence, the DAG must be flattened. The instruction scheduling phase gets the DAG and linearises it according to the dependency graph of nodes. The scheduling dependency can be seen in Figure \ref{fig:sunit}. Chain edges are used to show dependencies between instructions where one instruction cannot be placed before the other.

\subsection{Machine Instruction in SSA Form}
\begin{figure}
    \centering
    \includegraphics[width=0.9\textwidth]{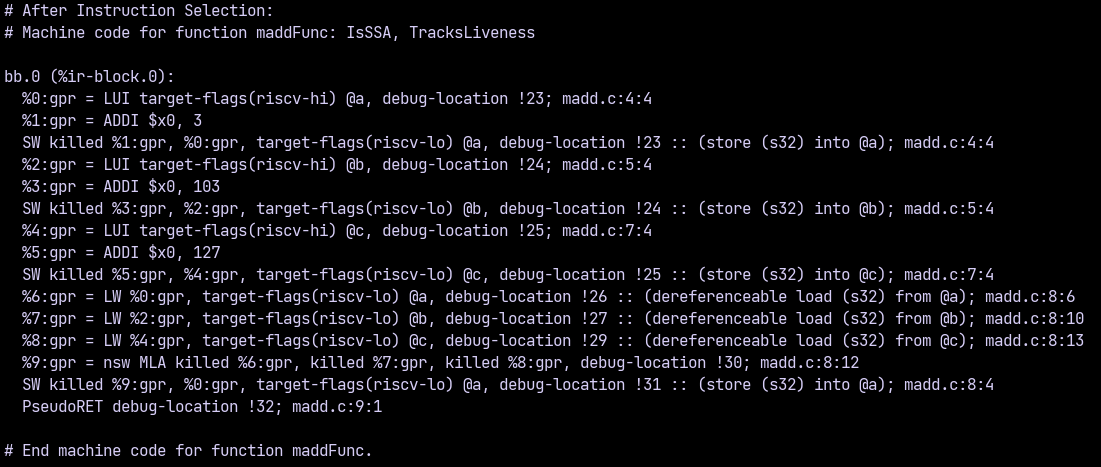}
    \caption{Machine Instruction before Register Allocation}
    \label{fig:mc_inst}
\end{figure}

The generated Machine Instruction as a result of scheduling is shown in Figure \ref{fig:mc_inst}. Because register allocation is not yet performed, the instructions are in SSA form. In SSA form, virtual registers are considered to be infinite unless some specific registers have to be used. In this case, '\$x0' is mentioned with ADDI instructions as they are hardwired zero in RISC-V. 

\clearpage
\section{Machine Code Instruction}
After register allocation, a Machine Code Instruction (MCInst) representation of the code is created. MCInst can be thought of as an IR of the lower-level code. It can be used to produce both an object file and an Assembly file. The generated Assembly is presented below:
\begin{lstlisting}[ caption=madd.s Assembly Output]
    maddFunc:           
# %
	addi	sp, sp, -16
.Ltmp0:
	sw	ra, 12(sp)     
	sw	s0, 8(sp)     
	addi	s0, sp, 16
	lui	a0, %
	li	a1, 3
	sw	a1, %
	lui	a1, %
	li	a2, 103
	sw	a2, %
	lui	a2, %
	li	a3, 127
	sw	a3, %
	lw	a3, %
	lw	a1, %
	lw	a2, %
	mla	a1, a3, a1 ,a2
	sw	a1, %
	lw	ra, 12(sp)   
	lw	s0, 8(sp)   
	addi	sp, sp, 16
	ret
\end{lstlisting}

\cleardoublepage
\clearpage
\chapter{ADDING CUSTOM INSTRUCTIONS}\label{ch:custom_instr}
The instruction selection system we focused on at the back end of the LLVM compiler is SelectionDAG among FastISel and GlobalIsel. SelectionDAG is the most mature Instruction Selection framework with more target support. However, shortly it is worth considering GlobalISel as it is developed recently as an alternative to SelectionDAG. The reasons to replace it are to make it faster, smaller, more testable and open to low-level optimizations.
\section{TableGen Reference}
TableGen is a domain-specific language used in the LLVM back end side to generate CPP header files. The purpose it serves is that it removes the redundancy of instruction declaration code which can be common to numerous architectures with minor differences. To maintain and scale the framework the minor differences are implemented level by level at a series of inheritance operations between TableGen classes. 

\par
 LLVM Static Compiler, LLC, is responsible for converting LLVM IR to Assembly codes. To add new instructions, changes are made in TableGen files and LLC is recompiled. During the compilation operation of the LLC program, TableGen records are created which declare every instruction’s encoding and describe its features. Referring to the records, DAGs are used in the process of instruction selection. 
 DAG is a graph structure that has no cycles and has directions on the edges.

 Operations or functions are represented as nodes in the DAG. They are critical parts of declaring the logic or pattern of the new instruction. 
\par

The operations represented on the DAG can be LLVM intrinsics as well as instructions. LLVM instructions resemble conventional assembly instructions, in contrast, LLVM intrinsics have higher level abstraction depending on their functionality. Their instruction generation may vary depending on the target hardware. It is possible to define a new complicated instruction either by combining simple LLVM instructions and higher-level intrinsics in the DAG level or by creating a new LLVM intrinsic which gets created at the Intermediate Level of the compilation process.

\section{RISC-V TableGen Classes}
The most general instruction class used for every target architecture is the “InstructionEncoding” TableGen class defined in llvm/include/llvm/Target/Target.td. This class holds the decoder method and size of instruction in addition to minor variables. It gets inherited by the generic “Instruction” class which is defined in the same class. This class holds input and output DAGs and information which is useful to the compiler and is generalizable to all architectures.
\par

The general class gets inherited by every target-specific class. In RISC-V’s case, the next stop of the instruction is the “RVInst” class which inherits from the general “Instruction” class and it resides in llvm/lib/Target/RISCV/RISCVInstrFormats.td TableGen file. It defines the general bit patterns of RISC-V instructions. For example, the opcode being the first 7 bits. It defines additional information like the assembly string pattern. This general class is inherited by every type of instruction of R, I, S, B, U, and J types. As a simple example, XOR instruction can be traced. As XOR is an R type, a register-register instruction, it continues its inheritance journey from “RVInstR”. It is common to R type instructions to have funct7, rs2, rs1, funct3, and rd format ordered from most significant bit (MSB) to least significant bit (LSB). These variables are assigned corresponding bit fields in the class. 
\par

The RISC-V formats mentioned are included in the llvm/lib/Target/RISCV/RISCVInstrInfo.td file which is in the same directory as the RISCVInstrFormats.td file. After inclusion, the “RVInstR” class gets inherited by the “ALU\_rr” class. The “ALU\_rr” class adds the commutability feature which means swapping source 1 and source 2 does not create a different result like in addition but not in subtraction. In the end, XOR’s record is defined by putting funct7, funct3 and assembly string manually in a single line with scheduling information added. 

\pagebreak
\section{Adding a New Instruction Using TableGen}\label{sec:MLA_add_section}

This section will guide the reader in introducing new instructions via TableGen.
Create a new TableGen file for custom additions and include it at the end of the RISCVInstrInfo.td file. We named it RISCVInstrInfoCrypt.td as it is going to be cryptography related.
\begin{lstlisting}[caption= Include file]
include "RISCVInstrInfoCrypt.td"
\end{lstlisting}

The specifications of the instruction will be added to the RISCVInstrInfoCrypt.td file.

\subsection{Introducing the ALU\_rrr TableGen Class}\label{sec:alurrr}

Here we created a new class of instruction named ALU\_rrr. MLA instruction requires three source registers and is defined to be ALU type so the specifications are:

\begin{lstlisting}[caption={ALU\_rrr class definition},label={lst:ALU_rrr}]
let hasSideEffects = 0, mayLoad = 0, mayStore = 0 in
class ALU_rrr<bits<2> funct2, bits<3> funct3, string opcodestr,
            bit Commutable = 0>
   : RVInstR4<funct2, funct3, OPC_OP, 
   (outs GPR:$rd), (ins GPR:$rs1, GPR:$rs2, GPR:$rs3),
             opcodestr, "$rd, $rs1, $rs2 ,$rs3"> {
 let isCommutable = Commutable;
}
\end{lstlisting}
The class is wrapped with three flags:
\begin{lstlisting}
let hasSideEffects = 0, mayLoad = 0, mayStore = 0 in
\end{lstlisting}

\begin{itemize}
    \item If the instruction has no side effect, \textit{hasSideEffects} will be 0.
    \item If there is no need or possibility to load data from memory, \textit{mayLoad} will be 0.
    \item If there is no need or possibility to store data from memory, \textit{mayStore} will be 0.
\end{itemize}

Here you can see class arguments:
\begin{lstlisting}
class ALU_rrr<bits<2> funct2, bits<3> funct3, string opcodestr,
            bit Commutable = 0>
\end{lstlisting}

Class is defined with ALU\_rrr name. Variables are defined. \textit{funct2} is a two-bit binary number as RVInstR4 is used which reserves 5 bits of \textit{funct7} for another register. \textit{funct3} is a three-bit binary number. \textit{opcodestr} is the string that will be shown in the assembly file. \textit{Commutable} is a zero bit which determines the importance of the order of the inputs.

\begin{lstlisting}
: RVInstR4<funct2, funct3, OPC_OP,
(outs GPR:$rd), (ins GPR:$rs1, GPR:$rs2, GPR:$rs3),
\end{lstlisting}

RVInstR4 instruction type is called from RISCVInstrFormats.td file. \textit{funct2}, \textit{funct3}, \textit{opcode}, output and inputs are given as arguments to the higher class in order.

\begin{lstlisting}
opcodestr, "$rd, $rs1, $rs2 ,$rs3"> {
 let isCommutable = Commutable;
}
\end{lstlisting}

Opcode string used for Assembly and activating commutability option.

\subsection{Introducing the MLA Instruction for Assembler Support}

The definition of the instruction can now be added using the ALU\_rrr class defined above and by choosing the correct scheduling variables. For the MLA instruction, it is:

\begin{lstlisting}
def MLA : ALU_rrr<0b10, 0b100, "mla">,
Sched<[WriteIMul, ReadIMul, ReadIMul]>;
\end{lstlisting}

MLA instruction is defined and ALU\_rrr instruction type is used. funct2,funct3, opcode string and schedules are sufficient to have the full definition of the instruction thanks to the custom ALU\_rrr class.

\subsection{Introducing the MLA Instruction for Pattern Matching Support}
Add the instruction’s pattern defining source to the target custom instruction transformation. For the MLA instruction, it is:

\begin{lstlisting}
def : Pat< (add (mul GPR:$src1, GPR:$src2), GPR:$src3),
(MLA GPR:$src1, GPR:$src2, GPR:$src3)>;
\end{lstlisting}

Note that it is possible to define more patterns to introduce optimizations. There can be multiple source patterns for the same target pattern. The target pattern can also be a tree of SDNodes containing the custom instruction.

\section{Adding Pattern Matching Support for New Instruction Using C++ in SelectionDAG}\label{sec:cpp}
TableGen aims to provide a declarative way to introduce new patterns for new instruction developers. However, not all instructions can be described in this scheme.  Although it is called "dag" as a keyword in TableGen, it expects a tree of instructions. For certain use cases, custom C++ can be the only way to match until the TableGen based system improves. It is also possible to use C++ in complex patterns together with TableGen, which can make the most of the pattern declarative and only the necessary part in imperative style.
\par
A domain that TableGen fails is matching a graph of instructions with dependant operands. As an example, we can think of an instruction having two operands of Load Instructions. If the Load instructions are from an array, they must be related to each other by an offset and it might need to be detected for certain patterns.

\begin{lstlisting}[caption={Minimal Subtree of Optimized S-box LLVM IR},linerange={1-5},label={lst:sbox-xor},language=llvm,style=nasm]
define void @sbox(ptr  %0)  {
  %2 = getelementptr inbounds [5 x i32], ptr %0, i32 0, i32 4
  %3 = load i32, ptr %2
  %4 = load i32, ptr %0
  %5 = xor i32 %4, %3
\end{lstlisting}

In Code \ref{lst:sbox-xor}, it can be seen that the XOR instruction is between the first element of the input struct and the fifth element of it. As the locations of elements matter, they must be matched by not only looking at Load instructions but also their operands. What we are looking for is to have the base of Load instruction to be equal and the offset operand of it to evaluate to 4, designating the fifth element of the struct. TableGen is not suitable for this operation and the source code of SelectionDAG's RISC-V backend should be analyzed to place the logic to pattern match this set of instructions.

The process for adding an instruction via C++ is as follows:
\begin{enumerate}
    \item Create a record declaration in TableGen to provide the Assembler support, ignoring the Pattern declaration.
    \item Observe the DAG in the debug output or dot file and locate the root of it. 
    \item Add a function in RISCVISelDAGToDAG.cpp file in the root instruction case. 
    \item Implement pattern matching and replacement with SDnode.
\end{enumerate}

SelectionDAG consists of numerous files but the most relevant ones to the developer can be few if the complexity of the pattern is small. The order of files will be from IR to Assembly. RISCVCodeGenPrepare.cpp file provides mechanisms for matching in IR form. This file exists mainly due to the limitation of SelectionDAG which is running per basic block. RISCVISelLowering.cpp contains the lowering of IR to SDnodes. It can decide on whether a type or expression should be legalized or expanded. RISCVISelDAGToDAG.cpp is the instruction selector in C++. Its implementation traverses the DAG from the root and runs the selection functions depending on the SDnode type which is parallel to instructions.

\begin{lstlisting}[caption={The corresponding Optimized and Legalized DAG of Code \ref{lst:sbox-xor}},linerange={3-3,4-4,8-9,13-13,16-16},label={lst:sbox-xor-dag},language=llvm,style=nasm]
  t0: ch,glue = EntryToken
  t2: i32,ch = CopyFromReg t0, Register:i32 %0
  t8: i32,ch = load<(load (s32) from %ir.0, !tbaa !7)> t0, t2, undef:i32
  t4: i32 = add nuw t2, Constant:i32<16>
  t7: i32,ch = load<(load (s32) from %ir.2, !tbaa !7)> t0, t4, undef:i32
  t9: i32 = xor t8, t7
\end{lstlisting}

In an attempt to match the three instructions in Code \ref{lst:sbox-xor}, the DAG in Code \ref{lst:sbox-xor-dag} is analyzed. The first remark is that "getelementptr" is converted to an add instruction which calculates the offset. Another remark is that the load instructions have the same base address in the first operand as expected pointing to the same node. If the pattern is large, we can introduce a new function which will contain the logic. The function's prototype should be added to the corresponding header file.

\begin{lstlisting}[caption={Introduction of New Function for Pattern Matching in C++},label={lst:sbox-iseldagtodag},language=C++]
void RISCVDAGToDAGISel::Select(SDNode *Node) {
    .
}
    .
switch (Opcode) {
    .
    .
    case ISD::XOR:{
        if (tryShrinkShlLogicImm(Node))
          return;
        if (selectSbox(Node))
          return;
        break;
    }
}
\end{lstlisting}

The C++ logic for matching this pattern is provided below.

\begin{lstlisting}[caption={C++ logic for Pattern Matching the DAG in Code \ref{lst:sbox-xor-dag}},label={lst:sbox-cpp},language=C++]
bool RISCVDAGToDAGISel::selectSBox(SDNode *Node) {
    SDValue LOAD0 = Node->getOperand(0);
    SDValue LOAD1 = Node->getOperand(1);
    //Check if there is a load pair
    if(LOAD0.getOpcode() != ISD::LOAD 
            || LOAD1.getOpcode() != ISD::LOAD)
      return false;
    SDValue LOAD0_op0 = LOAD0.getOperand(0);
    SDValue LOAD0_op1_offset = LOAD0.getOperand(1); // t0

    SDValue LOAD1_op0 = LOAD1.getOperand(0); 
    SDValue LOAD1_op1_offset = LOAD1.getOperand(1); 

    if(LOAD1_op1_offset.getOpcode() != ISD::ADD)
        return false;

    if(LOAD1_op0 != LOAD0_op0)
        return false;

   //Check if the addendum0 is the same as the second 
    SDValue LOAD1_op1_off_Addend0 = LOAD1_op1_offset.getOperand(0);
    if(LOAD1_op1_off_Addend0 != LOAD0_op1_offset)
        return false;

    SDValue LOAD1_op1_off_Addend1 = LOAD1_op1_offset.getOperand(1);

    auto *LOAD1_op1_offset_Addendum1C = 
        dyn_cast<ConstantSDNode>(LOAD1_op1_offset_Addendum1);
    if(!LOAD1_op1_offset_Addendum1C)
        return false;
    //Check if addendum1 is 16 more than first load offset
    if(LOAD1_op1_offset_Addendum1C->getZExtValue() != 16)
        return false;

    //Pattern is matched replace here

    return true;
}


\end{lstlisting}

The logic starts from the root of the DAG which in this case is XOR instruction or the ISD::XOR SDnode. Then we iterate through its leaves and check the distinctive features of the pattern. In this pattern, we are interested in the distance of offset addresses of Load instructions so we progressively approach them by assuming the pattern holds and quitting if not. Progressive checking is a common theme in LLVM and as Instruction Selection is one of the stages affecting the compilation times significantly, patterns should not be checked in a single if statement by logical combinations. 

The checks, if statements are doing can be summarised in steps. As the function only runs when the instruction selection finds an XOR SDNode, the root can be assumed as XOR safely.
\begin{enumerate}
    \item Check if both the operands are Load instructions.
    \item Check if the second Load instruction has an Add instruction in its second operand
    \item Check if the base offset of the first load and the first addendum of the second load are the same, as they should point to the beginning of the struct.
    \item Check if the second addendum is a constant.
    \item Check if the unsigned value of constant second addendum is equal to 16.
\end{enumerate}

At this point, it can be safely assumed that the only XOR that conforms to the pattern can be in this line of program. SelectionDAG provides more API to simply replace the nodes in that pattern with the custom instruction.

To interact with the DAG, SelectionDAG's API is used. We encourage the developers to read the source code and learn to use the public functions exposed by SelectionDAG in order to interact with the DAG most effectively. RISCVISelDAGToDAG.cpp file already has many instruction selection mechanisms in place which can be read through. 

\section{Discussion of Pattern Matching in Other Stages of the Compiler}\label{sec:patmatchdisc}
Pattern Matching can be assumed to mainly be an Instruction Selection problem where the pattern will be simply identified and replaced. However, when we take a look at the baseline problem any Compiler technology solves, it is to convert more familiar patterns in some language to a more unfamiliar pattern in machine language. Also, this conversion occurs in a large number of steps through optimization in IR form as discussed in Section \ref{sec:opt}, to DAG formation in SelectionDAG to MCInstr form down the pipeline. Their data structures can differ in representing the instructions which can make some pattern matching schemes to be more fragile than others. Also, as the lowering gets performed high-level information about the program is lost but the formation gets closer to the final output of Assembly. 

For simple cases where for example a combination of R-type instructions will be matched and replaced, Instruction Selection might be the most convenient stage to extend. However if the pattern requires the instruction selection to be performed already, pattern match can be done in the MC layer. On the contrary, if higher level information of the pattern is required, a pattern can be matched to an intrinsic function at the IR level. 

Another reason to consider different stages is that there can be multiple patterns mapped to the same instruction. Further optimization opportunities can rise in further stages. 

\subsection{Case Study: SH1ADD in SelectionDAG and MC Layer}

It was discussed that dealing with the lowered DAG to MC layer can provide more optimization opportunities. In this section, the case of "SH1ADD" instruction which is ratified in the RISC-V Zba extension will be analyzed. The instruction shifts rs1 left by one, adds rs2 and writes to rd. Its encoding and pattern in the standard implementation of LLVM in TableGen are as follows:

\begin{lstlisting}[%
caption={Instruction Encoding of the Instructions} ]
let Predicates = [HasStdExtZba] in {
def SH1ADD : ALU_rr<0b0010000, 0b010, "sh1add">,
             Sched<[WriteSHXADD, ReadSHXADD, ReadSHXADD]>;
def SH2ADD : ALU_rr<0b0010000, 0b100, "sh2add">,
             Sched<[WriteSHXADD, ReadSHXADD, ReadSHXADD]>;
def SH3ADD : ALU_rr<0b0010000, 0b110, "sh3add">,
             Sched<[WriteSHXADD, ReadSHXADD, ReadSHXADD]>;
} // Predicates = [HasStdExtZba]
\end{lstlisting}

As the SH2ADD and SH3ADD have similar implementations to SH1ADD their patterns will be stripped.

\begin{lstlisting}[%
, caption={Instruction Pattern of the Instructions}]
let Predicates = [HasStdExtZba] in {
def : Pat<(add (shl GPR:$rs1, (XLenVT 1)), non_imm12:$rs2),
          (SH1ADD GPR:$rs1, GPR:$rs2

// More complex cases use a ComplexPattern.
def : Pat<(add sh1add_op:$rs1, non_imm12:$rs2),
          (SH1ADD sh1add_op:$rs1, GPR:$rs2)>;)>;

def : Pat<(add (mul_oneuse GPR:$rs1, (XLenVT 6)), GPR:$rs2),
          (SH1ADD (SH1ADD GPR:$rs1, GPR:$rs1), GPR:$rs2)>;
def : Pat<(add (mul_oneuse GPR:$rs1, (XLenVT 10)), GPR:$rs2),
          (SH1ADD (SH2ADD GPR:$rs1, GPR:$rs1), GPR:$rs2)>;
def : Pat<(add (mul_oneuse GPR:$rs1, (XLenVT 18)), GPR:$rs2),
          (SH1ADD (SH3ADD GPR:$rs1, GPR:$rs1), GPR:$rs2)>;
\end{lstlisting}

We can observe that ComplexPattern's  are used to enable using C++ together with TableGen. The Complex Pattern's TableGen declarations are provided below:

\begin{minipage}{\linewidth}
\begin{lstlisting}[language=C++, caption={TableGen Declaration of ComplexPatterns}]

def sh1add_op : ComplexPattern<XLenVT, 1, 
                        "selectSHXADDOp<1>", [], [], 6>;

class binop_oneuse<SDPatternOperator operator>
    : PatFrag<(ops node:$A, node:$B),
              (operator node:$A, node:$B), [{
  return N->hasOneUse();
}]>;

def mul_oneuse : binop_oneuse<mul>;
\end{lstlisting}
\end{minipage}

"selectSHXADDOp" is a template function which provides the shift amount argument.

\begin{lstlisting}[language=C++, caption={Template Function of the ComplexPattern for "sh1add\_op"}]

bool selectSHXADDOp(SDValue N, unsigned ShAmt, SDValue &Val);
template <unsigned ShAmt> bool 
                      selectSHXADDOp(SDValue N, SDValue &Val) {
return selectSHXADDOp(N, ShAmt, Val);
}
\end{lstlisting}

The C++ logic can be found in RISCVISelDAGToDAG.cpp file, the implementation will be reduced to the patterns described in the comments:

\begin{lstlisting}[language=C++, caption={Implementation of the ComplexPattern for "sh1add\_op"}]
/// Look for various patterns that can be done with a SHL that can be 
/// folded into a SHXADD. \p ShAmt contains 1, 2, or 3 and is set based 
/// on which SHXADD we are trying to match.
bool RISCVDAGToDAGISel::selectSHXADDOp(SDValue N, unsigned ShAmt,
                                       SDValue &Val) {
  if (N.getOpcode() == ISD::AND 
                      && isa<ConstantSDNode>(N.getOperand(1))) {
    SDValue N0 = N.getOperand(0);

    bool LeftShift = N0.getOpcode() == ISD::SHL;
    if ((LeftShift || N0.getOpcode() == ISD::SRL) &&
        isa<ConstantSDNode>(N0.getOperand(1))) {
      uint64_t Mask = N.getConstantOperandVal(1);
      unsigned C2 = N0.getConstantOperandVal(1);

      unsigned XLen = Subtarget->getXLen();
      if (LeftShift)
        Mask &= maskTrailingZeros<uint64_t>(C2);
      else
        Mask &= maskTrailingOnes<uint64_t>(XLen - C2);

      // Look for (and (shl y, c2), c1) where c1 is a shifted mask 
      // with no leading zeros and c3 trailing zeros. We can use an 
      // SRLI by c2+c3 followed by a SHXADD with c3 for the X amount.
       ...
        // Look for (and (shr y, c2), c1) where c1 is a shifted mask 
        // with c2 leading zeros and c3 trailing zeros. We can use an
        // SRLI by C3 followed by a SHXADD using c3 for the X amount.
        ...
      }
    }
  }

  bool LeftShift = N.getOpcode() == ISD::SHL;
  if ((LeftShift || N.getOpcode() == ISD::SRL) &&
      isa<ConstantSDNode>(N.getOperand(1))) {
    SDValue N0 = N.getOperand(0);
    if (N0.getOpcode() == ISD::AND && N0.hasOneUse() &&
        isa<ConstantSDNode>(N0.getOperand(1))) {
      uint64_t Mask = N0.getConstantOperandVal(1);
      if (isShiftedMask_64(Mask)) {
        unsigned C1 = N.getConstantOperandVal(1);
        unsigned XLen = Subtarget->getXLen();
        unsigned Leading = XLen - llvm::bit_width(Mask);
        unsigned Trailing = llvm::countr_zero(Mask);
        // Look for (shl (and X, Mask), C1) where Mask has 32 leading 
        // zeros and C3 trailing zeros. 
        // If C1+C3==ShAmt we can use SRLIW+SHXADD.
        ...
        // Look for (srl (and X, Mask), C1) where Mask has 32 leading
        // zeros and C3 trailing zeros. 
        // If C3-C1==ShAmt we can use SRLIW+SHXADD.
        ...
      }
    }
  }
  return false;
}
\end{lstlisting}

Despite using TableGen and SelectionDAG, there is an optimization opportunity for "SH1ADD" in MCInst form. It is done in immediate materialization where a constant node in the DAG representation is not converted to instructions until needed. In the MCTargetDesc/RISCVMatInt.cpp file, an optimization for representing immediates is provided as follows:

\begin{lstlisting}[language=C++, caption={Immediate Materialization for "SH1ADD"}]
namespace llvm::RISCVMatInt {
InstSeq generateInstSeq(int64_t Val,
                              const FeatureBitset &ActiveFeatures) {
  RISCVMatInt::InstSeq Res;
  generateInstSeqImpl(Val, ActiveFeatures, Res);
  ...

// Perform optimization with SH*ADD in the Zba extension.
  if (Res.size() > 2 && ActiveFeatures[RISCV::FeatureStdExtZba]) {
    int64_t Div = 0;
    unsigned Opc = 0;
    RISCVMatInt::InstSeq TmpSeq;
    // Select the opcode and divisor.
    if ((Val %
      Div = 3;
      Opc = RISCV::SH1ADD;
    } else if ((Val %
      Div = 5;
      Opc = RISCV::SH2ADD;
    } else if ((Val %
      Div = 9;
      Opc = RISCV::SH3ADD;
    }
    // Build the new instruction sequence.
    if (Div > 0) {
      generateInstSeqImpl(Val / Div, ActiveFeatures, TmpSeq);
      TmpSeq.emplace_back(Opc, 0);
      if (TmpSeq.size() < Res.size())
        Res = TmpSeq;
    } else {
      // Try to use LUI+SH*ADD+ADDI.
      int64_t Hi52 = ((uint64_t)Val + 0x800ull) & ~0xfffull;
      int64_t Lo12 = SignExtend64<12>(Val);
      Div = 0;
      if (isInt<32>(Hi52 / 3) && (Hi52 %
        Div = 3;
        Opc = RISCV::SH1ADD;
      } else if (isInt<32>(Hi52 / 5) && (Hi52 %
        Div = 5;
        Opc = RISCV::SH2ADD;
      } else if (isInt<32>(Hi52 / 9) && (Hi52 %
        Div = 9;
        Opc = RISCV::SH3ADD;
      }
      // Build the new instruction sequence.
      if (Div > 0) {
        // For Val that has zero Lo12 (implies Val equals to Hi52) 
        // should has already been processed to LUI+SH*ADD
        // by previous optimization.
        generateInstSeqImpl(Hi52 / Div, ActiveFeatures, TmpSeq);
        TmpSeq.emplace_back(Opc, 0);
        TmpSeq.emplace_back(RISCV::ADDI, Lo12);
        if (TmpSeq.size() < Res.size())
          Res = TmpSeq;
      }
    }
  }

\end{lstlisting}

By using "SH1ADD" in immediate materialization, a single instruction can be used instead of two. It can be checked that in the standard testing suite, there is the following function which returns a 64-bit integer:
\begin{minipage}{\linewidth}
\begin{lstlisting}[language=llvm,style=nasm, caption={Function for Immediate Materialization}]
define i64 @PR54812() {
; RV64I-LABEL: PR54812:
; RV64I:       # %
; RV64I-NEXT:    lui a0, 1048447
; RV64I-NEXT:    addiw a0, a0, 1407
; RV64I-NEXT:    slli a0, a0, 12
; RV64I-NEXT:    ret
;
; RV64IZBA-LABEL: PR54812:
; RV64IZBA:       # %
; RV64IZBA-NEXT:    lui a0, 872917
; RV64IZBA-NEXT:    sh1add a0, a0, a0
; RV64IZBA-NEXT:    ret
;
  ret i64 -2158497792;
}
\end{lstlisting}
\end{minipage}
In the FileCheck lines which are explained in Section \ref{sec:llc_test}, we can see the Assembly lines that should be emitted. It can be observed that the desired immediate can be obtained with "sh1add" in fewer instructions.

\subsection{Case Study: ROR in SelectionDAG and IR Level}

In LLVM, rotation instruction which is shifting and feeding the carry back to the shift point is captured at the IR level. To match in IR level, a new intrinsic function is defined in TableGen. TableGen is not only used in instruction selection, it is used wherever a declarative form is better suited such as in IR or MLIR levels.

\begin{lstlisting}[language=C++, caption={Funnel Shift Intrinsic Definition}]
//===-------------------- Bit Manipulation Intrinsics ---------===//
//

// None of these intrinsics accesses memory at all.
let IntrProperties = [IntrNoMem, IntrSpeculatable, IntrWillReturn]
                                                                in {
  def int_bswap: DefaultAttrsIntrinsic<[llvm_anyint_ty],
                                                 [LLVMMatchType<0>]>;
  def int_ctpop: DefaultAttrsIntrinsic<[llvm_anyint_ty],
                                                 [LLVMMatchType<0>]>;
  def int_bitreverse: DefaultAttrsIntrinsic<[llvm_anyint_ty],
                                                 [LLVMMatchType<0>]>;
  def int_fshl : DefaultAttrsIntrinsic<[llvm_anyint_ty],
      [LLVMMatchType<0>, LLVMMatchType<0>, LLVMMatchType<0>]>;
  def int_fshr : DefaultAttrsIntrinsic<[llvm_anyint_ty],
      [LLVMMatchType<0>, LLVMMatchType<0>, LLVMMatchType<0>]>;
}
\end{lstlisting}

'fshr' is defined as funnel shift right intrinsic function \cite{llvmref-fshl}. It is matched in IR optimizations by InstCombine pass.

\begin{lstlisting}[language=C++, caption={Funnel Shift Right Pattern Matching}]
/// Match UB-safe variants of the funnel shift intrinsic.
static Instruction *matchFunnelShift(Instruction &Or,
                                              InstCombinerImpl &IC){
  unsigned Width = Or.getType()->getScalarSizeInBits();

  // First, find an or'd pair of opposite shifts:
  // or (lshr ShVal0, ShAmt0), (shl ShVal1, ShAmt1)
  BinaryOperator *Or0, *Or1;
  if (!match(Or.getOperand(0), m_BinOp(Or0)) ||
      !match(Or.getOperand(1), m_BinOp(Or1)))
    return nullptr;

  Value *ShVal0, *ShVal1, *ShAmt0, *ShAmt1;
  if (!match(Or0, m_OneUse(m_LogicalShift(m_Value(ShVal0),
       m_Value(ShAmt0)))) ||
      !match(Or1, m_OneUse(m_LogicalShift(m_Value(ShVal1), 
       m_Value(ShAmt1)))) ||
      Or0->getOpcode() == Or1->getOpcode())
    return nullptr;

  // Canonicalize to or(shl(ShVal0, ShAmt0), lshr(ShVal1, ShAmt1)).
  if (Or0->getOpcode() == BinaryOperator::LShr) {
    std::swap(Or0, Or1);
    std::swap(ShVal0, ShVal1);
    std::swap(ShAmt0, ShAmt1);
  }

  // Match the shift amount operands for a funnel shift pattern. 
  // This always matches a subtraction on the R operand.
  auto matchShiftAmount = [&](Value *L, Value *R, unsigned Width) 
                                                       -> Value * {
    // Check for constant shift amounts that sum to the bitwidth.
    ...

    return nullptr;
  };

  Value *ShAmt = matchShiftAmount(ShAmt0, ShAmt1, Width);
  bool IsFshl = true; // Sub on LSHR.
  if (!ShAmt) {
    ShAmt = matchShiftAmount(ShAmt1, ShAmt0, Width);
    IsFshl = false; // Sub on SHL.
  }
  if (!ShAmt)
    return nullptr;

  Intrinsic::ID IID = IsFshl ? Intrinsic::fshl : Intrinsic::fshr;
  Function *F = Intrinsic::getDeclaration(Or.getModule(), IID,
                                                   Or.getType());
  return CallInst::Create(F, {ShVal0, ShVal1, ShAmt});
}
\end{lstlisting}

The pattern matching API is provided by IR/PatternMatch.h file. After the pattern for rotation and the shift amount are matched the corresponding intrinsic function is called. 

The function is called in OR visiting function, so the root of the pattern is OR:

\begin{lstlisting}[language=C++, caption={Funnel Shift Right Pattern Function Called}]
Instruction *InstCombinerImpl::visitOr(BinaryOperator &I) {
  ...
  if (Instruction *Funnel = matchFunnelShift(I, *this))
    return Funnel;
  ...
  return nullptr;
}


\end{lstlisting}

The intrinsic function is converted to ROTR SDnode in the general SelectionDAGBuilder.cpp file.
\begin{lstlisting}[language=C++, caption={Funnel Shift Intrinsic converted to ROTL}]
/// Lower the call to the specified intrinsic function.
void SelectionDAGBuilder::visitIntrinsicCall(const CallInst &I,
                                             unsigned Intrinsic) {
  const TargetLowering &TLI = DAG.getTargetLoweringInfo();
  SDLoc sdl = getCurSDLoc();
  DebugLoc dl = getCurDebugLoc();
  SDValue Res;

  SDNodeFlags Flags;
  if (auto *FPOp = dyn_cast<FPMathOperator>(&I))
    Flags.copyFMF(*FPOp);

  switch (Intrinsic) {
  default:
    // By default, turn this into a target intrinsic node.
    visitTargetIntrinsic(I, Intrinsic);
    return;
  ...

  case Intrinsic::fshl:
  case Intrinsic::fshr: {
    bool IsFSHL = Intrinsic == Intrinsic::fshl;
    SDValue X = getValue(I.getArgOperand(0));
    SDValue Y = getValue(I.getArgOperand(1));
    SDValue Z = getValue(I.getArgOperand(2));
    EVT VT = X.getValueType();

    if (X == Y) {
      auto RotateOpcode = IsFSHL ? ISD::ROTL : ISD::ROTR;
      setValue(&I, DAG.getNode(RotateOpcode, sdl, VT, X, Z));
    } else {
      auto FunnelOpcode = IsFSHL ? ISD::FSHL : ISD::FSHR;
      setValue(&I, DAG.getNode(FunnelOpcode, sdl, VT, X, Y, Z));
    }
    return;
  }
  ...
}

\end{lstlisting}

The intrinsic functions can be defined as target-specific or target independent. "fshl" is a general intrinsic function and targets can either expand it by replacing it with its equivalent instructions or lower it directly to an instruction by legalizing it. RISC-V Zbb extension supports bitwise rotation so LLVM has the extension's implementation in the source. In RISCVISelLowering.cpp file the legalization of "ROTR" is managed regarding whether the extension is enabled or disabled.

\begin{lstlisting}[language=C++, caption={ROTR Legalization Conditional}]
RISCVTargetLowering::RISCVTargetLowering(const TargetMachine &TM,
                                         const RISCVSubtarget &STI)
    : TargetLowering(TM), Subtarget(STI) {
    ...
  if (Subtarget.hasStdExtZbb() || Subtarget.hasStdExtZbkb() ||
      Subtarget.hasVendorXTHeadBb()) {
    if (Subtarget.is64Bit())
      setOperationAction({ISD::ROTL, ISD::ROTR}, MVT::i32, Custom);
  } else {
    setOperationAction({ISD::ROTL, ISD::ROTR}, XLenVT, Expand);
  }
  ...
}
\end{lstlisting}

The "Custom" action is defined in TableGen in RISCVInstrInfoZb.td file as well as instruction encodings.

\begin{lstlisting}[language=C++, caption={ROR Encodings and Pattern}]
def riscv_rolw   : SDNode<"RISCVISD::ROLW",   SDT_RISCVIntBinOpW>;
def riscv_rorw   : SDNode<"RISCVISD::RORW",   SDT_RISCVIntBinOpW>;
...
let Predicates = [HasStdExtZbbOrZbkb] in {
def ROL   : ALU_rr<0b0110000, 0b001, "rol">,
            Sched<[WriteRotateReg, ReadRotateReg, ReadRotateReg]>;
def ROR   : ALU_rr<0b0110000, 0b101, "ror">,
            Sched<[WriteRotateReg, ReadRotateReg, ReadRotateReg]>;

def RORI  : RVBShift_ri<0b01100, 0b101, OPC_OP_IMM, "rori">,
            Sched<[WriteRotateImm, ReadRotateImm]>;
} // Predicates = [HasStdExtZbbOrZbkb]
...
let Predicates = [HasStdExtZbbOrZbkb] in {
def : PatGprGpr<shiftop<rotl>, ROL>;
def : PatGprGpr<shiftop<rotr>, ROR>;

def : PatGprImm<rotr, RORI, uimmlog2xlen>;
// There's no encoding for roli in the the 'B' extension as it can be
// implemented with rori by negating the immediate.
def : Pat<(rotl GPR:$rs1, uimmlog2xlen:$shamt),
          (RORI GPR:$rs1, (ImmSubFromXLen uimmlog2xlen:$shamt))>;
} // Predicates = [HasStdExtZbbOrZbkb]
\end{lstlisting}

As you can see when the pattern match logic is lifted up to the IR level the modifications in Instruction Selection are straightforward to implement.

\cleardoublepage
\clearpage
\chapter{NEW INSTRUCTIONS}\label{Ch9}
This chapter contains more examples and demonstrations of new instructions.

\section{SHLXOR Instruction}\label{sec:shlxor}
The purpose of SHLXOR instruction is to shift the first source operand one bit to the left and then XOR it with the second source operand. Then, the obtained result is stored in the destination register. By adding this instruction, we can perform this operation with a single instruction instead of using shift left and XOR instructions separately, making it more efficient.
Let’s give an example to make it clearer what this instruction does. RS1 and RS2 are source operands and RD is the output.
\par
RS1: 0x0101     RS2: 0xFFFF     RD: 0xFDFD
\par
0x0101 is shifted left by one and then XOR’ed with 0xFFFF, giving the result 0xFDFD.
As we mentioned, this new instruction requires two source registers and one destination register. Therefore, unlike the MLA instruction, we don’t need to create a new class to support it. There is already a class named ALU\_rr in the RISCVInstrinfo.td file that has two source and one destination register. Therefore, the new SHLXOR instruction is going to belong to the ALU\_rr class. The ALU\_rr class definition is given below.

\begin{lstlisting}
let hasSideEffects = 0, mayLoad = 0, mayStore = 0 in
class ALU_rr<bits<7> funct7, bits<3> funct3, string opcodestr,
bit Commutable = 0>
: RVInstR<funct7, funct3, OPC_OP, (outs GPR:$rd),
                                   (ins GPR:$rs1, GPR:$rs2),
opcodestr, "$rd, $rs1, $rs2"> {
let isCommutable = Commutable;
}
\end{lstlisting}

As we can see, encoding of this type of instruction consists of funct7, funct3, opcode, source registers, and the destination register. The encoding format and the other properties are described in the class. The source registers are described as inputs and the destination register is described as the output.
In the RISCVInstrInfoCrypt.td file, we add the definition of the SHLXOR instruction by using the ALU\_rr class. In this part, we define funct7, funct3, and the mnemonic of the new instruction as well as the scheduling.

\begin{lstlisting}
def SHLXOR : ALU_rr<0b0011000, 0b111, "shlxor">,
                        Sched<[WriteIALU, ReadIALU, ReadIALU]>;
\end{lstlisting}

Also in the same file, we define the instruction’s pattern. When we examine this definition, we can clearly see what the instruction performs and its pattern. In the inner parentheses, we can see the shifting of the first source operand by one bit. Then, the result of this shifting operation is used as an input for the XOR operation alongside the second source operand.

\begin{lstlisting}
def : Pat< (xor (shl GPR:$src1, (i32 1)), GPR:$src2),
                              (SHLXOR GPR:$src1, GPR:$src2)>;
\end{lstlisting}

After doing these, we can try it with a simple C code given below. 

\begin{lstlisting}[language=C++]
int a,b;

void shlxor() {
	a = 3;
	b = 5;
		
	a = b^(a<<1);
}
\end{lstlisting}

Let’s get an assembly output from this C code by running the following commands:

\begin{lstlisting}[language=Bash]
clang -S -target riscv32-linux-gnu -emit-llvm shlxor.c
<llvm-build-path>/build/bin/llc -mtriple=riscv32 shlxor.ll
\end{lstlisting}

This C code basically shifts the variable "a" by one bit and XOR’s it with the "b" variable. Then the result is stored in "a". We can see that, the register that stores the value "a" is both the first source register and the destination register. The assembly output is given below.
\begin{lstlisting}%
shlxor:                                 # @shlxor
.Lfunc_begin0:
	.loc	0 4 0                           # shlx.c:4:0
	.cfi_sections .debug_frame
	.cfi_startproc
# %
	addi	sp, sp, -16
	.cfi_def_cfa_offset 16
.Ltmp0:
	.loc	0 5 4 prologue_end              # shlx.c:5:4
	sw	ra, 12(sp)                      # 4-byte Folded Spill
	sw	s0, 8(sp)                       # 4-byte Folded Spill
	.cfi_offset ra, -4
	.cfi_offset s0, -8
	addi	s0, sp, 16
	.cfi_def_cfa s0, 0
	lui	a0, %
	li	a1, 3
	sw	a1, %
	.loc	0 6 4                           # shlx.c:6:4
	lui	a1, %
	li	a2, 5
	sw	a2, %
	.loc	0 9 6                           # shlx.c:9:6
	lw	a1, %
	.loc	0 9 9 is_stmt 0                 # shlx.c:9:9
	lw	a2, %
	.loc	0 9 7                           # shlx.c:9:7
	shlxor	a1, a2, a1
	.loc	0 9 4                           # shlx.c:9:4
	sw	a1, %
	.loc	0 10 1 is_stmt 1                # shlx.c:10:1
	lw	ra, 12(sp)                      # 4-byte Folded Reload
	lw	s0, 8(sp)                       # 4-byte Folded Reload
	addi	sp, sp, 16
	ret
\end{lstlisting}

In the assembly output, we can see the SHLXOR instruction in line 28. a1 and a2 are the sources and a1 is also the destination as we can see.

\begin{figure}[h!]
    \centering
    \includegraphics[scale= 0.3]{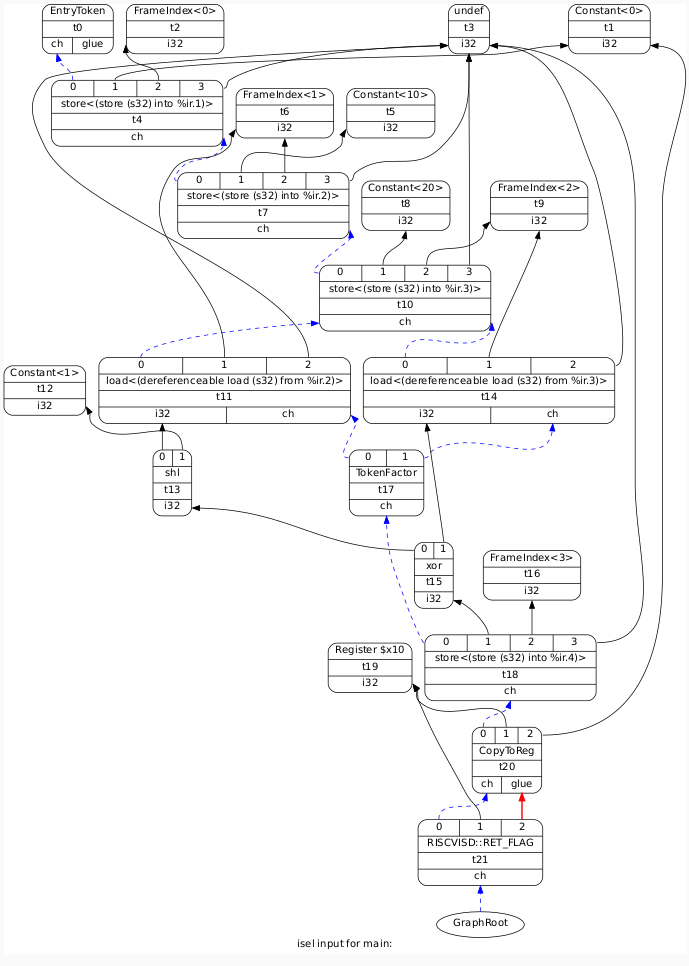}
    \caption{DAG before SHLXOR is added}
    \label{fig:shlxor_before}
\end{figure}

In addition to that, we can check the DAG in order to see the effect of our newly added instruction. We can compare the DAGs before and after the new instruction is added. We can observe the DAG before SHLXOR is added in Figure \ref{fig:shlxor_before}. In this DAG, shift left (shl) and xor instructions can be seen separately

\begin{figure}[h!]
    \centering
    \includegraphics[scale= 0.3]{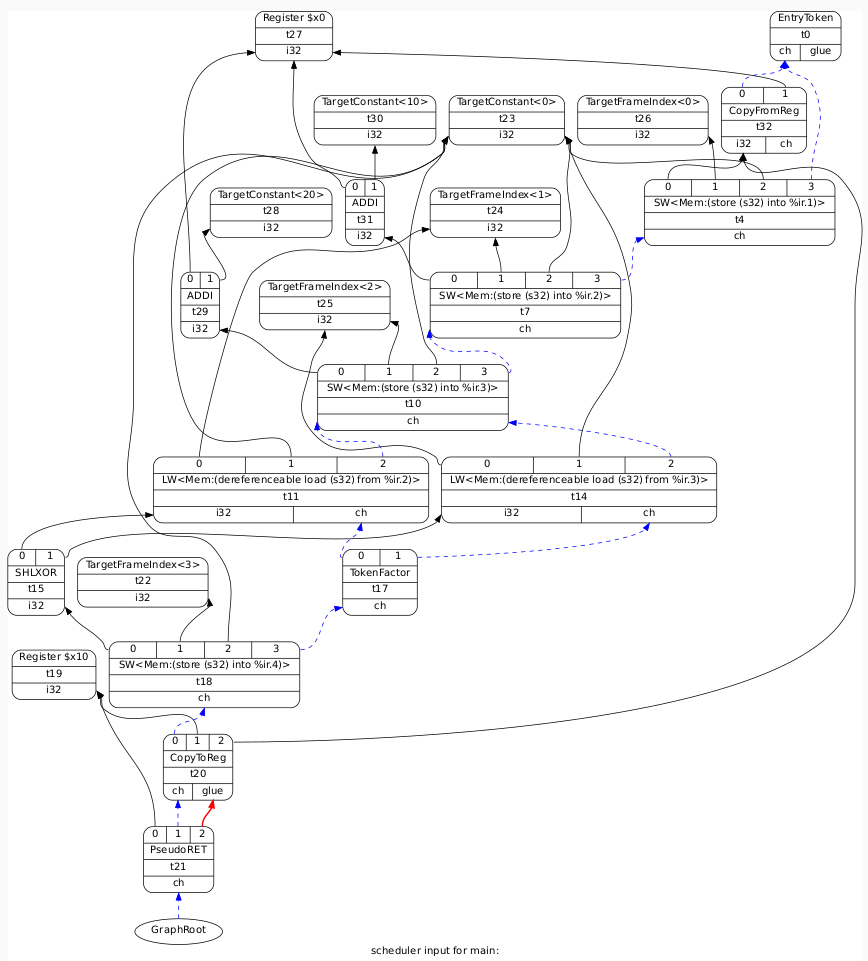}
    \caption{DAG after SHLXOR is added}
    \label{fig:shlxor_after}
\end{figure}

The DAG after we add the SHLXOR instruction can be seen in Figure \ref{fig:shlxor_after}. In this DAG, instead of two separate instructions, a single SHLXOR instruction can be seen.

\section{RORI Instruction}\label{sec:rori}
One of the Instructions that we worked on is RORI instruction. The purpose of this instruction is to take the operand and rotate it to the right by the amount of the immediate value. This is different from shifting right using an immediate. When shifting a number to the right, the LSBs are deleted and the MSBs are either zero or sign extended. On the other hand, when a number is rotated right, the LSBs that are pushed out are not deleted but written into the most significant bits. We want to add an instruction that performs this operation.

First of all, we tried pattern matching and added the definition and pattern of our new instruction to the InstrInfoCrypt.td file. However, the RORI instruction was not observed when we checked the assembly output created by using a simple C code that implements the rotation operation. This is because the pattern can have different combinations and may not match what we expect. Therefore, the instruction cannot be recognized and we can’t see it in the assembly output.

Realizing that, we looked for other options and tried to make use of intrinsics and builtin functions. We added the definitions to the InstrInfoCrypt.td file.

\begin{lstlisting}
def ROTI : ALU_ri<0b101, "roti">;
\end{lstlisting}

It is ALU\_ri type because one of the operands is an immediate value and we only need one register for this instruction.

\begin{lstlisting}
def : Pat<(rotr GPR:$rs1, simm12:$imm12),
(ROTI GPR:$rs1, simm12:$imm12)>;
\end{lstlisting}

We used rotr here because we wanted to make use of the builtin function \_\_builtin\_rotateright32. "rotr" is defined in the RISVIselLowering.cpp file.

\begin{lstlisting}[language=C++]
if (Subtarget.hasStdExtZbb() || Subtarget.hasStdExtZbkb()) {
if (Subtarget.is64Bit())
setOperationAction({ISD::ROTL, ISD::ROTR}, MVT::i32, Custom);
} else {
setOperationAction({ISD::ROTL, ISD::ROTR}, XLenVT, Expand); 
}
\end{lstlisting}

However, we need to change the “Expand” to “Legal” here otherwise we will not see the ROTI instruction in the assembly output.  This way, we are legalizing the action. If we don’t do this, we will see the fshr (funnel shift right) intrinsic in the .ll file but ROTI won’t make it into the assembly output. After doing these, we can try it with a simple C code. As mentioned before, we used a builtin rotate function in the C code given below to guarantee the generation of "fshl" llvm intrinsic function. 

\begin{lstlisting}[language=C++]
int a;

void ROT() {
	a = 15;
	
	a = __builtin_rotateright32(a,2);	
}
\end{lstlisting}

This code rotates 15 to the right by 2 bits. We can obtain the .ll file by running the following command.

\begin{lstlisting}[language=Bash]
clang -S -target riscv32-linux-gnu -emit-llvm roti.c
\end{lstlisting}

The simplified contents of the .ll file is given below.

\begin{minipage}{\linewidth}
\begin{lstlisting}[language=llvm,style=nasm]
@a = dso_local global i32 0

define void @ROT() {
  store i32 15, i32* @a
  store i32 %
  ret void
}
\end{lstlisting}
\end{minipage}

The fshr intrinsic is visible in the sixth line. After that, the assembly output can be obtained by running the following command. 

\begin{lstlisting}[language=Bash]
<llvm-build-path>/build/bin/llc -mtriple=riscv32 roti.ll$
\end{lstlisting}

The obtained assembly output is given below.

\begin{lstlisting}
ROT:                                    # @ROT
.Lfunc_begin0:
	.loc	0 3 0                           # roti.c:3:0
	.cfi_sections .debug_frame
	.cfi_startproc
# \%bb.0:
	addi	sp, sp, -16
	.cfi_def_cfa_offset 16
.Ltmp0:
	.loc	0 4 4 prologue_end              # roti.c:4:4
	sw	ra, 12(sp)                      # 4-byte Folded Spill
	sw	s0, 8(sp)                       # 4-byte Folded Spill
	.cfi_offset ra, -4
	.cfi_offset s0, -8
	addi	s0, sp, 16
	.cfi_def_cfa s0, 0
	lui	a0, \%hi(a)
	li	a1, 15
	sw	a1, \%lo(a)(a0)
	.loc	0 11 30                         # roti.c:11:30
	lw	a1, \%lo(a)(a0)
	.loc	0 11 6 is_stmt 0                # roti.c:11:6
	roti	a1, a1, 2
	.loc	0 11 4                          # roti.c:11:4
	sw	a1, \%lo(a)(a0)
	.loc	0 14 1 is_stmt 1                # roti.c:14:1
	lw	ra, 12(sp)                      # 4-byte Folded Reload
	lw	s0, 8(sp)                       # 4-byte Folded Reload
	addi	sp, sp, 16
	ret
\end{lstlisting}

The ROTI instruction can be seen in the assembly output in line 23. As expected, it uses one register as both  the destination and the source alongside an immediate value.

 After further investigation, we realized that rotation instruction was already implemented in bit manipulation extension for RISC-V and in the RISCVInstrInfoZb.td file. This file contains the instruction extensions for bit manipulations. These instructions operate on the bits of the data and RORI is one of those instructions. However, in order to utilize this extension, we need to add some flags to the command while running Clang in order to get an assembly output from the C code we write. The simple C code for RORI is given below .

\begin{lstlisting}[language=C++]
#define XLEN 32
#include <stdint.h>
#define uint_xlen_t uint32_t

uint_xlen_t rotimm(uint_xlen_t rs1){
   uint_xlen_t a = 0;
   a = ((rs1>>2) | (rs1<<(XLEN-2)));
   return a;
}
\end{lstlisting}

Here, we use 32 as the length because our target is 32-bit. The rotation is implemented in the seventh line. After that, we run the following command with additional flags as mentioned before.

\begin{lstlisting}[language=Bash]
clang --target=riscv32 -O -S rori.c -march=rv32imaczbb
\end{lstlisting}

Here, -O defines the level of optimization. -S is used for getting an assembly file as an output. rori.c is the name of our simple C code. -march=rv32imaczbb designates that we want to utilize the Zbb subgroup of the bit manipulation extension. The assembly output is given below.%

\begin{lstlisting}
rotimm:
	rori	a0, a0, 2
	ret
\end{lstlisting}

This way, we managed to successfully obtain RORI instruction in the assembly output.

\section{NAXOR Instruction}\label{sec:naxor}

The S-box algorithm includes a certain pattern that is used repeatedly. NOT-AND-XOR pattern is used five times in an s-box cycle. This pattern is lowered into one instruction using TableGen. The definition and specifications of the pattern are added to RISCVInstrInfoCrypt.td file. After matching this pattern 15 rows of the assembly file are reduced to one single instruction. 

\begin{lstlisting}
def NAXOR : ALU_rrr<0b11, 0b100, "naxor">,
Sched<[WriteIMul, ReadIMul, ReadIMul]>;
\end{lstlisting}

Since there are three variables in this instruction, custom ALU\_rrr class is used which is explained in Section \ref{sec:alurrr}. "11" and "100" base 2 numbers are used for funct2 and funct3.

\begin{lstlisting}
def : Pat< (xor (and (not GPR:$src1), GPR:$src2), GPR:$src3),
                           (NAXOR GPR:$src1, GPR:$src2, GPR:$src3)>;
\end{lstlisting}

\begin{figure}
    \centering
    \includegraphics[scale=0.3]{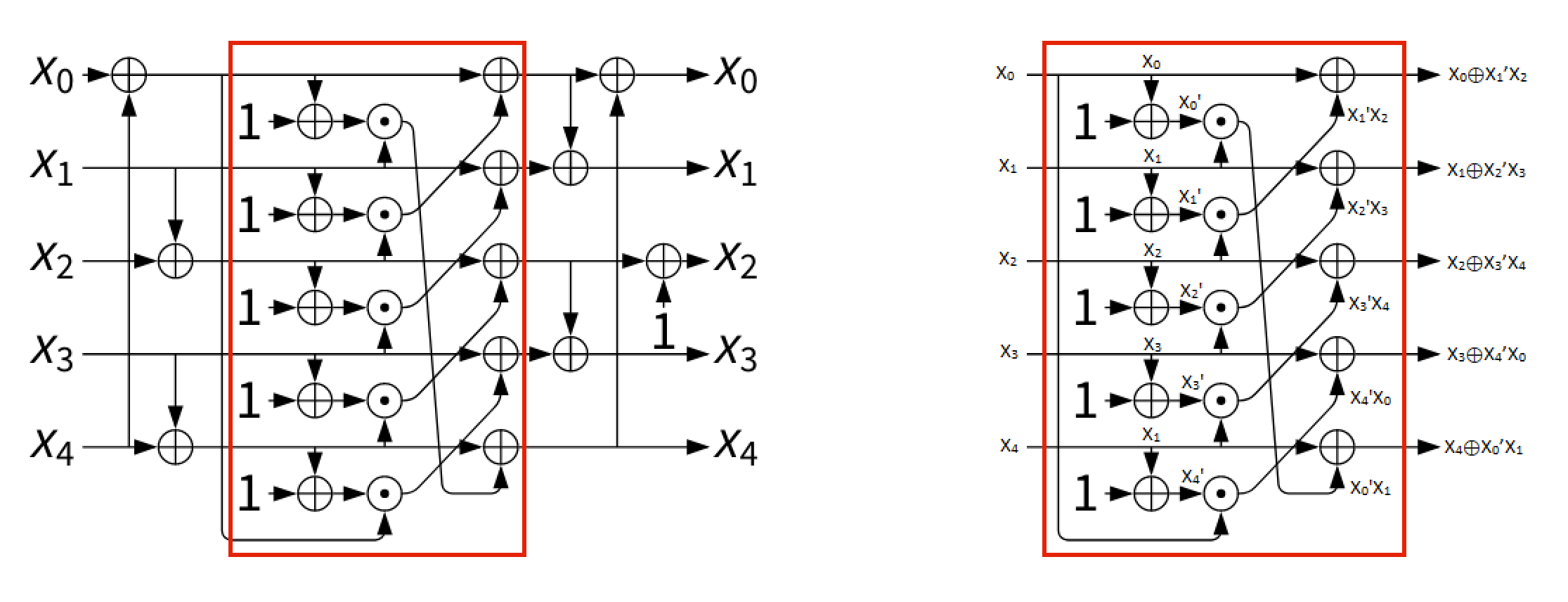}
    \caption{NAXOR patterns in s-box algorithm}
    \label{fig:sbox_naxor_pattern}
\end{figure}

This pattern is repeated 5 times as emphasized in Figure \ref{fig:sbox_naxor_pattern}. NAXOR instruction reduces 15 instructions into 5 instructions.

\begin{figure}
    \centering
    \includegraphics[scale=0.2]{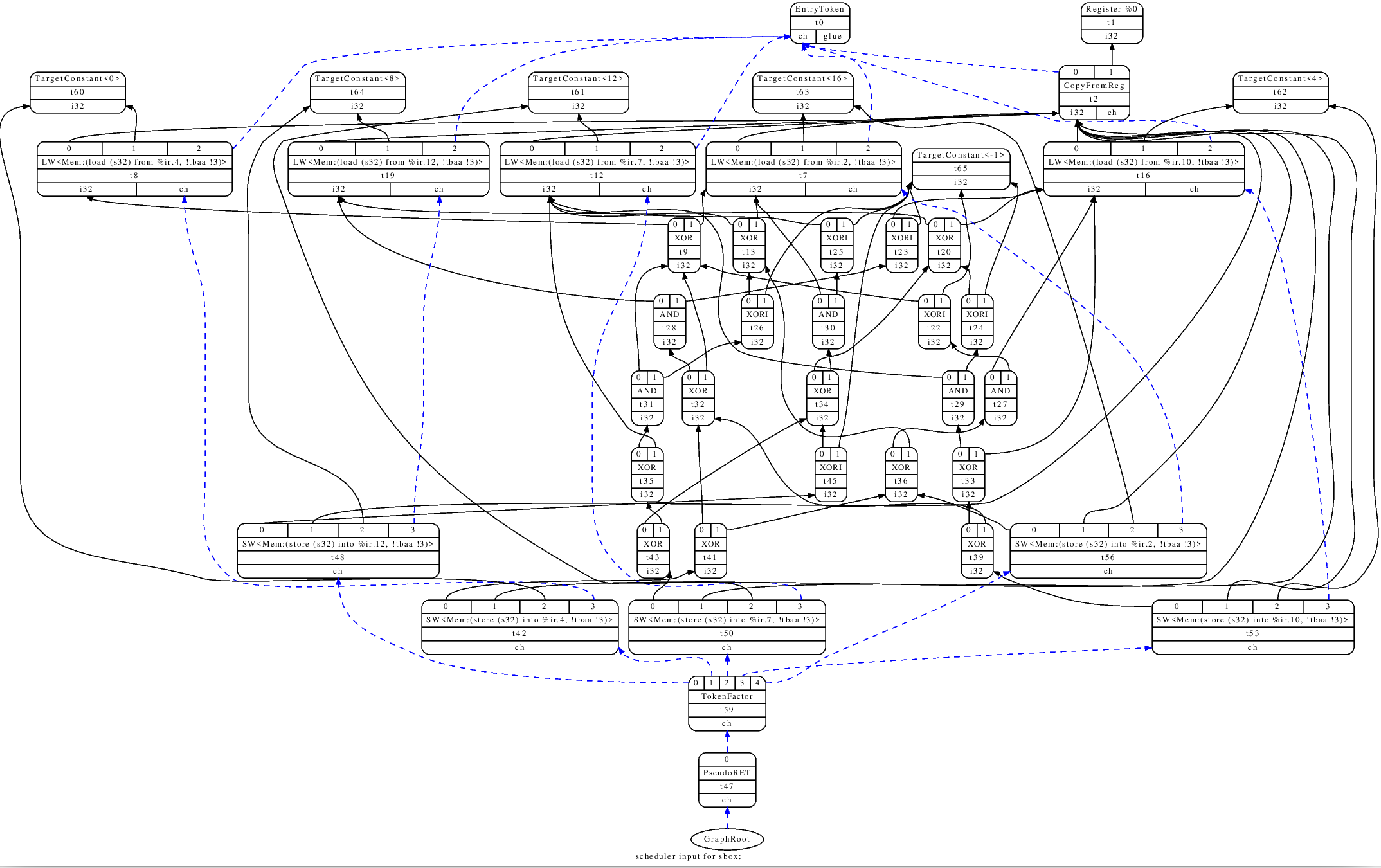}
    \caption{DAG diagram output for the s-box algorithm before scheduling}
    \label{fig:naxor_sched_diagram}
\end{figure}

\begin{figure}
    \centering
    \includegraphics[scale=0.35]{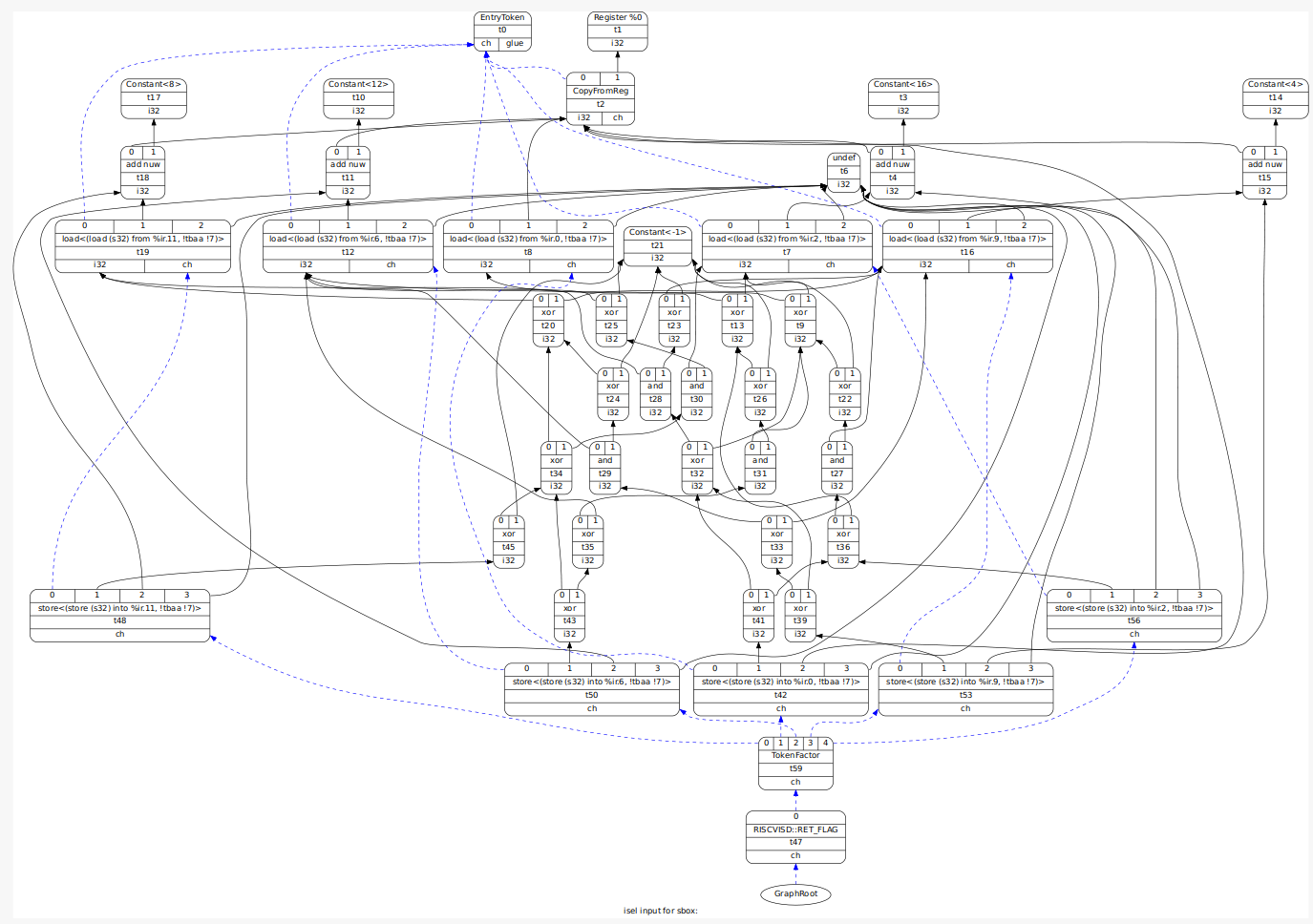}
    \caption{DAG diagram output for the s-box algorithm}
    \label{fig:naxor_dag_diagram}
\end{figure}

\begin{minipage}{\linewidth}
\begin{lstlisting}[caption={C code input for the S-box algorithm},language=C, label={lst:sbox-c2}]
typedef struct {
  int x[5];
} ascon_state_t;

void sbox(ascon_state_t state) {
     int t0, t1, t2, t3, t4;
     state.x[0] ^= state.x[4];    
     state.x[4] ^= state.x[3];    
     state.x[2] ^= state.x[1];
     t0  = state.x[0];    
     t1  = state.x[1];    
     t2  = state.x[2];    
     t3  = state.x[3];    
     t4  = state.x[4];
     t0 =~ t0;    
     t1 =~ t1;    
     t2 =~ t2;    
     t3 =~ t3;    
     t4 =~ t4;
     t0 &= state.x[1];   
     t1 &= state.x[2];    
     t2 &= state.x[3];    
     t3 &= state.x[4];    
     t4 &= state.x[0];
     state.x[0] ^= t1;   
     state.x[1] ^= t2;    
     state.x[2] ^= t3;    
     state.x[3] ^= t4;    
     state.x[4] ^= t0;
     state.x[1] ^= state.x[0];
     state.x[0] ^= state.x[4]; 
     state.x[3] ^= state.x[2]; 
     state.x[2] =~ state.x[2];

     return;
}

\end{lstlisting}
\end{minipage}

\begin{minipage}{\linewidth}
\begin{lstlisting}[caption={Optimized S-box LLVM IR}, language=llvm, style=nasm]
define void @sbox(ptr  %0)  {
  %2 = getelementptr inbounds [5 x i32], ptr %0, i32 0, i32 4
  %3 = load i32, ptr %2
  %4 = load i32, ptr %0
  %5 = xor i32 %4, %3
  %6 = getelementptr inbounds [5 x i32], ptr %0, i32 0, i32 3
  %7 = load i32, ptr %6
  %8 = xor i32 %7, %3
  %9 = getelementptr inbounds [5 x i32], ptr %0, i32 0, i32 1
  %10 = load i32, ptr %9
  %11 = getelementptr inbounds [5 x i32], ptr %0, i32 0, i32 2
  %12 = load i32, ptr %11
  %13 = xor i32 %12, %10
  %14 = xor i32 %5, -1
  %15 = xor i32 %10, -1
  %16 = xor i32 %13, -1
  %17 = xor i32 %7, -1
  %18 = xor i32 %8, -1
  %19 = and i32 %10, %14
  %20 = and i32 %12, %15
  %21 = and i32 %7, %16
  %22 = and i32 %3, %17
  %23 = and i32 %5, %18
  %24 = xor i32 %20, %5
  %25 = xor i32 %21, %10
  %26 = xor i32 %13, %22
  %27 = xor i32 %23, %7
  %28 = xor i32 %19, %8
  store i32 %28, ptr %2
  %29 = xor i32 %25, %24
  store i32 %29, ptr %9
  %30 = xor i32 %24, %28
  store i32 %30, ptr %0
  %31 = xor i32 %26, %27
  store i32 %31, ptr %6
  %32 = xor i32 %26, -1
  store i32 %32, ptr %11
  ret void
}
\end{lstlisting}
\end{minipage}

\begin{figure}
    \centering
    \includegraphics[scale=0.19]{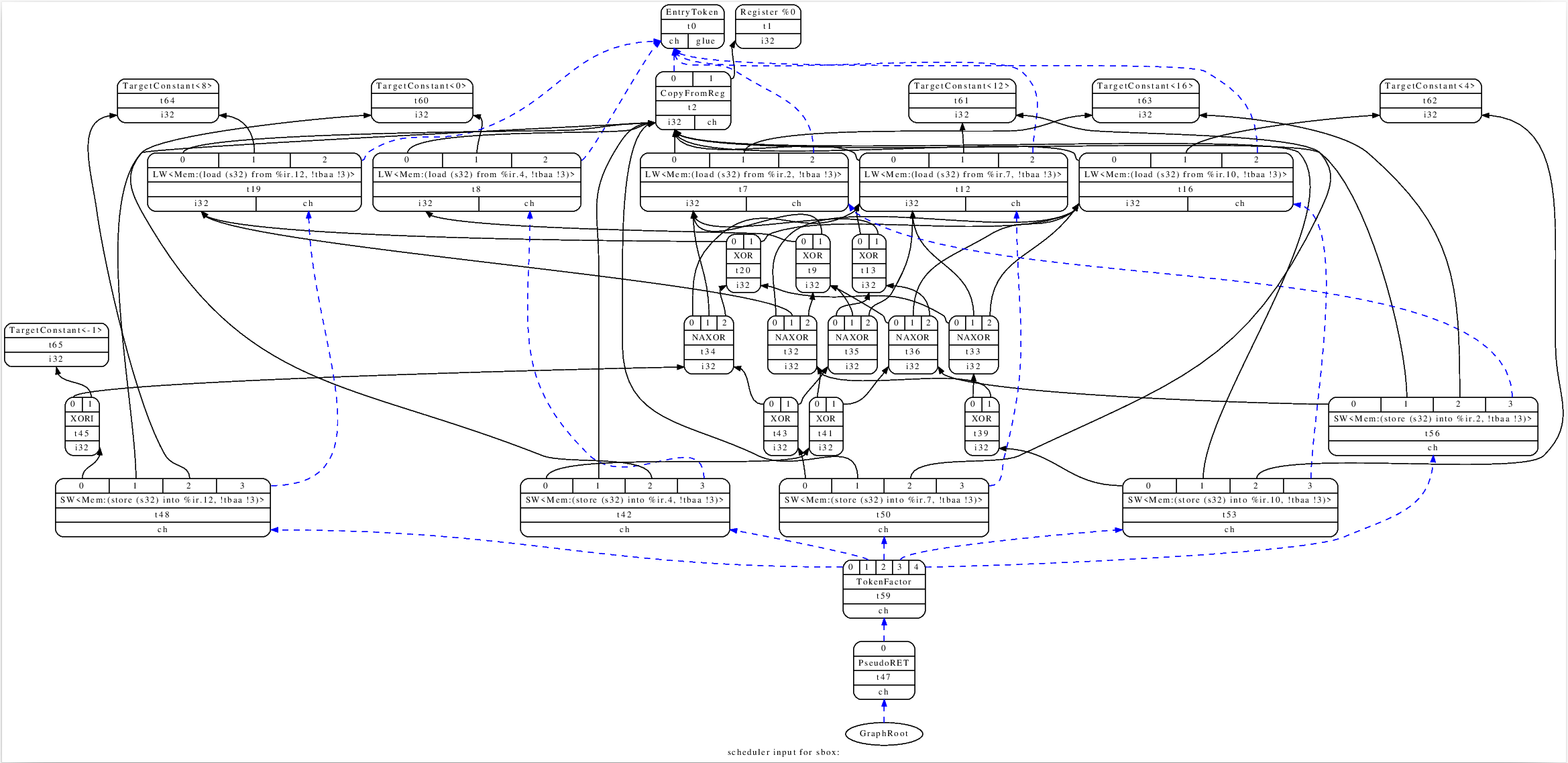}
    \caption{DAG diagram output for the s-box algorithm after NAXOR instruction is matched}
    \label{fig:naxor_match_diagram}
\end{figure}

\begin{minipage}[t]{.4\linewidth}
\begin{lstlisting}[caption={Assembly output without NAXOR instruction}]
sbox:            
	lw	a1, 16(a0)
	lw	a2, 0(a0)
	lw	a3, 12(a0)
	lw	a4, 4(a0)
	lw	a5, 8(a0)
	xor	a2, a2, a1
	xor	a6, a3, a1
	xor	a7, a5, a4
	not	t0, a2
	not	t1, a4
	not	t2, a7
	not	t3, a3
	not	t4, a6
	and	t0, a4, t0
	and	a5, a5, t1
	and	t1, a3, t2
	and	a1, a1, t3
	and	t2, a2, t4
	xor	a2, a2, a5
	xor	a4, t1, a4
	xor	a1, a7, a1
	xor	a3, t2, a3
	xor	a5, t0, a6
	sw	a5, 16(a0)
	xor	a4, a4, a2
	sw	a4, 4(a0)
	xor	a2, a2, a5
	sw	a2, 0(a0)
	xor	a3, a3, a1
	sw	a3, 12(a0)
	not	a1, a1
	sw	a1, 8(a0)
	ret
\end{lstlisting}
\end{minipage}

\begin{minipage}[t]{.4\linewidth}
\begin{lstlisting}[caption={Assembly output with NAXOR instruction}]
sbox:         
	lw	a1, 16(a0)
	lw	a2, 0(a0)
	lw	a3, 12(a0)
	lw	a4, 4(a0)
	lw	a5, 8(a0)
	xor	a2, a2, a1
	xor	a6, a3, a1
	xor	a7, a5, a4
	naxor	a5, a4, a5 ,a2
	naxor	t0, a7, a3 ,a4
	naxor	a1, a3, a1 ,a7
	naxor	a3, a6, a2 ,a3
	naxor	a2, a2, a4 ,a6
	sw	a2, 16(a0)
	xor	a4, t0, a5
	sw	a4, 4(a0)
	xor	a2, a2, a5
	sw	a2, 0(a0)
	xor	a3, a3, a1
	sw	a3, 12(a0)
	not	a1, a1
	sw	a1, 8(a0)
	ret
\end{lstlisting}
\end{minipage}

15 lines of not, and, xor operations are reduced to 5 NAXOR instructions.

\section{LXR Instruction}\label{sec:lxr}

 LXR instruction covers an XOR operation of two loads from independent addresses. ALU\_rr class is used. Note that in order to match loads with dependant addresses, C++ logic must be implemented as discussed in Section \ref{sec:cpp}.

\begin{lstlisting}
let mayLoad = 1 in{
def LXR : ALU_rr<0b0011011, 0b101, "lxr">,
Sched<[WriteIALU, ReadIALU, ReadIALU]>;
}
\end{lstlisting}

ALU\_rr class is used and 0011011, 101 base 2 numbers are used for funct7 and funct3. \textit{mayLoad} flag is 1 to enable the load instruction in the pattern.

\begin{lstlisting}
def : Pat< (xor (load GPR:$rs1),(load GPR:$rs2)),
(LXR GPR:$rs1,GPR:$rs2)>;
\end{lstlisting}

LXR instruction covers the xor operation of two loaded numbers.

\begin{figure}
    \centering
    \includegraphics[scale=0.25]{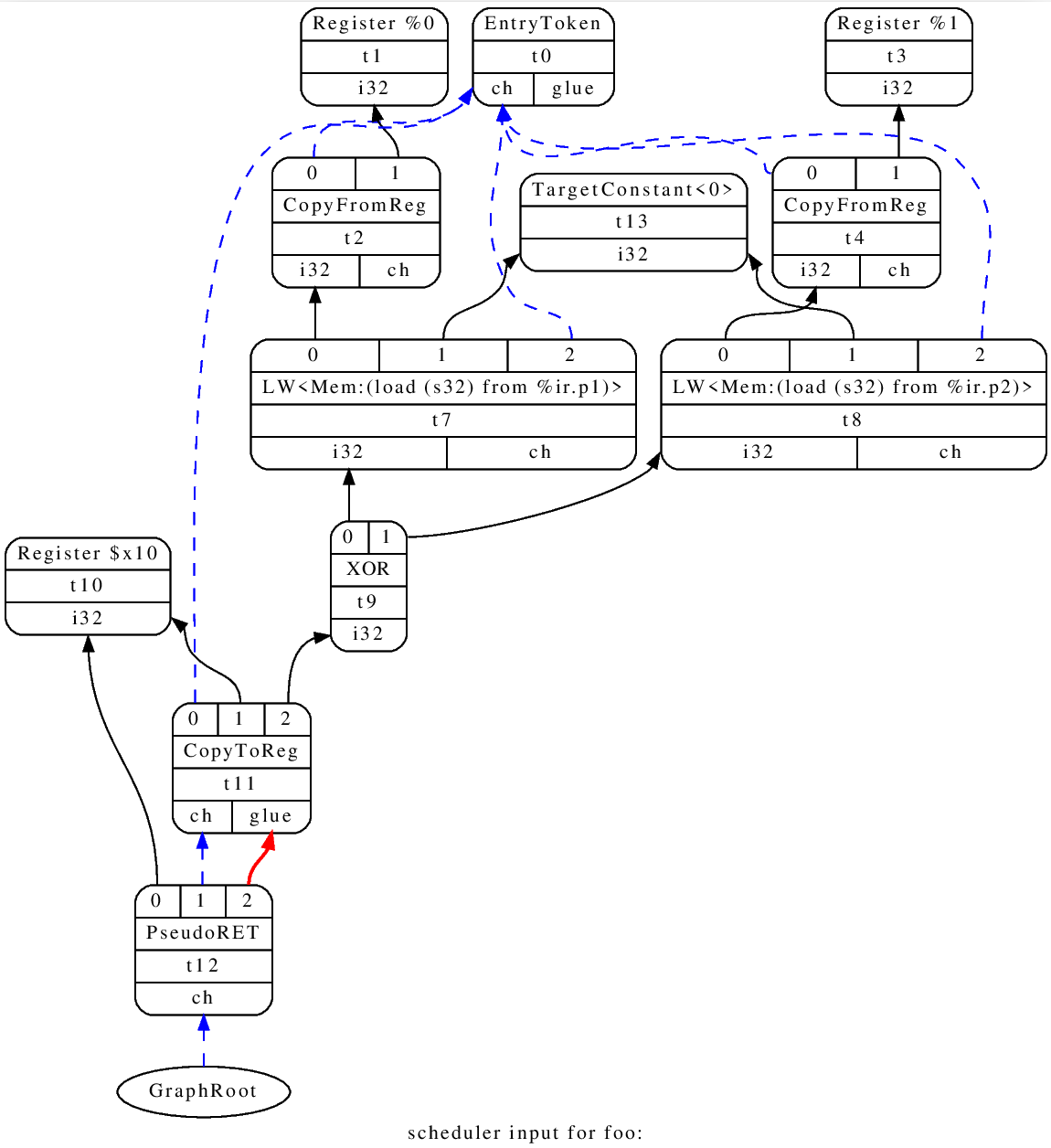}
    \caption{DAG diagram for the example LXR algorithm before scheduling}
    \label{fig:lxr_sched_diagram}
\end{figure}

\begin{figure}
    \centering
    \includegraphics[scale=0.4]{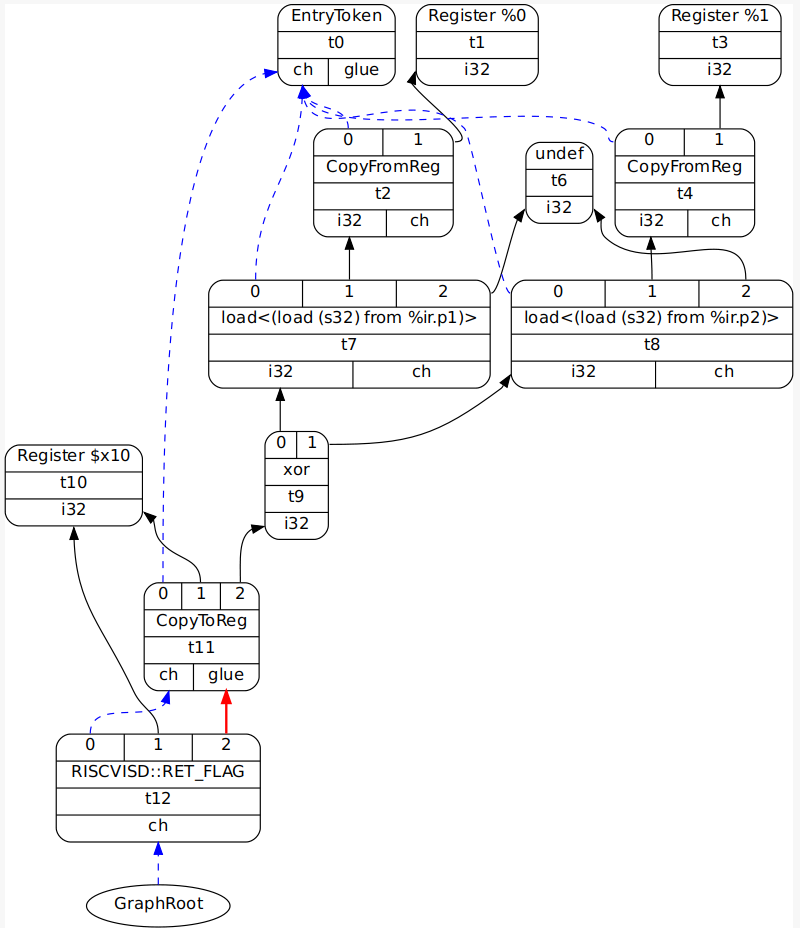}
    \caption{DAG diagram output for the example LXR algorithm}
    \label{fig:lxr_dag_diagram}
\end{figure}

\begin{minipage}{\linewidth}
\begin{lstlisting}[caption={IR code input for LXR algorithm}, language=llvm, style=nasm]
    define i32 @foo(ptr %
        ret i32 %
      }
\end{lstlisting}
\end{minipage}

\begin{lstlisting}[caption= Assembly output without LXR instruction]
      foo:                 
          lw	a0, 0(a0)
          lw	a1, 0(a1)
          xor	a0, a0, a1
          ret
\end{lstlisting}

\begin{figure}
    \centering
    \includegraphics[scale=0.25]{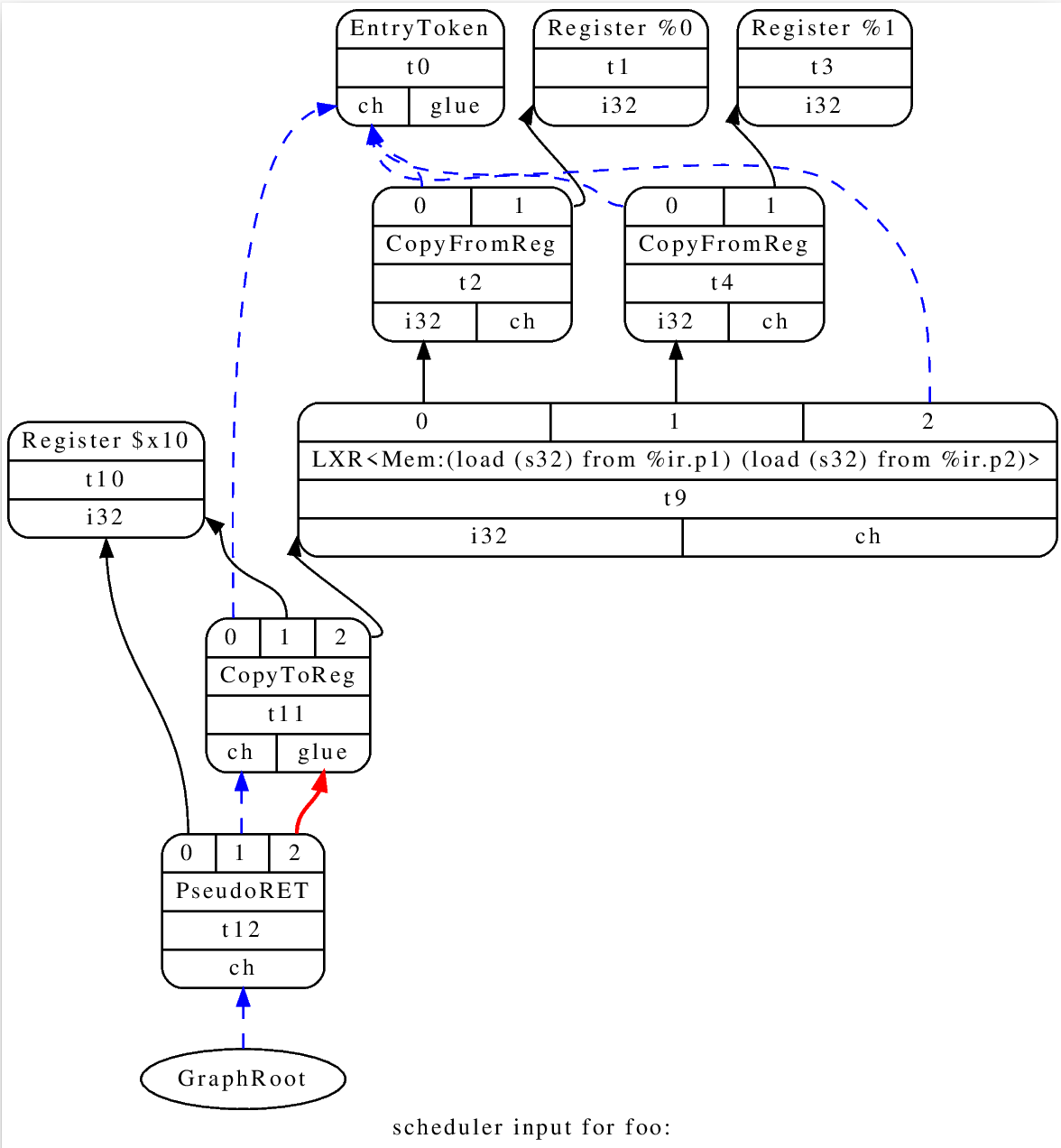}
    \caption{DAG diagram output for the example LXR algorithm after LXR instruction is matched}
    \label{fig:lxr_match}
\end{figure}

\begin{lstlisting}[caption= Assembly output with LXR instruction]
foo:               
	lxr	a0, a0, a1
	ret
\end{lstlisting}

Two loads and one xor instructions are reduced to LXR instruction.

\cleardoublepage
\clearpage
\chapter{TESTING}\label{Ch10}
Regression testing is a core part of LLVM because of its size and active development. To make sure newly added features don’t break the already present functionality it is a must to both build functionality and its corresponding tests.

LLVM-lit coordinates the testing procedure. The comment lines that start with “RUN” call other programs via LLVM-lit. LLVM-lit also gives the output of a program to another program as an input.

FileCheck, as the name implies, controls the checking process. It basically compares the file and the corresponding lines of the output. It is used with CHECK-* command.

In this chapter SHLXOR instruction which was introduced in Section \ref{sec:shlxor} will be tested. Its instruction encoding and Assembler support are tested with MC tests. Its pattern matching support is tested with LLC tests.   
\section{MC Test}
LLVM-MC is an abstracted assembler and object file emitter integrated with the compiler \cite{Lattner2010Apr}.

Before adding a new instruction, its encoding should be designed. LLVM-MC is going to be used to get the encoding results. A wrong input to TableGen class may result in a different encoding than expected. With an MC test, the expectation and result will be compared. 

\begin{minipage}{\linewidth}
\begin{lstlisting}[caption={MC Test File}, label={lst:mc_test_file} ]
# RUN: llvm-mc %
# RUN: 	| FileCheck -check-prefixes=CHECK-ASM,CHECK-ASM-AND-OBJ %
# RUN: llvm-mc -filetype=obj -triple=riscv32 < %
# RUN: 	| llvm-objdump -M no-aliases -d -r - \ 
# RUN: 	| FileCheck -check-prefixes=CHECK-ASM-AND-OBJ %

# CHECK-ASM-AND-OBJ: shlxor s2, s2, s8 
# CHECK-ASM: encoding: [0x33,0x59,0x89,0x81] 
shlxor s2, s2, s8
\end{lstlisting}
\end{minipage}

The first RUN command sequence results in the output given in Code \ref{lst:result_of_first_run_command}.

\begin{minipage}{\linewidth}
\begin{lstlisting}[language=sh, caption={Result of first "RUN" command}, label={lst:result_of_first_run_command} ]
$ <llvm-build-path>/build/bin/llvm-mc \\
<llvm-build-path>/llvm/test/MC/RISCV/crypt.s \\ 
-triple=riscv32 -riscv-no-aliases -show-encoding

             . text

             shlxor s2, s2, s8 # encoding: [0x33,0x59, 0x89, 0x81]
\end{lstlisting}
\end{minipage}

In Code \ref{lst:mc_test_file}, FileCheck is checking both the Assembly encoding and the string as it was provided with the prefixes: CHECK-ASM, CHECK-ASM-AND-OBJ.

The second RUN sequence with only the llvm-mc part gives out an ELF object which itself isn’t useful. The console output can be seen in Code \ref{lst:result_of_second_run_command}.

\begin{lstlisting}[language=sh, caption={Result of second "RUN" command}, label={lst:result_of_second_run_command} ]
$ <llvm-build-path>/build/bin/llvm-mc \\
<llvm-build-path>/llvm/test/MC/RISCV/crypt.s \\ 
-filetype=obj -triple=riscv32 

ELF`4(3Y.text.strtab.symtab48%
\end{lstlisting}
By using the pipe ‘|’ operator similar to Shell usage, we can tell LLVM-lit to feed another program with the output of a program. The test commands with shell pipe are demonstrated in Code \ref{lst:use_of_pipe_operator}.

\begin{lstlisting}[language=sh, caption={Use of pipe operator in Shell}, label={lst:use_of_pipe_operator} ]
$ <llvm-build-path>/build/bin/llvm-mc -filetype=obj -triple=riscv32 <
<llvm-build-path>/llvm/test/MC/RISCV/crypt.s | \\
<llvm-build-path>/build/bin/llvm-objdump -M no-aliases -d -r -


<stdin>: file format elf32-littleriscv

Disassembly of section .text:

00000000 <.text>:
       0: 33 59 89 81 = shlxor s2, s2, s8
\end{lstlisting}

Observing the outputs directly through the shell indicates what FileCheck is looking for in the standard output. Here in the second RUN sequence, FileCheck is provided only with CHECK-ASM-AND-OBJ and therefore it does not check the CHECK-ASM line. It seems that the object dump resulted in the correct string.

In Code \ref{lst:output_for_successfully_passed_test} we can observe that the test is passed:

\begin{minipage}{\linewidth}
\begin{lstlisting}[language=sh, caption={Output for successfully passed test}, label={lst:output_for_successfully_passed_test} ]
$ <llvm-build-path>/build/bin/llvm-lit \\
<llvm-build-path>/llvm/test/MC/RISCV/crypt.s -v

-- Testing: 1 tests, 1 workers --
PASS: LLVM :: MC/RISCV/crypt.s (1 of 1)

Testing Time: 0.07s
Passed: 1
\end{lstlisting}
\end{minipage}

Another complexity LLVM-lit handles is the path of our compiled binaries. The test case is free of the paths and \%s placeholders are populated by LLVM-lit.

\section{LLC/CodeGen Test}\label{sec:llc_test}
LLC tests are more familiar since they check how an LLVM IR file produces Assembly strings. Instead of manually checking whether an instruction is emitted in the output, this tool can be used and batch testing can be done.

An example is given in Code \ref{lst:minimal_func} of the minimal LLVM IR code implemented producing the SHLXOR instruction we want. As can be seen from its signature, it is a function taking two 32-bit integers and returning one. It takes one input, shifts it left by one and assigns it to a variable \%1. \%1 is then XOR’ed with the second input and the result is returned.

\begin{lstlisting}[language=llvm,style=nasm, caption={Minimal shlxor producing LLVM IR}, label={lst:minimal_func} ]
; RUN: llc -mtriple=riscv32 -verify-machineinstrs < %
; RUN: | FileCheck %


define i32 @shlxor(i32 %
ret i32 %
} 
\end{lstlisting}

A utility script as its usage is shown in Code \ref{lst:command_for_using_utility_script}, can be used to generate the expected result and insert it to the test file.

\begin{lstlisting}[language=sh, caption={Command for using utility script}, label={lst:command_for_using_utility_script} ]
$ <llvm-build-path>/utils/update_llc_test_checks.py \\
--llc-binary <llvm-build-path>/build/bin/llc shlxor.ll
\end{lstlisting}

The LLC test file is populated with FileCheck lines as seen in Code \ref{lst:llc_test_file}.

\begin{minipage}{\linewidth}
\begin{lstlisting}[language=llvm,style=nasm, caption={The final LLC test file}, label={lst:llc_test_file} ]
NOTE: Assertions have been autogenerated by utils/update_llc_test_checks.py
; RUN: llc -mtriple=riscv32 -verify-machineinstrs < %
; RUN: | FileCheck %


define i32 @shlxor(i32 %
; RV32R-LABEL: shlxor:
; RV32R:       # %
; RV32R-NEXT:    shlxor    a0, a0, a1
; RV32R-NEXT:    ret
ret i32 %
} 
\end{lstlisting}
\end{minipage}

We can observe that our newly added SHLXOR instruction is recognized and placed in the check lines. When LLVM-lit is run with the test file, the test is passed as the new instruction is implemented. Output for a successful test is given in Code \ref{lst:output_for_passed_ll_test}.

\begin{lstlisting}[language=sh, caption={Output for successfully passed test}, label={lst:output_for_passed_ll_test} ]
$ <llvm-build-path>/build/bin/llvm-lit shlxor.ll -v


-- Testing: 1 tests, 1 workers --
PASS: LLVM :: CodeGen/RISCV/shlxor.ll (1 of 1)

Testing Time: 0.08s
Passed: 1
\end{lstlisting}

This way a large number of instructions can be tested and the effect of our modifications can be analyzed by running the tests.

\cleardoublepage
\chapter{REALISTIC CONSTRAINTS AND CONCLUSIONS}\label{Ch6}

In this thesis, the process of modifying a compiler backend extension to support a custom extended processor is presented from various aspects. LLVM compiler infrastructure is used as the compiler design environment. We explained RISC-V standard extensions and the ASCON encryption algorithm for background information. 

Our findings indicate that custom instruction design is a critical task. RISC-V standard extensions should be considered and analyzed. Instructions should be designed by considering both hardware and software. Our analysis of the S-box indicates that as the patterns get larger or contain high-level information, different compiler stages such as the middle-end should be considered as well. This work presents various pattern matching schemes and example implementations that the reader will have an intuition about the process of adding any custom instruction to the compiler. 

In Chapter \ref{Ch2}, the common structure of the compilers is presented to give general knowledge about the compilers and expand the reader’s view on compiler structure. Following this in Chapter \ref{ch:Ch3}, LLVM compiler infrastructure is explained. Clang frontend, LLVM optimizer and LLVM RISC-V backend are described. LLVM optimizer is described with a case study tracing the optimizations on LLVM IR. In Chapter \ref{ch:riscv}, RISC-V standard extensions are presented. It is emphasized that using standard extensions can reduce the necessary compiler support workload for hardware developers. In Chapter \ref{Chascon}, the ASCON algorithm is presented briefly to provide background information about the target hardware. In Chapter \ref{Ch4}, LLVM RISC-V backend was analyzed by tracing the transformations of a simple high-level code being compiled down to assembly instructions. We explained how assembly codes are generated through LLVM’s compilation steps. In Chapter \ref{ch:custom_instr}, we showed how to add an instruction to an LLVM backend using TableGen and C++. We discussed that adding an instruction can be designed in two steps, assembler support and pattern matching. We emphasized that assembler support can be implemented in Instruction Selection however pattern matching can be covered in different regions of the compiler. In Chapter \ref{Ch9}, we presented our collection of custom instructions with two new instructions proposed, LXR and NAXOR for the ASCON application. In Chapter \ref{Ch10}, we explained how to validate a newly added instruction and set up testing infrastructure. We believe that this work will inspire future work in the related area.
\section{Practical Application of This Project}
This project can be useful for increasing the efficiency of applications that require the frequent use of specific instructions. Cryptography applications with RISC-V may be one of these.

\section{Realistic Constraints}
LLVM is a huge infrastructure and while working with it, sometimes it may be hard to find what you are looking for. Also, there aren’t many sources or documentation to find solutions to the specific problems that we encounter which sometimes slows down the progress.

\subsection{Social, Environmental, and Economic Impact}
The end product is going to help the custom processor to be programmed by a high-level programming language. It will make the programming of the custom processor a more efficient process and encourage the use of the custom processor. Because of this efficiency, the energy and time costs would be reduced during the programming of the processor. Also, using a custom processor for handling a problem is faster and requires less power. Therefore, encouraging the use of one would be another benefit of the end product.

\subsection{Cost Analysis}
Open-source tools and programs were used on our computers during the project. Therefore, it wasn’t costly for us.

\pagebreak
\subsection{Standards}
LLVM project is very selective in the technologies they use. Latest versions of C++ and build tools with software engineering principles are followed. Instructions abide by the RISC-V instruction set standard.

\subsection{Health and Safety Concerns}
Since we are working in software area, there is no possible risk of harm to users.

\section{Future Work and Recommendations}
For the pattern matching process, we recommend future work to focus on LLVM IR transformations and optimizations. As discussed in Section \ref{sec:patmatchdisc}, LLVM IR provides flexibility to perform source matching and transformation and thus pattern matching. For more complicated patterns MLIR is recommended to be focused on. Despite it being under active development, MLIR provides higher level pattern matching potential.

For LLVM RISC-V backend, we recommend working on GlobalISel which is developed to replace SelectionDAG. GlobalISel may have fewer limitations than SelectionDAG so it can be useful for complex patterns. GlobalISel has a modular pass-based structure which is easier to work with than the monolithic SelectionDAG.

\cleardoublepage

\bibliographystyle{thesis_itubib}      %
\bibliography{thesis_bib}			   %

\eklerkapak{}
\singlespacing

\begin{itemize}[leftmargin=3.3cm,itemsep=-0.4em,labelsep=1.5mm] %
\item [\textbf{APPENDIX A.1 :}]Installation of Software
\item [\textbf{APPENDIX A.2 :}]Unoptimized S-box IR Code
\item [\textbf{APPENDIX A.3 :}]Creating Assembly File From C File
\item [\textbf{APPENDIX A.4 :}]Adding the Crypt extension to the LLVM
\end{itemize}

\newpage

\eklerbolum{0}
\chapter{APPENDICES}
\section*{APPENDIX A.1}
\vglue6pt
\renewcommand{\theequation}{A.1.\arabic{equation}}
\setcounter{equation}{0}
\section{Installation of Software}

As the LLVM codebase is large and has many options while building from source, finding the right options that our computers can handle easily was both essential to get started and critical as it decides the time it takes to see a change in code to get compiled. For this purpose, we accumulated the commands and created a tutorial that we can use in the future.

\begin{lstlisting}[language=Bash, caption={Clone Repository and Install Necessary Packages}]
git clone https://github.com/llvm/llvm-project
cd llvm-project
mkdir build
cd build	
sudo apt install cmake, ninja-build, clang, lld	
\end{lstlisting}

\begin{lstlisting}[language=Bash, caption={CMake Configuration We Used}]
cmake -S ../llvm . -G Ninja -DCMAKE_BUILD_TYPE="Debug"  \
-DBUILD_SHARED_LIBS=True -DLLVM_USE_SPLIT_DWARF=True  \
-DLLVM_BUILD_TESTS=True   -DCMAKE_C_COMPILER=clang \
-DCMAKE_CXX_COMPILER=clang++ -DLLVM_TARGETS_TO_BUILD="all" \
-DLLVM_EXPERIMENTAL_TARGETS_TO_BUILD="RISCV" -DLLVM_ENABLE_LLD=ON	
\end{lstlisting}
With this command, we are choosing the type as debug. Shared\_libs=TRUE causes all libraries to be built shared instead of static libraries. ..SPLIT\_DWARF is set to True to minimize memory usage at link time. We want to use clang as the C compiler. Therefore, it is specified in the command as DCMAKE\_C\_COMPILER=clang. In addition to that, we want to use lld as the linker instead of gold, so we specify that as well.
This configuration is the most efficient in terms of memory and disk usage among our previous attempts at building LLVM from source.
\noindent
\begin{minipage}[t]{\linewidth}
\begin{lstlisting}[language=Bash, caption={To build from scratch or to rebuild files with change, automatically}]
ninja
\end{lstlisting}
\end{minipage}

\begin{minipage}[t]{\linewidth}
\begin{lstlisting}[language=Bash, caption={To build llc only which is the binary we modify}]
ninja llc	
\end{lstlisting}
\end{minipage}

While running ninja, CPU and ram usage significantly increases. All available cores are used capacity. This may prevent doing other tasks while running ninja. In order to prevent this one may opt to use the following command instead. It allows you to choose how many cores are going to be utilized.
\begin{lstlisting}[language=Bash]
	ninja llc -j<number of cores to use>
\end{lstlisting}

\newpage
\section*{APPENDIX A.2}
\vglue6pt
\renewcommand{\theequation}{A.2.\arabic{equation}}
\setcounter{equation}{0}
\section{Unoptimized S-box IR Code}
The output of the unoptimized S-box function is provided below. The C code used to produce this LLVM IR is in Code \ref{lst:sbox-c}

\begin{lstlisting}[language=llvm,style=nasm, label={lst:unopt-sbox}]
; ModuleID = 's-box.c'
source_filename = "s-box.c"
target datalayout = "e-m:e-p:32:32-i64:64-n32-S128"
target triple = "riscv32-unknown-linux-gnu"

%struct.ascon_state_t = type { [5 x i32] }

; Function Attrs: nounwind uwtable
define dso_local void @sbox(ptr noundef %state) #0 {
entry:
  %state.indirect_addr = alloca ptr, align 4
  %t0 = alloca i32, align 4
  %t1 = alloca i32, align 4
  %t2 = alloca i32, align 4
  %t3 = alloca i32, align 4
  %t4 = alloca i32, align 4
  store ptr %state, ptr %state.indirect_addr, align 4, !tbaa !7
  call void @llvm.lifetime.start.p0(i64 4, ptr %t0) #2
  call void @llvm.lifetime.start.p0(i64 4, ptr %t1) #2
  call void @llvm.lifetime.start.p0(i64 4, ptr %t2) #2
  call void @llvm.lifetime.start.p0(i64 4, ptr %t3) #2
  call void @llvm.lifetime.start.p0(i64 4, ptr %t4) #2
  %x = getelementptr inbounds %struct.ascon_state_t, ptr %state, i32 0, i32 0
  %arrayidx = getelementptr inbounds [5 x i32], ptr %x, i32 0, i32 4
  %0 = load i32, ptr %arrayidx, align 4, !tbaa !11
  %x1 = getelementptr inbounds %struct.ascon_state_t, ptr %state, i32 0, i32 0
  %arrayidx2 = getelementptr inbounds [5 x i32], ptr %x1, i32 0, i32 0
  %1 = load i32, ptr %arrayidx2, align 4, !tbaa !11
  %xor = xor i32 %1, %0
  store i32 %xor, ptr %arrayidx2, align 4, !tbaa !11
  %x3 = getelementptr inbounds %struct.ascon_state_t, ptr %state, i32 0, i32 0
  %arrayidx4 = getelementptr inbounds [5 x i32], ptr %x3, i32 0, i32 3
  %2 = load i32, ptr %arrayidx4, align 4, !tbaa !11
  %x5 = getelementptr inbounds %struct.ascon_state_t, ptr %state, i32 0, i32 0
  %arrayidx6 = getelementptr inbounds [5 x i32], ptr %x5, i32 0, i32 4
  %3 = load i32, ptr %arrayidx6, align 4, !tbaa !11
  %xor7 = xor i32 %3, %2
  store i32 %xor7, ptr %arrayidx6, align 4, !tbaa !11
  %x8 = getelementptr inbounds %struct.ascon_state_t, ptr %state, i32 0, i32 0
  %arrayidx9 = getelementptr inbounds [5 x i32], ptr %x8, i32 0, i32 1
  %4 = load i32, ptr %arrayidx9, align 4, !tbaa !11
  %x10 = getelementptr inbounds %struct.ascon_state_t, ptr %state, i32 0, i32 0
  %arrayidx11 = getelementptr inbounds [5 x i32], ptr %x10, i32 0, i32 2
  %5 = load i32, ptr %arrayidx11, align 4, !tbaa !11
  %xor12 = xor i32 %5, %4
  store i32 %xor12, ptr %arrayidx11, align 4, !tbaa !11
  %x13 = getelementptr inbounds %struct.ascon_state_t, ptr %state, i32 0, i32 0
  %arrayidx14 = getelementptr inbounds [5 x i32], ptr %x13, i32 0, i32 0
  %6 = load i32, ptr %arrayidx14, align 4, !tbaa !11
  store i32 %6, ptr %t0, align 4, !tbaa !11
  %x15 = getelementptr inbounds %struct.ascon_state_t, ptr %state, i32 0, i32 0
  %arrayidx16 = getelementptr inbounds [5 x i32], ptr %x15, i32 0, i32 1
  %7 = load i32, ptr %arrayidx16, align 4, !tbaa !11
  store i32 %7, ptr %t1, align 4, !tbaa !11
  %x17 = getelementptr inbounds %struct.ascon_state_t, ptr %state, i32 0, i32 0
  %arrayidx18 = getelementptr inbounds [5 x i32], ptr %x17, i32 0, i32 2
  %8 = load i32, ptr %arrayidx18, align 4, !tbaa !11
  store i32 %8, ptr %t2, align 4, !tbaa !11
  %x19 = getelementptr inbounds %struct.ascon_state_t, ptr %state, i32 0, i32 0
  %arrayidx20 = getelementptr inbounds [5 x i32], ptr %x19, i32 0, i32 3
  %9 = load i32, ptr %arrayidx20, align 4, !tbaa !11
  store i32 %9, ptr %t3, align 4, !tbaa !11
  %x21 = getelementptr inbounds %struct.ascon_state_t, ptr %state, i32 0, i32 0
  %arrayidx22 = getelementptr inbounds [5 x i32], ptr %x21, i32 0, i32 4
  %10 = load i32, ptr %arrayidx22, align 4, !tbaa !11
  store i32 %10, ptr %t4, align 4, !tbaa !11
  %11 = load i32, ptr %t0, align 4, !tbaa !11
  %not = xor i32 %11, -1
  store i32 %not, ptr %t0, align 4, !tbaa !11
  %12 = load i32, ptr %t1, align 4, !tbaa !11
  %not23 = xor i32 %12, -1
  store i32 %not23, ptr %t1, align 4, !tbaa !11
  %13 = load i32, ptr %t2, align 4, !tbaa !11
  %not24 = xor i32 %13, -1
  store i32 %not24, ptr %t2, align 4, !tbaa !11
  %14 = load i32, ptr %t3, align 4, !tbaa !11
  %not25 = xor i32 %14, -1
  store i32 %not25, ptr %t3, align 4, !tbaa !11
  %15 = load i32, ptr %t4, align 4, !tbaa !11
  %not26 = xor i32 %15, -1
  store i32 %not26, ptr %t4, align 4, !tbaa !11
  %x27 = getelementptr inbounds %struct.ascon_state_t, ptr %state, i32 0, i32 0
  %arrayidx28 = getelementptr inbounds [5 x i32], ptr %x27, i32 0, i32 1
  %16 = load i32, ptr %arrayidx28, align 4, !tbaa !11
  %17 = load i32, ptr %t0, align 4, !tbaa !11
  %and = and i32 %17, %16
  store i32 %and, ptr %t0, align 4, !tbaa !11
  %x29 = getelementptr inbounds %struct.ascon_state_t, ptr %state, i32 0, i32 0
  %arrayidx30 = getelementptr inbounds [5 x i32], ptr %x29, i32 0, i32 2
  %18 = load i32, ptr %arrayidx30, align 4, !tbaa !11
  %19 = load i32, ptr %t1, align 4, !tbaa !11
  %and31 = and i32 %19, %18
  store i32 %and31, ptr %t1, align 4, !tbaa !11
  %x32 = getelementptr inbounds %struct.ascon_state_t, ptr %state, i32 0, i32 0
  %arrayidx33 = getelementptr inbounds [5 x i32], ptr %x32, i32 0, i32 3
  %20 = load i32, ptr %arrayidx33, align 4, !tbaa !11
  %21 = load i32, ptr %t2, align 4, !tbaa !11
  %and34 = and i32 %21, %20
  store i32 %and34, ptr %t2, align 4, !tbaa !11
  %x35 = getelementptr inbounds %struct.ascon_state_t, ptr %state, i32 0, i32 0
  %arrayidx36 = getelementptr inbounds [5 x i32], ptr %x35, i32 0, i32 4
  %22 = load i32, ptr %arrayidx36, align 4, !tbaa !11
  %23 = load i32, ptr %t3, align 4, !tbaa !11
  %and37 = and i32 %23, %22
  store i32 %and37, ptr %t3, align 4, !tbaa !11
  %x38 = getelementptr inbounds %struct.ascon_state_t, ptr %state, i32 0, i32 0
  %arrayidx39 = getelementptr inbounds [5 x i32], ptr %x38, i32 0, i32 0
  %24 = load i32, ptr %arrayidx39, align 4, !tbaa !11
  %25 = load i32, ptr %t4, align 4, !tbaa !11
  %and40 = and i32 %25, %24
  store i32 %and40, ptr %t4, align 4, !tbaa !11
  %26 = load i32, ptr %t1, align 4, !tbaa !11
  %x41 = getelementptr inbounds %struct.ascon_state_t, ptr %state, i32 0, i32 0
  %arrayidx42 = getelementptr inbounds [5 x i32], ptr %x41, i32 0, i32 0
  %27 = load i32, ptr %arrayidx42, align 4, !tbaa !11
  %xor43 = xor i32 %27, %26
  store i32 %xor43, ptr %arrayidx42, align 4, !tbaa !11
  %28 = load i32, ptr %t2, align 4, !tbaa !11
  %x44 = getelementptr inbounds %struct.ascon_state_t, ptr %state, i32 0, i32 0
  %arrayidx45 = getelementptr inbounds [5 x i32], ptr %x44, i32 0, i32 1
  %29 = load i32, ptr %arrayidx45, align 4, !tbaa !11
  %xor46 = xor i32 %29, %28
  store i32 %xor46, ptr %arrayidx45, align 4, !tbaa !11
  %30 = load i32, ptr %t3, align 4, !tbaa !11
  %x47 = getelementptr inbounds %struct.ascon_state_t, ptr %state, i32 0, i32 0
  %arrayidx48 = getelementptr inbounds [5 x i32], ptr %x47, i32 0, i32 2
  %31 = load i32, ptr %arrayidx48, align 4, !tbaa !11
  %xor49 = xor i32 %31, %30
  store i32 %xor49, ptr %arrayidx48, align 4, !tbaa !11
  %32 = load i32, ptr %t4, align 4, !tbaa !11
  %x50 = getelementptr inbounds %struct.ascon_state_t, ptr %state, i32 0, i32 0
  %arrayidx51 = getelementptr inbounds [5 x i32], ptr %x50, i32 0, i32 3
  %33 = load i32, ptr %arrayidx51, align 4, !tbaa !11
  %xor52 = xor i32 %33, %32
  store i32 %xor52, ptr %arrayidx51, align 4, !tbaa !11
  %34 = load i32, ptr %t0, align 4, !tbaa !11
  %x53 = getelementptr inbounds %struct.ascon_state_t, ptr %state, i32 0, i32 0
  %arrayidx54 = getelementptr inbounds [5 x i32], ptr %x53, i32 0, i32 4
  %35 = load i32, ptr %arrayidx54, align 4, !tbaa !11
  %xor55 = xor i32 %35, %34
  store i32 %xor55, ptr %arrayidx54, align 4, !tbaa !11
  %x56 = getelementptr inbounds %struct.ascon_state_t, ptr %state, i32 0, i32 0
  %arrayidx57 = getelementptr inbounds [5 x i32], ptr %x56, i32 0, i32 0
  %36 = load i32, ptr %arrayidx57, align 4, !tbaa !11
  %x58 = getelementptr inbounds %struct.ascon_state_t, ptr %state, i32 0, i32 0
  %arrayidx59 = getelementptr inbounds [5 x i32], ptr %x58, i32 0, i32 1
  %37 = load i32, ptr %arrayidx59, align 4, !tbaa !11
  %xor60 = xor i32 %37, %36
  store i32 %xor60, ptr %arrayidx59, align 4, !tbaa !11
  %x61 = getelementptr inbounds %struct.ascon_state_t, ptr %state, i32 0, i32 0
  %arrayidx62 = getelementptr inbounds [5 x i32], ptr %x61, i32 0, i32 4
  %38 = load i32, ptr %arrayidx62, align 4, !tbaa !11
  %x63 = getelementptr inbounds %struct.ascon_state_t, ptr %state, i32 0, i32 0
  %arrayidx64 = getelementptr inbounds [5 x i32], ptr %x63, i32 0, i32 0
  %39 = load i32, ptr %arrayidx64, align 4, !tbaa !11
  %xor65 = xor i32 %39, %38
  store i32 %xor65, ptr %arrayidx64, align 4, !tbaa !11
  %x66 = getelementptr inbounds %struct.ascon_state_t, ptr %state, i32 0, i32 0
  %arrayidx67 = getelementptr inbounds [5 x i32], ptr %x66, i32 0, i32 2
  %40 = load i32, ptr %arrayidx67, align 4, !tbaa !11
  %x68 = getelementptr inbounds %struct.ascon_state_t, ptr %state, i32 0, i32 0
  %arrayidx69 = getelementptr inbounds [5 x i32], ptr %x68, i32 0, i32 3
  %41 = load i32, ptr %arrayidx69, align 4, !tbaa !11
  %xor70 = xor i32 %41, %40
  store i32 %xor70, ptr %arrayidx69, align 4, !tbaa !11
  %x71 = getelementptr inbounds %struct.ascon_state_t, ptr %state, i32 0, i32 0
  %arrayidx72 = getelementptr inbounds [5 x i32], ptr %x71, i32 0, i32 2
  %42 = load i32, ptr %arrayidx72, align 4, !tbaa !11
  %not73 = xor i32 %42, -1
  %x74 = getelementptr inbounds %struct.ascon_state_t, ptr %state, i32 0, i32 0
  %arrayidx75 = getelementptr inbounds [5 x i32], ptr %x74, i32 0, i32 2
  store i32 %not73, ptr %arrayidx75, align 4, !tbaa !11
  call void @llvm.lifetime.end.p0(i64 4, ptr %t4) #2
  call void @llvm.lifetime.end.p0(i64 4, ptr %t3) #2
  call void @llvm.lifetime.end.p0(i64 4, ptr %t2) #2
  call void @llvm.lifetime.end.p0(i64 4, ptr %t1) #2
  call void @llvm.lifetime.end.p0(i64 4, ptr %t0) #2
  ret void
}

; Function Attrs: nocallback nofree nosync nounwind willreturn memory(argmem: readwrite)
declare void @llvm.lifetime.start.p0(i64 immarg, ptr nocapture) #1

; Function Attrs: nocallback nofree nosync nounwind willreturn memory(argmem: readwrite)
declare void @llvm.lifetime.end.p0(i64 immarg, ptr nocapture) #1

attributes #0 = { nounwind uwtable "no-trapping-math"="true" "stack-protector-buffer-size"="8" "target-cpu"="generic-rv32" "target-features"="+32bit,+a,+c,+d,+f,+m,+relax,-e,-experimental-zca,-experimental-zcb,-experimental-zcd,-experimental-zcf,-experimental-zfa,-experimental-zihintntl,-experimental-ztso,-experimental-zvfh,-h,-save-restore,-svinval,-svnapot,-svpbmt,-v,-xtheadba,-xtheadbb,-xtheadbs,-xtheadcmo,-xtheadcondmov,-xtheadfmemidx,-xtheadmac,-xtheadmemidx,-xtheadmempair,-xtheadsync,-xtheadvdot,-xventanacondops,-zawrs,-zba,-zbb,-zbc,-zbkb,-zbkc,-zbkx,-zbs,-zdinx,-zfh,-zfhmin,-zfinx,-zhinx,-zhinxmin,-zicbom,-zicbop,-zicboz,-zicsr,-zifencei,-zihintpause,-zk,-zkn,-zknd,-zkne,-zknh,-zkr,-zks,-zksed,-zksh,-zkt,-zmmul,-zve32f,-zve32x,-zve64d,-zve64f,-zve64x,-zvl1024b,-zvl128b,-zvl16384b,-zvl2048b,-zvl256b,-zvl32768b,-zvl32b,-zvl4096b,-zvl512b,-zvl64b,-zvl65536b,-zvl8192b" }
attributes #1 = { nocallback nofree nosync nounwind willreturn memory(argmem: readwrite) }
attributes #2 = { nounwind }

!llvm.module.flags = !{!0, !1, !2, !3, !4, !5}
!llvm.ident = !{!6}

!0 = !{i32 1, !"wchar_size", i32 4}
!1 = !{i32 1, !"target-abi", !"ilp32d"}
!2 = !{i32 8, !"PIC Level", i32 2}
!3 = !{i32 7, !"PIE Level", i32 2}
!4 = !{i32 7, !"uwtable", i32 2}
!5 = !{i32 8, !"SmallDataLimit", i32 8}
!6 = !{!"clang version 17.0.0 (git@github.com:Eymay/llvm-project.git cf23cb0fcdfc70aa489332bb12056e53d1385ea4)"}
!7 = !{!8, !8, i64 0}
!8 = !{!"any pointer", !9, i64 0}
!9 = !{!"omnipotent char", !10, i64 0}
!10 = !{!"Simple C/C++ TBAA"}
!11 = !{!12, !12, i64 0}
!12 = !{!"int", !9, i64 0}
\end{lstlisting}

\newpage

\section*{APPENDIX A.3}
\vglue6pt
\renewcommand{\theequation}{A.2.\arabic{equation}}
\setcounter{equation}{0}

\section{Creating Assembly File From C File}

LLVM consists of many flexible libraries which allows the users to use different libraries with their preferred  options. To create RISC-V assembly from c code, Clang and LLC are used with the following commands. \\

Clang is the C compiler front-end which is mainly used with the LLVM back-end. Clang is used in this project to produce LLVM IR code. The following command produces a .ll file in the current directory. 

\begin{lstlisting}[language=Bash]
-clang -S -target riscv32-linux-gnu -emit-llvm foo.c
\end{lstlisting}

-S option provides to run only preprocess and compilation steps. \\
-target option specifies the 32-bit RISC-V target architecture. \\
-emit-llvm is for targeting the LLVM back-end. \\

LLC is the LLVM compiler back-end which converts LLVM IR into native machine code for a specific target architecture. The following command produces a .s file for RISC-V architecture in the current directory. 

\begin{lstlisting}[language=Bash]
llvm-project/build/bin/llc -debug-only=isel -view-isel-dags -mtriple=riscv32 lxr.ll
\end{lstlisting}

-debug-only=isel option gives the debug information during the DAG lowering process.\\
-view-isel-dags option prints the DAG image of the IR code.\\
-view-sched-dags option can be used instead of -view-isel-dags, if the non-scheduled DAG wants to be shown.\\
-mtriple=riscv32 defines the 32-bit RISC-V target architecture.\\
\newpage

\section*{APPENDIX A.4}
\vglue6pt
\renewcommand{\theequation}{A.2.\arabic{equation}}
\setcounter{equation}{0}
\section{Adding the Crypt extension to the LLVM}
To add our extension to LLVM, a new file named RISCVInstrInfoCrypt.td should be created in ../llvm-project/llvm/lib/Target path and code \ref{lst:crypt_file} should be pasted in this file.

\begin{lstlisting}[caption={RISCVInstrInfoCrypt.td file},label={lst:crypt_file}]
	// Instruction class templates
	

	let hasSideEffects = 0, mayLoad = 0, mayStore = 0 in
	class ALU_rrr<bits<2> funct2, bits<3> funct3, string opcodestr,
				 bit Commutable = 0>
		: RVInstR4<funct2, funct3, OPC_OP, (outs GPR:$rd), (ins GPR:$rs1, GPR:$rs2, GPR:$rs3),
				  opcodestr, "$rd, $rs1, $rs2 ,$rs3"> {
	  let isCommutable = Commutable;
	}
	
	// Instructions
	
	
	def MLA     : ALU_rrr<0b10, 0b100, "mac">,
				  Sched<[WriteIMul, ReadIMul, ReadIMul]>;
	
	
	def NAXOR     : ALU_rrr<0b11, 0b100, "naxor">,
				  Sched<[WriteIMul, ReadIMul, ReadIMul]>;
	
	
	def SHLXOR  : ALU_rr<0b0011000, 0b111, "shlxor">,
			   Sched<[WriteIALU, ReadIALU, ReadIALU]>;
	
	
	let mayLoad = 1 in{
	def LXR  : ALU_rr<0b0011011, 0b101, "lxr">,
			   Sched<[WriteIALU, ReadIALU, ReadIALU]>;
	}
	
	// Instruction infos
	
	
	def : Pat<  (add (mul GPR:$src1, GPR:$src2), GPR:$src3),
				(MLA GPR:$src1, GPR:$src2, GPR:$src3)>;
	
	def : Pat<  (xor (and (not GPR:$src1), GPR:$src2), GPR:$src3),
				(NAXOR GPR:$src1, GPR:$src2, GPR:$src3)>;
	
	def : Pat<  (xor (shl GPR:$src1, (i32 1)), GPR:$src2),
				(SHLXOR GPR:$src1, GPR:$src2)>;
	
	def : Pat<  (xor (load GPR:$rs1),(load GPR:$rs2)),
				(LXR GPR:$rs1,GPR:$rs2)>;
	
\end{lstlisting}

After the file is created properly, code \ref{lst:includecrypt} should be added at the end of the ../llvm-project/llvm/lib/Target/RISCV/RISCVInstrInfo.td file.
\begin{lstlisting}[caption={Include line},label={lst:includecrypt}]
include "RISCVInstrInfoCrypt.td"
\end{lstlisting}
The new extension will be ready to use after building the LLVM.
\newpage

\ozgecmisbir{\vspace*{10mm}
\setlength{\TPHorizModule}{10pt}
\setlength{\TPVertModule}{10pt}
\begin{textblock}{1}(43.25,15.75) %
	\begin{figure}[p]
		\includegraphics[scale=0.05,keepaspectratio=true]{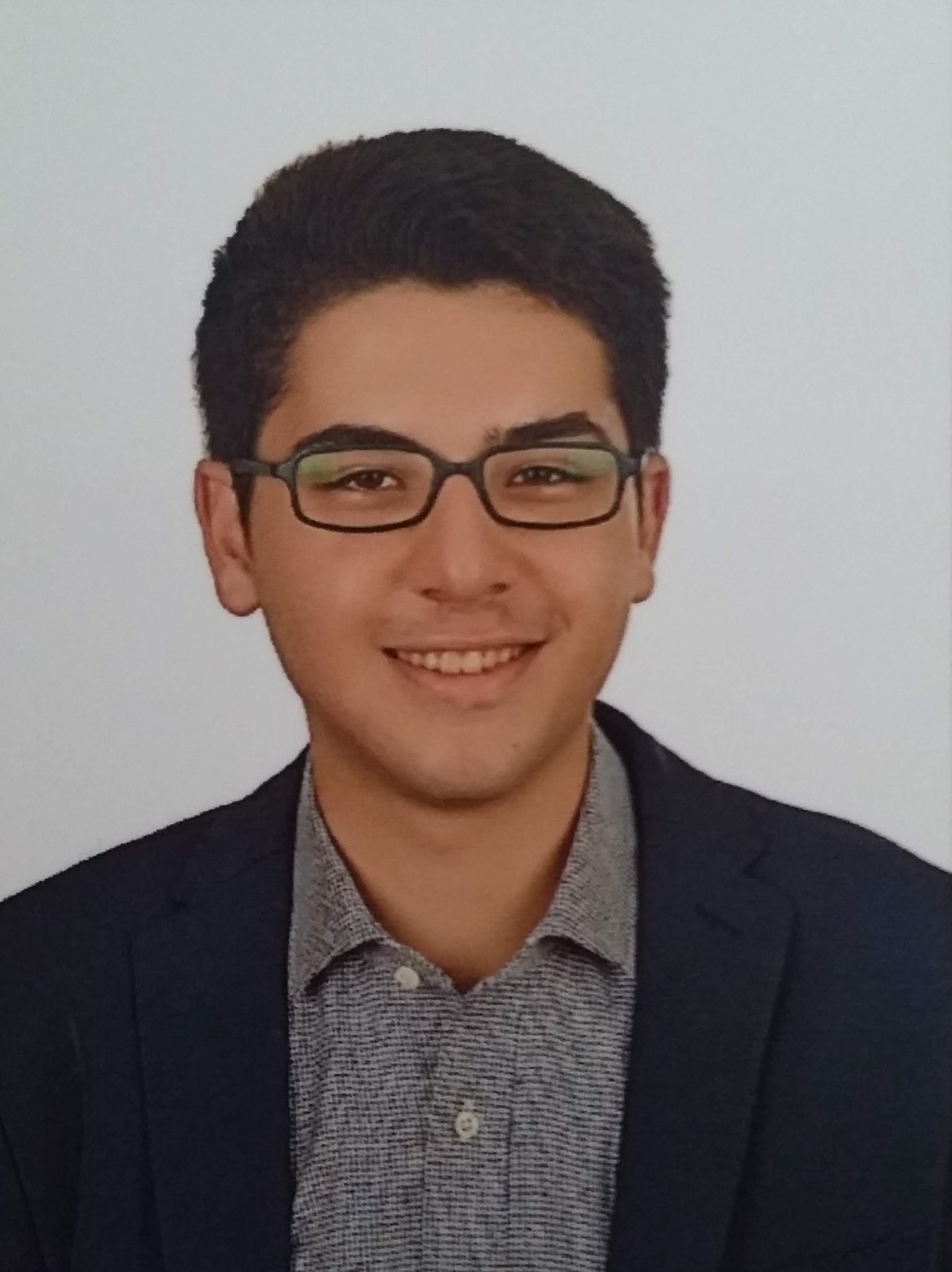}
	\end{figure}	
\end{textblock}
\vspace*{20mm}
\textbf{\makebox[4cm]{Name Surname}\makebox[1.6cm]{\hfill \textbf{:}}}\hspace{0.225em} Mehmet Eymen Ünay \\ %

\textbf{Place and Date of Birth\makebox[0.735cm]{\hfill \textbf{:}}}\hspace{0.225em} Manchester / 18.05.2000 \\ %

\textbf{E-Mail\makebox[3.685cm]{\hfill \textbf{:}}}\hspace{0.225em} eymenunay@outlook.com \\ %

\vspace{5mm}

\renewcommand\labelitemi{\normalsize$\bullet$} 			%

\textbf{EDUCATION\makebox[2.41cm]{\hfill \textbf{:}}}  	%
\vspace{-3mm}

\begin{itemize}[leftmargin=5.15cm,itemsep=-0.25em,labelsep=2mm] %
	\item [$\bullet$ \hspace{1em}\textbf{B.Sc.} \hspace{6.85em} \textbf{:}] 2023, Istanbul Technical University, Faculty of Electrical and Electronics, Department of Electronics and Communication Engineering
	\item [$\bullet$ \hspace{1em}\textbf{B.Sc.} \hspace{6.85em} \textbf{:}] 2024, Istanbul Technical University, Faculty of Computer and Informatics Engineering, Department of Computer Engineering
\end{itemize}

\textbf{PROFESSIONAL EXPERIENCE AND REWARDS:}   
\vspace{-3mm}
\begin{itemize}[leftmargin=0.7cm,itemsep=-0.25em,labelsep=5mm] %
	\item 2021 Summer Intern at Altınay Robotics
	\item 2022 Summer Intern at TÜBİTAK BİLGEM
	\item since then Research Scholar at TÜBİTAK BİLGEM
\end{itemize}

} 
\ozgecmisiki{\vspace*{10mm}
\setlength{\TPHorizModule}{10pt}
\setlength{\TPVertModule}{10pt}
\begin{textblock}{1}(39,15.75) %
	\begin{figure}[p]
		\includegraphics[scale=0.2,keepaspectratio=true]{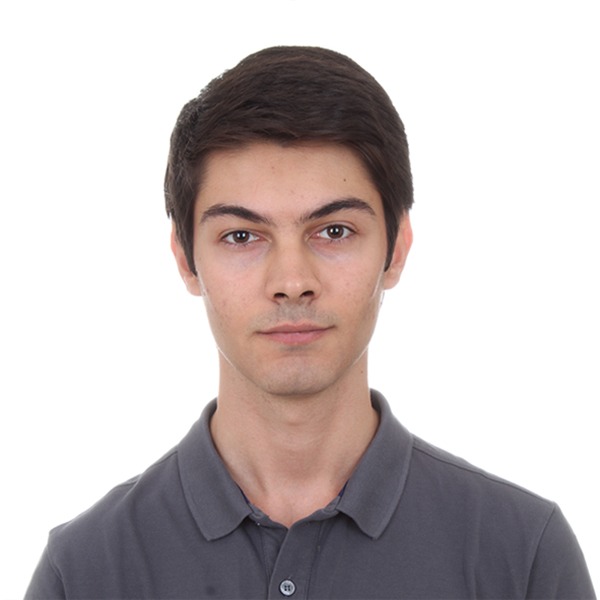}
	\end{figure}	
\end{textblock}
\vspace*{20mm}
\textbf{\makebox[4cm]{Name Surname}\makebox[1.6cm]{\hfill \textbf{:}}}\hspace{0.225em} Bora İnan \\ %

\textbf{Place and Date of Birth\makebox[0.735cm]{\hfill \textbf{:}}}\hspace{0.225em} Istanbul / 07.02.2000\\ %

\textbf{E-Mail\makebox[3.685cm]{\hfill \textbf{:}}}\hspace{0.225em} borainan0@gmail.com\\ %

\vspace{5mm}

\renewcommand\labelitemi{\normalsize$\bullet$} 			%

\textbf{EDUCATION\makebox[2.41cm]{\hfill \textbf{:}}}  	%
\vspace{-3mm}

\begin{itemize}[leftmargin=5.15cm,itemsep=-0.25em,labelsep=2mm] %
	\item [$\bullet$ \hspace{1em}\textbf{B.Sc.} \hspace{6.85em} \textbf{:}] 2023, Istanbul Technical University, Faculty of Electrical and Electronics, Department of Electronics and Communication Engineering
\end{itemize}

\textbf{PROFESSIONAL EXPERIENCE AND REWARDS:}   
\vspace{-3mm}
\begin{itemize}[leftmargin=0.7cm,itemsep=-0.25em,labelsep=5mm] %
	\item 24.06.2022-29.07.2022 Internship at TUSAŞ
	\item 08.08.2022-13.09.2022 Internship at ASELSAN
	\item 15.03.2023-currently Part-time working student at TUSAŞ
\end{itemize}

} 
\ozgecmisuc{\vspace*{10mm}
\setlength{\TPHorizModule}{10pt}
\setlength{\TPVertModule}{10pt}
\begin{textblock}{1}(43.25,15.75) %
	\begin{figure}[p]
		\includegraphics[scale=0.3,keepaspectratio=true]{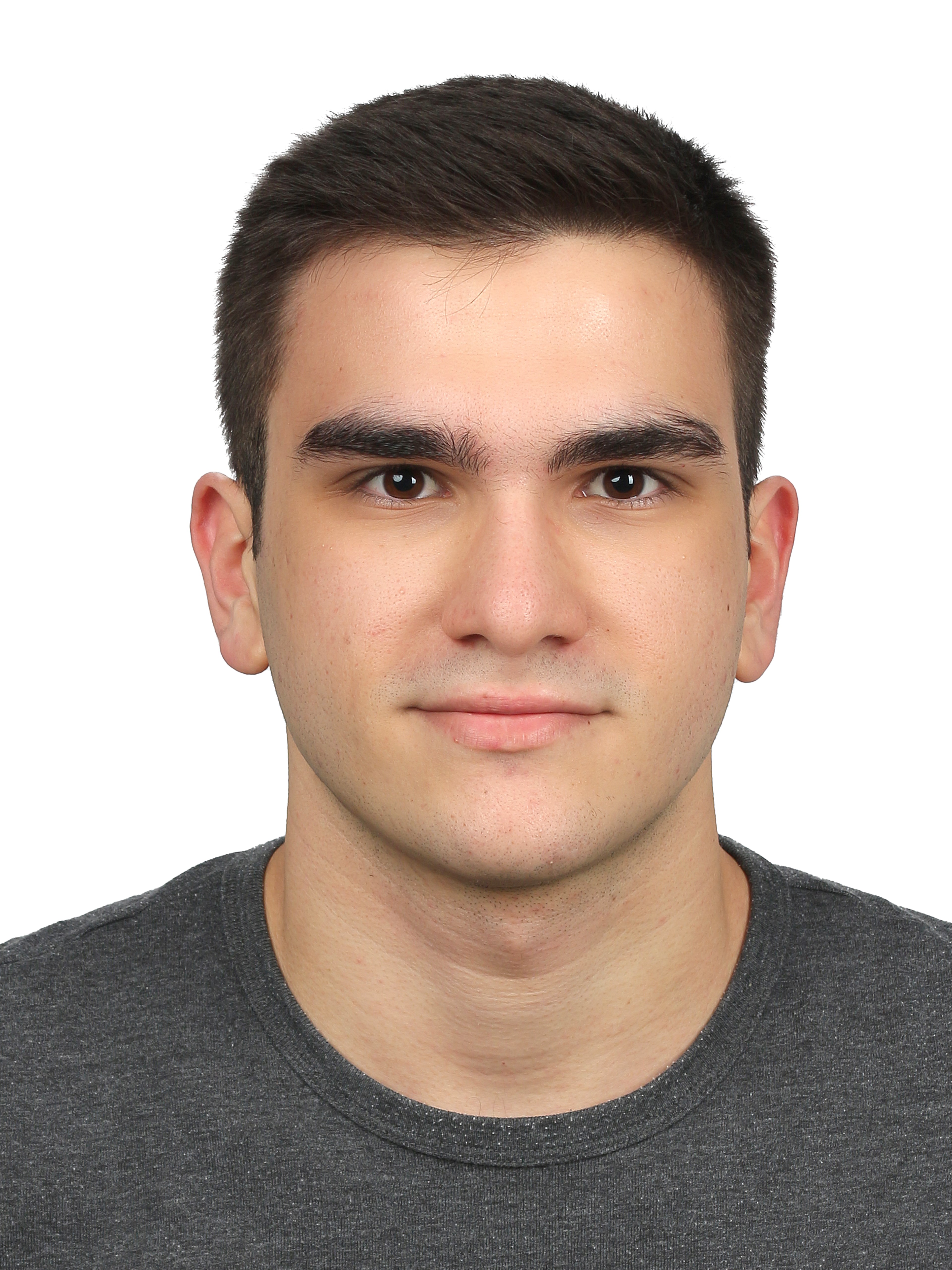}
	\end{figure}	
\end{textblock}
\vspace*{20mm}
\textbf{\makebox[4cm]{Name Surname}\makebox[1.6cm]{\hfill \textbf{:}}}\hspace{0.225em} Emrecan Yiğit \\ %

\textbf{Place and Date of Birth\makebox[0.735cm]{\hfill \textbf{:}}}\hspace{0.225em} Istanbul / 06.05.2002  \\ %

\textbf{E-Mail\makebox[3.685cm]{\hfill \textbf{:}}}\hspace{0.225em} emrecanyigit11@gmail.com \\ %

\vspace{5mm}

\renewcommand\labelitemi{\normalsize$\bullet$} 			%

\textbf{EDUCATION\makebox[2.41cm]{\hfill \textbf{:}}}  	%
\vspace{-3mm}

\begin{itemize}[leftmargin=5.15cm,itemsep=-0.25em,labelsep=2mm] %
	\item [$\bullet$ \hspace{1em}\textbf{B.Sc.} \hspace{6.85em} \textbf{:}] 2023, Istanbul Technical University, Faculty of Electrical and Electronics, Department of Electrical Engineering
\end{itemize}

\textbf{PROFESSIONAL EXPERIENCE AND REWARDS:}   
\vspace{-3mm}
\begin{itemize}[leftmargin=0.7cm,itemsep=-0.25em,labelsep=5mm] %
	\item 20.06.2022-31.12.2022 Internship at Bogazici University smart and autonomous laboratory
	\item 08.05.2023-currently long term intern at Renesas Electronics
\end{itemize}

\vspace{-3mm}

} 

\edge{\newpage %
\thispagestyle{empty} %
\def\sirtyili{2023} %
\def\studentname{} %
\def\thesisthickness{25mm} %

\hspace*{75mm}
\begin{tikzpicture}[remember picture,overlay]
{\rotatebox[origin=c,x=23.35mm,y=-247.75mm]{90}{\draw [line width=0.01mm, black, dashed] (0mm,0mm) rectangle node{\normalsize \studentname} (65mm,\thesisthickness);}}

{\rotatebox[origin=c,x=23.35mm,y=-247.75mm]{90}{\draw [line width=0.01mm, black ,dashed, text width=190mm, align=center] (67mm,0mm) rectangle node{\normalsize \Baslikspacing \Baslikbir~\Baslikiki~\Baslikuc} (193mm+65mm+2mm,\thesisthickness);}}

{\rotatebox[origin=c,x=23.35mm,y=-247.75mm]{90}{\draw [line width=0.01mm, black ,dashed] (193mm+65mm+4mm,0mm) rectangle node{\normalsize \sirtyili} (296.5mm,\thesisthickness);}}

{\rotatebox[origin=c,x=23.35mm,y=-247.75mm]{90}{\draw[black,line width=1mm] (64.5mm,0mm) -- (64.5mm,\thesisthickness);
\draw[black,line width=1mm] (67.3mm,0mm) -- (67.3mm,\thesisthickness);
}}

{\rotatebox[origin=c,x=23.35mm,y=-247.75mm]{90}{\draw[black,line width=1mm] (193mm+64.5mm+2mm,0mm) -- (193mm+64.5mm+2mm,\thesisthickness);
\draw[black,line width=1mm] (193mm+64.5mm+5mm,0mm) -- (193mm+64.5mm+5mm,\thesisthickness);
}}
\draw [line width=0.01mm, black, dashed] (0mm,-205mm) -- (0mm,-208mm);
\draw [line width=0.01mm, black, dashed] (-\thesisthickness,-205mm) -- (-\thesisthickness,-208mm);
\draw [line width=0.01mm, black, dashed] (0mm,-3.25mm) -- (0mm,-13mm);
\draw [line width=0.01mm, black, dashed] (-\thesisthickness,-1mm) -- (-\thesisthickness,-13mm);
\end{tikzpicture}
}%

\end{document}